\documentclass[12pt,a4paper]{article}
\usepackage{jheppub}
\usepackage{chemarrow}
\usepackage{amsmath,amssymb,amsfonts,mathtools}
\usepackage{float}
\usepackage{subfig}
\usepackage{scalefnt}
\usepackage{wrapfig}
\usepackage{fancybox}
\usepackage{ifsym}
\newcommand{\bea}{\begin{eqnarray}\displaystyle}
\newcommand{\eea}{\end{eqnarray}}
\newcommand{\nn}{\nonumber}
\newlength{\arrow}
\newcommand{\figref}[1]{Fig.~\protect\ref{#1}}
\newcommand{\sq}[2]{\mbox{{\raisebox{.06in}{$#1$}}}\underset{\mbox{$#2$}}{\mbox{\LARGE{$\square$}}}}
{\setlength{\fboxsep}{15pt}
\setlength{\mylength}{\linewidth}%
\addtolength{\mylength}{-2\fboxsep}%
\addtolength{\mylength}{-2\fboxrule}%
\Sbox
\minipage{\mylength}%
\setlength{\abovedisplayskip}{0pt}%
\setlength{\belowdisplayskip}{0pt}%
\equation}%
{\endequation\endminipage\endSbox
\[\fbox{\TheSbox}\]}

\begin{document}
\title{M-Strings}
\author[\ast]{Babak Haghighat,}
\author[\S]{Amer Iqbal,}
\author[\dag]{Can Koz\c{c}az,}
\author[\ast]{Guglielmo Lockhart,}
\author[\ast]{and \\Cumrun Vafa}
\affiliation[\ast]{Jefferson Physical Laboratory, Harvard University, Cambridge, MA 02138, USA}

\affiliation[\S]{Department of Physics, LUMS School of Science \& Engineering, U-Block, D.H.A, Lahore, Pakistan.}
\affiliation[\S]{Department of Mathematics, LUMS School of Science \& Engineering, U-Block, D.H.A, Lahore, Pakistan.}

\affiliation[\dag]{International School of Advanced Studies (SISSA),
via Bonomea 265, 34136 Trieste, Italy and INFN, Sezione di Trieste}

\abstract{M2 branes suspended between adjacent parallel M5 branes lead to light strings, the `M-strings'.
In this paper we compute the elliptic genus of M-strings, twisted by maximally allowed symmetries
that preserve 2$d$ $(2,0)$ supersymmetry.  In a codimension one subspace of parameters this reduces to the elliptic
genus of the $(4,4)$ supersymmetric $A_{n-1}$ quiver theory in 2$d$.
 We contrast the elliptic genus of $N$  M-strings with the $(4,4)$ sigma model on the $N$-fold symmetric product of ${\mathbb R}^4$.  For $N=1$ they are the same,
but for $N>1$ they are close, but not identical.  Instead the elliptic genus of $(4,4)$ $N$ M-strings is the same as the elliptic genus of $(4,0)$ sigma
models on the $N$-fold symmetric product of $\mathbb{R}^4$, but where the right-moving fermions couple to a modification of the tangent bundle.
This construction arises from a dual $A_{n-1}$ quiver 6$d$ gauge theory with $U(1)$ gauge groups.
 Moreover we compute the elliptic genus of domain walls which separate different numbers of M2 branes on the two sides of the wall.}

\maketitle

\section{Introduction}
The SCFTs with the maximal amount of supersymmetry in the highest dimension are the (2,0)
theories in $d=6$.  Despite the unique status they enjoy, and despite
the fact that they have been instrumental in constructing lower dimensional
theories, they remain among the least understood theories.  This is mainly
related to the fact that we do not have a Lagrangian description of these theories.
Moreover, if we go slightly away from the conformal point
we get a theory of interacting almost tensionless strings.   Clearly a deeper
understanding of these strings is called for.  One goal of the present
paper is to take a step in this direction.  In particular we focus
on the SCFT arising from $N$ coincident M5 branes, and study the M2 branes
suspended between the M5 branes when we separate them, which leads to strings
on their boundaries.  We will call these strings `M-strings', as they involve basic
M-theory ingredients for their definition.

If we consider two parallel M5 branes, and consider one M2 brane suspended between them
clearly the moduli space of the M2 brane is labeled by its transverse position on the M5 brane it ends on,
which is a copy of ${\mathbb R}^4$.  So at least the IR degrees of freedom on this string should
correspond to the $(4,4)$ supersymmetric sigma model on ${\mathbb R}^4$.  Moreover if we
consider $n$ M2 branes stretched between 2 M5 branes, one would naively expect
the IR degrees of freedom to correspond to the choice of $n$ points on ${\mathbb R}^4$, modulo
the action of the permutation group on the points, i.e. to a $(4,4)$ supersymmetric sigma model on
$$\mbox{Sym}^n({\mathbb R}^4)=({\mathbb R}^4)^n/S_n.$$
This space is singular and one can ask whether the target space is smoothed
out at coincident points.  If the target space is smoothed out, as in the Hilbert scheme of $n$-points
on $\mathbb{R}^4$, then this would give
us an effective way to compute at least supersymmetry protected quantities for this theory. However, as argued in a related context in \cite{Witten:1997yu} this is not necessarily the case (not even the B-field
on the vanishing $\mathbb{P}^1$'s is turned on as in the orbifold points), and one expects that the relevant theory should be the one corresponding to the singular target space, which is infinitely far
away from the smoothed out points.  This in particular raises
the question of whether at least for BPS quantities one may use the smoothed out target space
to perform such computations.  A surprising result we find is that this is not possible for $n>1$.
Instead we find a related sigma model with $(4,0)$ supersymmetry on the smoothed out space
(the Hilbert scheme of n-points on ${\mathbb{C}}^2$)
which has the same elliptic genus as the suspended M2 branes.  The right-moving fermions
couple, instead of the tangent bundle, to a bundle $V=E\oplus E^*$ where $E$ is the tautological bundle on the Hilbert scheme.

From the viewpoint of the M2 brane worldvolume theory, ending on an M5
brane corresponds to a boundary condition on the theory \cite{Berman:2009xd}, as is
familiar in the context of D-branes.  More generally we will be considering
a number $N_L$ of M2 branes suspended on an M5 brane from the left and a
number $N_R$ of M2 branes suspended from the right.  This can be viewed
as a domain wall which separates $(N_L,N_R)$ M2 branes.   In addition we need to choose
a vacuum for each M2 brane, which in turn is labeled by the
partition $\nu_L$ of $N_L$ for the left-vacuum and $\nu_R$ of $N_R$ for
the right vacuum \cite{Lin:2004nb,Gomis:2008vc, Kim:2010mr} . Thus the theory living on the $2d$ domain wall is labeled
by $D_{\nu_L\,\nu_R}$.
One main computational result of this paper is the supersymmetric
partition function of the theory $D_{\nu_L\, \nu_R}$ on $T^2$.
More precisely we consider the elliptic genus of this theory, including
twisting by maximal allowed symmetries consistent with $(2,0)$ supersymmetry as we go around the cycles of $T^2$.
In some limit (turning off some chemical potentials corresponding to turning off the `mass for the adjoint') this computation can also be viewed, using a dual type IIB description,
as elliptic genus of $A_N$ quiver ${\cal N}=(4,4)$ supersymmetric theories in $d=2$, which
can be computed using the recent works \cite{Benini:2013nda,Gadde:2013dda}.  We check our computations
against these results in this limit and find agreement.

The main tool we use is the relation between the refined topological
string partition function and the degeneracy of BPS states in 5$d$ \cite{Gopakumar:1998ii,Gopakumar:1998jq,Hollowood:2003cv,Iqbal:2007ii}, which
we apply to the M5 brane SCFT
compactified on $S^1$.   Apart from the Kaluza-Klein reduction of the 6$d$ mode, the suspended M2 branes wrapped around $S^1$ are the only other states contributing to BPS states, and we are thus able
to extract the partition function, or more precisely the elliptic genus of the strings obtained
from suspended M2 branes.
 Reversing this, one can recover the full
refined topological string partition function in terms of the elliptic
genera of the M-strings.  This in turn can be used to compute
the index of M5 branes \cite{Lockhart:2012vp,Kim:2012qf}.

The organization of this paper is as follows:  In section 2 we review the
relation between M5 brane CFT and ${\cal N}=2^{*}$ supersymmetric Yang-Mills
in 5$d$, and toric realization of it.  In section 3 we show how to use this
setup to compute the partition function of it on twisted $S^1\times \mathbb{R}^4$
(including modular properties and symmetries) using refined topological
strings as well as the instanton calculus.  Moreover we explain the relation between the partition function
for the $U(N)$ case and the BPS degeneracies.  We compute this using two dual
5$d$ theories:  one in terms of the ${\cal N}=2^{*}$ $U(N)$ theory, and the other in terms
of a dual $A_{N-1}$ quiver theory in 6 dimensions with $U(1)^{N-1}$ gauge group.
This latter perspective turns out to be particularly important for our purposes.
In section 4 we contrast some expectations of BPS degeneracies based on
generalities about suspended M2 branes with the actual results we obtain using
the topological strings.  We interpret our computations and explain what they
tell us about M-strings, including their relation to elliptic genus of quiver theories.  Furthermore we interpret our results as leading to the partition function of
domain walls separating M2 branes.  Moreover we discuss the fact that new
additional bound states between M-strings arise when we
compactify the M5 brane theory on the circle, which cannot be
viewed as bound states before compactification.  Furthermore, we show that in a particular
limit (where the mass term is turned off) the result agrees with that of the dual type IIB description
involving the elliptic genus of $(4,4)$ quiver theories.  In section
5 we explain the relation of our results with the computation
of the superconformal index for $N$ coincident M5 branes (i.e. the partition
function on $S^1\times S^5$) as well as their partition function on $T^2\times S^4$,
which can be viewed as the partition function of a quantum deformation of $A_{N-1}$ Toda
theories on $T^2$.  In section 6 we conclude by suggesting some directions for future research.
Some technical aspects of the computations are discussed in appendices A, B, C, and D.


\section{Parallel M5 branes on $S^1$ and $S^1\times S^1$ and Suspended M2 Branes}
In this section we discuss some general aspects of parallel M5 branes including their
twisted compactifications on $S^1$ and $S^1\times S^1$.  The twisted compactification
on $S^1$ leads to a theory with the same IR degrees of freedom as ${\cal N}=2^*$ in 5 dimensions,
where the mass of the adjoint field is given by the twist parameter.  The further compactification
on the circle can be used to twist the left-over 4 dimensions of the M5 brane.
  We also discuss
general aspects of M2 branes suspended between the parallel M5 branes.  Furthermore
we discuss various dualities which map this to related systems, and in particular to compactifications
of M-theory on elliptically fibered geometries, which we will use in the following
section to compute the partition function of M5 branes using the refined topological strings.

\subsection{Basics of M-strings}

Consider $N$ parallel and coincident $M5$ branes.  This is believed to lead to a $(2,0)$
superconformal theory in six dimensions usually called the $(2,0)$ $A_{N-1}$ theory.
The choice for the terminology is because the same system is believed to arise when
considering type IIB string theory in the presence of $A_{N-1}$ singularities.  The latter
viewpoint generalizes it to the $D$ and $E$ versions of the $(2,0)$ theory.

This theory has $Osp(2,6|4)$ as the superconformal group whose bosonic part is
\bea
Spin(2,6)\times Spin(5)\subset Osp(2,6|4)\,.
\eea
$Spin_{R}(5)$ is the global R-symmetry group of this theory and is the double cover of $SO(5)$ which is the rotation group of the space transverse to the M5 branes.

On the worldvolume of a single M5 brane we have the tensor multiplet of the $(2,0)$ theory which consists of:\\
$\bullet$ an antisymmetric 2-form $B$ such that its field strength $H=dB$ is self-dual, $\star H=H$\\
$\bullet$ four symplectic Majorana-Weyl fermions in the $({\bf 4},{\bf 4})$ of $Spin(1,5)\times Spin_{R}(5)$\\
$\bullet$ five scalar fields giving the transverse fluctuations of the M5 brane.

If we compactify the six dimensional $(2,0)$ theory described above on a circle it gives a theory with 16 real supercharges in five dimensions, the ${\cal N}=2$ Super Yang-Mills in five dimensions. Since we will be discussing the M2 branes suspended between M5 branes let us fix the worldvolume and the transverse directions of the M5/M2 branes. We denote the coordinates of $\mathbb{R}^{1,10}$ as $X^{I}$, $I=0,1,2,\cdots 10$, then
\begin{gather*}\nn
\mbox{\small \bf The worldvolume of coincident M5 branes has coordinates}\\ \nn X^{0},X^{1},X^{2},X^{3},X^{4},X^{5}
\end{gather*}
The space transverse to the coincident M5 brane worldvolume is $\mathbb{R}^5$ and is acted upon by the R-symmetry group $Spin_{R}(5)$. We can pick a direction in $\mathbb{R}^5$ and separate the coincident M5 branes along this direction. We choose the $X^{6}$ coordinate to separate the branes. This breaks the global $Spin_{R}(5)$ symmetry to $Spin_{R}(4)$ acting on the coordinates $X^{7},X^{8},X^{9},X^{10}$. It is important to note that $Spin_{R}(4)$ does not act on the M5 brane worldvolume coordinates. For later convenience we denote the position of the M5 branes in the $X^6$ direction as $a_{i}$, $i=1,2,\cdots ,N$.

We can now introduce M2 branes ending on M5 branes with the boundary of the M2 brane inside the M5 brane coupling to the 2-form $B$. We can introduce multiple M2 branes for each pair of M5 branes extending in the $X^{6}$ direction. We consider the worldvolume of M2 branes such that
\begin{gather*}\nn
\mbox{\small \bf The worldvolume of an M2 brane suspended between $(i,j)$ M5 branes}\\\nn
X^{0},X^{1},X^{6}\,\,\,\mbox{with}\,\,\,\,\,a_{i}\leq X^{6}\leq a_{j}\,
\end{gather*}
The boundary of the M2 brane given by the coordinates $(X^{0},X^{1})$ is a string inside the M5 brane, which we
call the M-string. The presence of this string breaks the M5 brane worldvolume Lorentz group $Spin(1,5)$ to $Spin(1,1)\times Spin(4)$, where $Spin(1,1)$ is the Lorentz group on the string and $Spin(4)$ acts on the space transverse to the string inside the M5 brane.

From our choice of the worldvolume coordinates of the M5/M2 branes and the string it is easy to see that the supersymmetries preserved by the string are given by
\bea
\Gamma^{016}\epsilon=\epsilon\,,\,\,\,\Gamma^{012345}\epsilon=\epsilon\,,
\eea
where $\epsilon$ is the 32-component spinor, $\Gamma^{I_{1}I_{2}\cdots I_{k}}=\Gamma^{I_{1}}\Gamma^{I_{2}}\cdots\Gamma^{I_{k}}$ and $\Gamma^{I}$ are the $32\times 32$ eleven dimensional Gamma matrices. Since in eleven dimensions $\Gamma^{0}\Gamma^{1}\Gamma^{2}\cdots \Gamma^{10}=1$ the above two conditions imply that
\bea\label{susy}
\Gamma^{2345}\epsilon=\Gamma^{01}\epsilon\,,\,\,\,\,\,\Gamma^{789(10)}\epsilon=\Gamma^{01}\epsilon\,.
\eea
Hence the chirality under $Spin(1,1)$, chirality under $Spin_{R}(4)$ and chirality under $Spin(4)\subset Spin(1,5)$ of the preserved supersymmetries on the string are the same. Since the M2/M5 brane configuration breaks $\frac{1}{4}$ of the 32 supersymmetries therefore on the string world sheet we have a $(p,q)$ supersymmetric theory with $p+q=8$. By taking a specific form of the eleven dimensional Gamma matrices it is easy to show that the theory on the string has $(4,4)$ supersymmetry. It then follows from Eq.(\ref{susy}) that preserved supercharges $Q^{\dot{\alpha}\,\dot{a}}_{-\frac{1}{2}}$ and $Q^{\alpha\,a}_{+\frac{1}{2}}$ where $\alpha,\dot{\alpha}=1,2$ denote the chiral/antichiral spinor of $Spin_{R}(4)$ and $a,\dot{a}=1,2$ denote the chiral/anti-chiral spinor of $Spin(4)\subset Spin(1,5)$ are in the representation,
\bea\label{super2}
({\bf 2},{\bf 1},{\bf 2},{\bf 1})_{+\frac{1}{2}}\oplus ({\bf 1},{\bf 2},{\bf 1},{\bf 2})_{-\frac{1}{2}}
\eea
of $Spin(4)\times Spin_{R}(4)\times Spin(1,1)$. The $\pm\frac{1}{2}$ denote the chirality with respect to $Spin(1,1)$.

The above supercharges can be organized in terms of representations of $Spin(8)\supset Spin(4)\times Spin_{R}(4)$ as well and it will be useful for later purposes to do so. Consider a number of coincident M2 branes in $\mathbb{R}^{1,10}$ with worldvolume along $X^{0},X^{1}$ and $X^{6}$. Then the transverse space is $\mathbb{R}^{8}$ and the global symmetry of the theory on the M2 branes is given by $Spin(8)$. Now introducing M5 branes, separated along $X^6$ as before and M2 branes ending on them, breaks $Spin(8)$ to $Spin(4)\times Spin_{R}(4)$. Notice that the preserved supercharges form a positive chirality spinor of $Spin(8)$, i.e. they are in ${\bf 8}_{s}$. The chirality for $Spin(8)$ is determined by $\Gamma_{9}\equiv \Gamma^{2}\Gamma^{3}\Gamma^{4}\Gamma^{5}\Gamma^{7}\Gamma^{8}\Gamma^{9}\Gamma^{10}$ and therefore it follows from Eq.(\ref{susy}) that
\bea
\Gamma_{9}\epsilon=(\Gamma^{01})^2\epsilon=\epsilon,
\eea
and hence preserved supersymmetries form a positive chirality spinor of $Spin(8)$. If we denote by $\alpha_{1}=e_{1}-e_{2},\alpha_{2}=e_{2}-e_{3},\alpha_{3}=e_{3}-e_{4}$ and $\alpha_{4}=e_{3}+e_{4}$ the simple roots of $Spin(8)$ then the $(4,4)$ supercharges are in ${\bf 8}_{s}$ with highest weight vector \footnote{The $Spin(4)\times Spin(4)_{R}$ subgroup of $Spin(8)$ mentioned above corresponds to the simple roots $\{e_{1}-e_{2},e_{1}+e_{2},e_{3}-e_{4},e_{3}+e_{4}$\}\,.}
\bea\nn
\frac{e_{1}+e_{2}+e_{3}+e_{4}}{2}\,.
\eea
The weight vectors for the $(4,4)$ supercharges are given by
\bea\label{super}
&&\scriptstyle{({\bf 1},{\bf 2},{\bf 1},{\bf 2})_{-\frac{1}{2}}:\,\frac{e_{1}+e_{2}+e_{3}+e_{4}}{2}\,,\frac{e_{1}+e_{2}-e_{3}-e_{4}}{2}\,,\frac{-e_{1}-e_{2}+e_{3}+e_{4}}{2}
\,,\frac{-e_{1}-e_{2}-e_{3}-e_{4}}{2}}\\\nn
&&\scriptstyle{({\bf 2},{\bf 1},{\bf 2},{\bf 1})_{+\frac{1}{2}}:\,\frac{e_{1}-e_{2}+e_{3}-e_{4}}{2}\,,\frac{e_{1}-e_{2}-e_{3}+e_{4}}{2}\,,
\frac{-e_{1}+e_{2}-e_{3}+e_{4}}{2}\,,\frac{-e_{1}+e_{2}+e_{3}-e_{4}}{2}}\,.
\eea

\subsection{Compactification on $S^1$}

Next, consider compactifying the M5 branes on a circle.
  Recall that
\begin{gather*}
\mbox{\small \bf The worldvolume of M5 branes has coordinates}\\ \nn X^{0},X^{1},X^{2},X^{3},X^{4},X^{5}
\end{gather*}
and that the M5 branes are separated in the $X^6$ direction. Now consider compactifying $X^1$ to a circle of radius $R_1$. More generally we can introduce
a partial breaking of the supersymmetry by making the $\mathbb{R}^4$ transverse to the M5 branes fibered over $S^1$. We will denote this $\mathbb{R}^4$ spanned by $(X^{7},X^{8},X^{9},X^{10})$ by  $\mathbb{R}_{\perp}^4$. In particular identifying $\mathbb{R}_{\perp}^{4}\simeq \mathbb{C}^2$ with coordinates $(w_{1},w_{2})$ and consider a rotation of the two complex planes as we go around the circle:
\bea
U(1)_{m}: \,\,\,(w_1,w_2)\rightarrow (e^{2\pi i m}\,w_1,e^{-2\pi i\,m}w_2)\,.
\eea
The resulting theory in 5$d$ is a mass deformation of the maximally supersymmetric Yang-Mills theory, by addition of a mass
term to the adjoint, which we have informally called the `${\cal N}=2^*$ theory in 5$d$' (borrowing the terminology from the
more familiar 4$d$ case).  The radius $R_1$ of the $S^1$ is identified with the gauge coupling of the Yang-Mills theory as follows
\begin{equation}
	R_1 = \frac{g_{YM}^2}{4\pi^2}.
\end{equation}
The 5$d$ theory has charged particles in its spectrum which carry instanton number which is identified
with the momentum around the $S^1$:
\begin{equation}
	\frac{k}{R_1} = -\frac{1}{8g_{YM}^2} \int d^4 x \ \textrm{tr}(F\wedge F ),
\end{equation}
From the point of view of the six-dimensional theory these particles arise as M-strings wrapped around the $S^1$.  If we consider $l$ M-strings wrapped around $S^1$ and carrying a momentum of $k$ units along $S^1$ its BPS mass is given by
\begin{equation} \label{eq:5dstates}
	M = l R_1 \delta_{ij} + \frac{k}{R_1}, \quad k, l \in \mathbb{Z},
\end{equation}
where $\delta_{ij}$ is the separation between
the M5 branes which gives the tension of the M-string stretched between $i$ and $j$ M5 branes.

We can also ask how the twisting by $m$ around $S^1$ affects the theory as seen by the M-string wrapped around $S^1$.  The $U(1)_{m}$ is embedded in the $SU(2)_{L}$ of $Spin_{R}(4)\subset Spin(8)$ of the M2 brane theory. It is easy to see that this choice of the $U(1)_{m}$ leaves the negative chirality supercharges of Eq.(\ref{super}) invariant but not the positive chirality ones. Hence the resulting theory has broken the $(4,4)\mapsto (4,0)$ supersymmetric theory on the worldsheet in the $X^{0},X^{1}$ directions.

\begin{figure}[here!]
  \centering
	\includegraphics[width=0.8\textwidth]{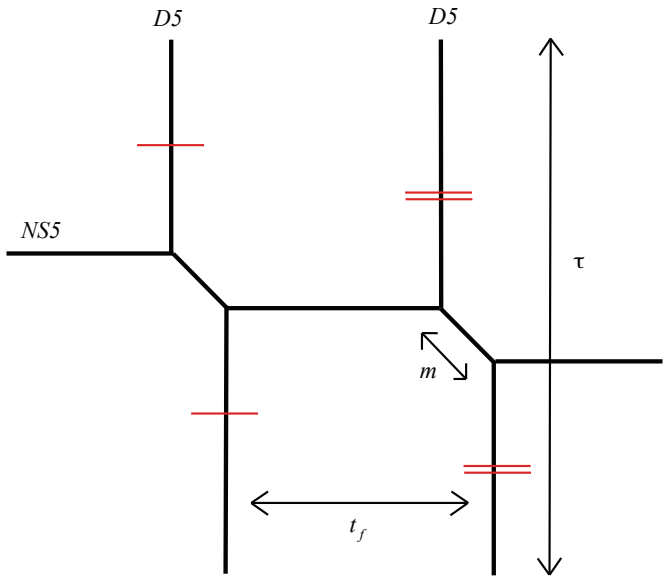}
  \caption{The brane and toric geometry. The red line marks mean to identify the toric legs or branes with each other and therefore describe a compactified direction which is associated to the gauge coupling $\tau$ in the gauge theory. The length of the $(1,1)$ branes is associated to the mass of the $\mathcal{N}=2^*$ theory. Last but not least the separation of the branes maps to the Coulomb branch parameter $t_f$ of the gauge theory.}
  \label{fig:toricg}
\end{figure}

As already mentioned we can view the 6$d$ theory as coming from type IIB theory with an $A_{N-1}$
singularity.  Compactifying this on a circle and using the duality between M-theory and type IIB
we can view this as compactification of M-theory on a threefold with geometry $T^2\times A_{N-1}$.
The duality between type IIB and M-theory identifies the K\"{a}hler class $t_e^M$ of $T^2$
with
$$t_e^M={1\over R_1}.$$
Moreover the twisting by the mass parameter can be viewed as blowing up ${\mathbb P}^1$ \cite{Hollowood:2003cv}.
This is the geometric analog of giving mass to the adjoint field in the brane construction \cite{Witten:1997sc}.
The blow up parameter $t_m^M$ is identified with
$$t_m^M={m\over R_1}.$$
The geometry of the blow-up is a local Calabi-Yau and is given by the periodic toric diagram \cite{Aganagic:2003db,Hollowood:2003cv} in \figref{fig:toricg}  where we have specialized to the case of the $U(2)$ theory which corresponds to two M5 branes. There is a dual description of the same system \cite{Leung:1997tw} in terms of the $(p,q)$ web of 5-branes
\cite{Aharony:1997bh}.  The picture is the same as the one of the toric diagram, only one has to associate the toric legs with branes of type IIB as is shown in \figref{fig:toricg}.

In the massless case, where one has the maximally supersymmetric gauge theory, the NS5 brane is extended along an $\mathbb{R}^6$ subspace of $\mathbb{R}^{10}$ while the D5 branes have the geometry $\mathbb{R}^5 \times S^1$ and intersect the NS5 branes transversally such that they have five dimensions in common. Then the gauge theory is living on the intersection and its rank is specified by the number of D5 branes. Note furthermore, that the  compactified direction of the D5 branes is perpendicular to the NS5 brane. The gauge theory is then living on the intersection of these branes. Now let us deform the theory by introducing mass as shown in \figref{fig:toricg}.  To simplify matters we will take the gauge group to be $U(2)$ for the moment. In this case the Calabi-Yau is the canonical bundle over a surface $D$ which is an elliptic fibration over $\mathbb{P}^1$, that is locally we have $D \cong T^2\times \mathbb{P}^1_f$. The torus arises from the compactified direction of the brane system with size $t_e^M$ and the size of the $\mathbb{P}^1_f$ is the Coulomb branch parameter of the gauge theory of size $t_f^M$ that is the separation of the D5-branes in the brane-picture.  This is related to the separation between the M5 branes (which is proportional
to the tension $\delta$ of the M2 brane string) times $R_1$:
$$t_f^M=R_1\cdot \delta$$
 Moreover, there is yet a third K\"ahler class coming from a singular elliptic fibre over the discriminant locus. The singular fibre is a degeneration of the $T^2$ into two spheres and thus adds another K\"ahler class corresponding to the size of one of the $\mathbb{P}^1$'s. This size determines the mass of the adjoint hypermultiplet in five dimensions, i.e. it is identified with $t_m^M$.

\subsection{Compactification on $S^1\times S^1$}
We can also consider a further compactification on another $S^1$ which we take to be the $X^{0}$ direction.
In trying to connect this geometry to topological string \cite{Gopakumar:1998ii,Gopakumar:1998jq,Hollowood:2003cv,Dijkgraaf:2006um} or $\Omega$-background
\cite{Nekrasov:2002qd} we fiber the space-time $\mathbb{R}^4$ over
this circle.  In other words we twist the  $\mathbb{R}^{4}\times \mathbb{R}^{4}_{\perp}$  by the action of $U(1)\times U(1)$ as we go around the
circle in the $X^0$ direction:
\bea
U(1)_{\epsilon_{1}}\times U(1)_{\epsilon_{2}}&:&\,(z_{1},z_{2})\mapsto (e^{2\pi i\,\epsilon_{1}}\,z_{1},e^{2\pi i\,\epsilon_{2}}\,z_{2})\,,\\\nn
&:& (w_{1},w_{2})\mapsto (e^{-\frac{\epsilon_{1}+\epsilon_{2}}{2}}\,w_{1},e^{-\frac{\epsilon_{1}+\epsilon_{2}}{2}}\,w_{2})
\eea
Note that in the unrefined case where $\epsilon_1+\epsilon_2=0$ to preserve the symmetry we do not need to rotate
$ \mathbb{R}^{4}_{\perp}$.

Again we can ask what the suspended M2 brane theory sees if it is wrapped around the $X^{0},X^{1}$ directions.
\begin{figure}[here!]
	\begin{center}
		\includegraphics[width=.5\textwidth]{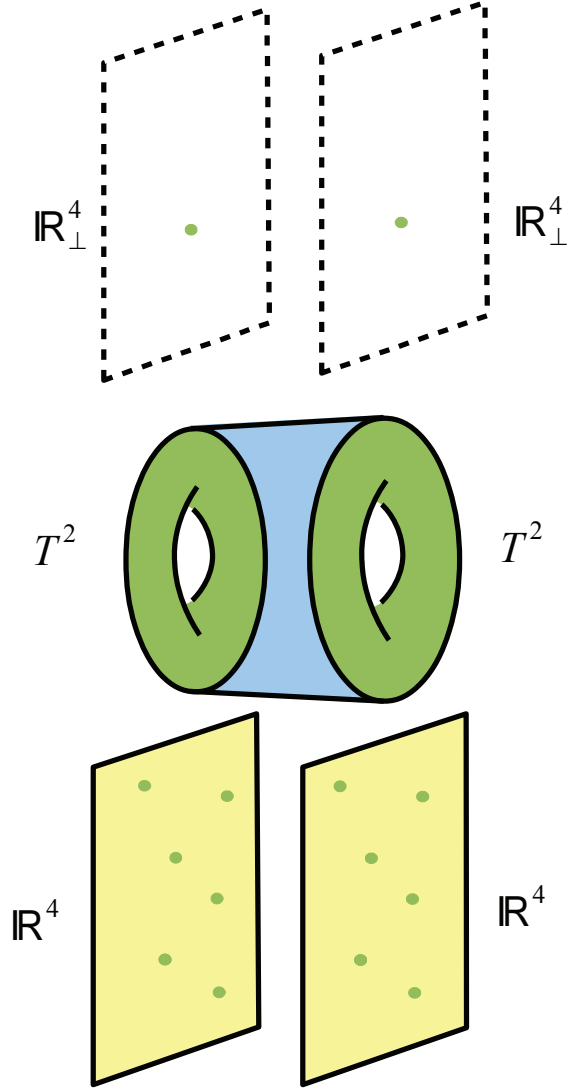}
	\end{center}
	\caption{The system of M2 and M5 branes. The M5 branes are depicted in yellow whereas the M2 brane is blue. They intersect at the torus $T^2$ which is depicted in green.}
\label{fig:M2M5}
\end{figure}
The M2 branes as well as the M5 branes will then be all at a fixed point in $\mathbb{R}^4_{\perp}$ and the M5 branes are extended along $T^2 \times \mathbb{R}^4$. Furthermore, the M2 branes will intersect the M5 branes along $T^2$ and appear point-like in $\mathbb{R}^4$. This configuration is shown schematically in \figref{fig:M2M5}. As these points can be separated in $\mathbb{R}^4$ it is natural to conceive that the effective worldvolume theory of $n$ M2 branes admits a description in terms of the Hilbert scheme on $n$ points on $\mathbb{R}^4$ as will be described in detail in section \ref{sec:sigmamodel}.

 The $Spin(8)$ weight vector corresponding to the $U(1)_{\epsilon_1}\times U(1)_{\epsilon_2}$  is $(\epsilon_{1}e_{1}+\epsilon_{2}e_{2}-\frac{\epsilon_{1}+\epsilon_{2}}{2}(e_{3}+e_{4}))$. For the unrefined case corresponding to $\epsilon_{1}+\epsilon_{2}=0$ the above action leaves the $(4,0)$ supercharges invariant. However, for $\epsilon_{1}+\epsilon_{2}\neq 0$ it breaks $(4,0)\mapsto (2, 0)$ with surviving supercharges corresponding to the $Spin(8)$ weights given by Eq.(\ref{super}),
\bea
\pm\,\frac{e_{1}+e_{2}+e_{3}+e_{4}}{2}\,.
\eea

In general we will be interested in the compactification on a generic torus $T^2$ with complex structure $\tau$. In the case where the torus is rectangular $\tau$ can be identified as the ratio of the radii of the circle from six to five and the one from five to four as follows,
\begin{equation}
	\tau = i \frac{R_0}{R_1}.
\end{equation}

Upon further compactification to four dimensions the K\"{a}hler parameters get complexified in the
type IIA setup.  Moreover all the K\"{a}hler parameters of M-theory get rescaled by a factor
of $R_0$ as we go to the type IIA description,
$$t_i^M\rightarrow t^{II}_i=iR_0 t_i^M.$$
These are the parameters we will be using, and in particular we get
K\"ahler parameters which can be identified with the gauge theory parameters as follows
\begin{eqnarray}
	\textrm{Vol}_{\mathbb{C}}(T^2) & = & \tau, \quad Q_{\tau} = e^{2\pi i\tau}, \nonumber \\
	\textrm{Vol}_{\mathbb{C}}(\mathbb{P}^1_f) & = & t_f, \quad Q_f = e^{2\pi i\,t_f}, \nonumber \\
	\textrm{Vol}_{\mathbb{C}}(\mathbb{P}^1_m) & = & t_m=m\tau , \quad Q_m = e^{2\pi i t_m}.
\end{eqnarray}

\begin{figure}[here!]
  \centering
  \subfloat[]
{\label{fig:torus1}\includegraphics[width=0.4\textwidth]{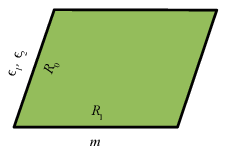}}
   \raisebox{0.9in}{$\Longrightarrow$}
  \subfloat[]
{\label{fig:torus2}\includegraphics[width=0.4\textwidth]{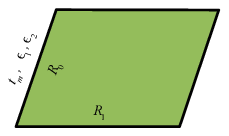}}
  \caption{The torus $T^2$ and its cycles.  In (a), the rectangular torus is depicted. While going around the circle with radius $R_1$ one twists with the mass rotation, and along the circle $R_0$ one introduces the $\epsilon_i$ rotations. In (b), the same geometry is depicted where we use the holomorphy of the
  result to move the twisting along the A-cycle by $m$ to twisting around the B-cycle by $t_m$.}
  \label{fig:torus}
\end{figure}

Thus from the viewpoint of the original M5 branes, we have compactified on a torus
with complex structure $\tau$, where the A-cycle of $T^2$ is twisted by $m$ and the B-cycle
of the torus is twisted by $(\epsilon_1,\epsilon_2)$.  Since we would be ultimately interested
in computing the elliptic genus of the M2 branes stretched between the M5 branes and
wrapped on $T^2$ and the twistings can be viewed as coupling to $U(1)$ background fields,
the dependence of the amplitudes for each of the twistings will appear in the combination:
$$z=\theta_B+\tau \theta_A$$
where $(\theta_A,\theta_B)$ denote the twist parameters around the two cycles.
Thus for the mass term we have
$$(\theta_A, \theta_B)=(m,0)$$ which is equivalent to
$$(\theta_A,\theta_B)=(0,m\tau)=(0,t_m)$$
and for the $\epsilon_i$ we have the twists
$$(\theta_A,\theta_B)=(0,\epsilon_i)$$
This suggests that we can think of all the twistings to be around the B-cycle as long as we use our type IIA parameterization of $t_m$.   For simplicity of notation later in this paper we replace $t_m$ with $m$, when we discuss partition functions.   We summarise the geometry of the torus $T^2$ and its relation to the parameters of the gauge theory in \figref{fig:torus}.

Let us now come to the identification of states. From the discussion preceding Eq.(\ref{eq:5dstates}) it is clear that self-dual string solutions which wrap the whole $T^2$ appear as instanton solutions in the four dimensional gauge theory. There will be also magnetically charged states (which shall be not of relevance here and which we only include for the purpose of completeness of the discussion) which arise from the string which does not wrap the first $S^1$. Electric-magnetic duality of the 4$d$ $\mathcal{N}=4$ theory then corresponds to $SL(2,\mathbb{Z})$ transformations of the $T^2$. A string which wraps the $T^2$ $l$ times and has Kaluza-Klein momentum $k$ then gives rise to a BPS degeneracy which can be counted with the topological string on the elliptic Calabi-Yau. Furthermore, such strings can have non-trivial charge $q_m$ under the rotation induced by $m$. Their degeneracies $d(l,k,q_m)$ appear in the free energy of the topological string in the form
\begin{equation}
	d(l,k,q_m)\, Q_{\tau}^k Q_f^l Q_m^{q_m}.
\end{equation}
The task of the following sections will be to compute these degeneracies in the presence of the $\epsilon_i$
rotations and obtain a closed formula for it in terms of the refined topological string partition function.
More precisely, the partition function of M-theory in this background is by definition the partition function
of the refined topological string on the corresponding Calabi-Yau threefold:

\bea
Z^{M-theory}((\mathbb{R}^4_\perp \times \underbrace{\mathbb{R}^4) \ltimes T^2_{\epsilon_1,\epsilon_2,m}}_{\mbox{\normalsize{$N {\rm M5_{t_f^i}}$}}}\times \mathbb{R})=Z_{top}^{refined}(\epsilon_1,\epsilon_2)(CY_{N,m,t_f^i})
\eea

Moreover the degeneracy of BPS states is known to be computed by the topological vertex and its refinement \cite{Aganagic:2003db,Gopakumar:1998ii,Gopakumar:1998jq,Hollowood:2003cv,Iqbal:2007ii}, which in this
case, as we will discuss in section 4, consists mainly of the suspended M2 branes wrapped on $T^2$.  We thus use this correspondence to compute the twisted elliptic genus of suspended M2 branes.

\subsubsection{Special values of parameters}

As already discussed, for generic values of $m,\epsilon_1,\epsilon_2$ the suspended
M2 branes lead to a $(2,0)$ supersymmetric system on $T^2$.  We can ask whether there
are any special values of these parameters and in particular what happens
to supersymmetry on the M-strings at these special values.

As already noted, in the unrefined limit where $\epsilon_1+\epsilon_2=0$ the
supersymmetry gets enhanced to $(4,0)$.  We can also ask if there are special values of $m$.
For $m=\pm \frac{\epsilon_{1}-\epsilon_{2}}{2}$ there is supersymmetry enhancement to $(2,2)$.
The $Spin(8)$ holonomy is
\bea(\epsilon_1,\epsilon_2,-\epsilon_1,-\epsilon_2),\eea
 (up to the
permutation of the last two factors) and  the preserved charges are given by,
\bea\nn
m&=&\frac{\epsilon_{1}-\epsilon_{2}}{2}:\,\,\pm\,\frac{e_{1}+e_{2}+e_{3}+e_{4}}{2}\,,\,\pm\,\frac{e_{1}-e_{2}-e_{3}+e_{4}}{2}\,,\\
m&=&-\frac{\epsilon_{1}-\epsilon_{2}}{2}:\,\,\pm\,\frac{e_{1}+e_{2}+e_{3}+e_{4}}{2}\,,\,\pm\,\frac{e_{1}-e_{2}+e_{3}-e_{4}}{2}\,.
\eea
The consequence of this enhancement is that the elliptic genus of suspended M2 branes should be a constant independent of the moduli $\tau$ of $T^2$.  There is also another limit in which the partition function simplifies and a different set of BPS states contribute. This limit is given by $m\mapsto \pm\frac{\epsilon_{1}+\epsilon_{2}}{2}$.
In this case the supersymmetry is still $(2,0)$ so a priori
nothing should have simplified, except that the center of mass degree of freedom of the string acquires
additional zero modes.  This is because in this case the holonomy becomes
\bea
m&=&\frac{\epsilon_{1}+\epsilon_{2}}{2}:\,\,(\epsilon_1,\epsilon_2,0,-(\epsilon_1+\epsilon_2))\,,\\\nn
m&=&-\frac{\epsilon_{1}+\epsilon_{2}}{2}:\,\,(\epsilon_1,\epsilon_2,-(\epsilon_1 + \epsilon_2),0)\,,
\eea
and a single M2 brane acquires a fermionic zero mode, due to the $0$ direction in the holonomy twist
(as we will review below in more detail).
We can modify the computation of the elliptic genus in this limit to get a non-zero answer by computing instead
a modified index
\bea
\mbox{Tr}\Big((-1)^{F}F_{L}\,q^{L_{0}}\,\bar{q}^{\bar{L}_{0}}\Big)\,,
\eea
to absorb the zero mode from this single fermion zero mode and obtain a non-trivial answer even in this limit.   This is somewhat similar to what one sees in the context of topologically
twisted ${\cal N}=4$ Yang-Mills in 4$d$ \cite{Vafa:1994tf}
 where the $U(N)$ theory gives a vanishing partition function due to fermionic zero modes, but stripping off the $U(1)$
leads to a non-vanishing partition function for $SU(N)$ theories.

To summarize, the M-strings enjoy a $(4,4)$ supersymmetry.
If we turn on generic $m,\epsilon_1,\epsilon_2$ on $T^2$  the supersymmetry is broken to $(2,0)$ and
we would be computing a non-trivial elliptic genus.  If $\epsilon_1+\epsilon_2=0$ we have
$(4,0)$ supersymmetry.   If we tune $m=\pm (\epsilon_1-\epsilon_2)/2$ the supersymmetry
is enhanced to $(2,2)$ and the elliptic genus becomes a constant.  If $m=\pm (\epsilon_1+\epsilon_2)/2$ the
supersymmetry is still $(2,0)$ but the partition function vanishes due to a fermionic zero mode associated to the `center
of mass mode'.  The fermionic zero mode can be eliminated in this case by insertion of suitable operators leading
to a non-trivial function of $\tau$.  We summarise our discussion in the following table.

\begin{table}[h]
\begin{center}
\begin{tabular}{|l|c|}
    \hline
      \textbf{Values} & \textbf{SUSY}\\
    \hline
      $\epsilon_i = 0,~m=0$                     & $(4,4)$\\
      $\epsilon_1+\epsilon_2 = 0$ & $(4,0)$ \\
      $m=\pm\frac{\epsilon_1-\epsilon_2}{2}$ & $(2,2)$ \\
      $m=\pm\frac{\epsilon_1 + \epsilon_2}{2}$ & ~$(2,0)^*$ \\
      $m\neq 0, \epsilon_i \neq 0$ & $(2,0)$\\
    \hline
\end{tabular}
\end{center}
\caption{Enhanced symmetry configurations. The case $(2,0)^*$ is a configuration with $(2,0)$ supersymmetry but an extra fermionic zero mode which leads to the vanishing of the $U(1)$ part of the elliptic genus.}
\end{table}

\subsection{Quiver realization of the suspended M2 branes}

There is a dual description of this system \cite{Witten:1995zh} which generalizes it to $D$ and $E$ superconformal theories\footnote{There is another dual description given by M5 brane wrapped on $\mathbb{P}^{1}\times T^2$ \cite{Minahan:1998vr}. The tension of the string in this case is given by the size of the $\mathbb{P}^1$. This description can also be generalized to $ADE$ case by wrapping M5 brane on a chain of $\mathbb{P}^{1}\times T^2$ where the chain of $\mathbb{P}^{1}$'s is given by the Dynkin diagram of the $ADE$ group. The $(1,0)$ tensionless string obtained by M5 brane wrapping $\frac{1}{2}K3$ is dual to M2 branes ending on M5 branes in the presence of the ``end of the world" M9 branes. It would be interesting to see if the techniques of this paper can be applied to this case.}.
This corresponds to type IIB theory in the presence of $ADE$ singularity.
The duality between the $A$-series and M5 branes follows from the fact that $A_{N-1}$ singularities in type IIB is dual to $N$ NS5 branes
for type IIA strings \cite{Ooguri:1995wj}.  By lifting the NS5 branes to M-theory we see that this is equivalent to $N$ M5 branes
where one of the five transverse directions to the 5-brane is compactified on $S^1$.  Therefore, when we consider
separated branes, the rotation symmetry is reduced from $SO(5)_R\rightarrow SO(3)$.  Thus this realization
has the slight disadvantage that not all the symmetries are manifest.  In particular we cannot twist by the mass
parameter as we go down on the circle from 6 to 5 dimensions.

 The $ADE$ singularity is given by $\mathbb{C}^{2}/\Gamma$ where $\Gamma\subset SU(2)$ is one of the discrete subgroups of $SU(2)$, which are in one to one correspondence with the $ADE$ Dynkin diagrams, for which $\mathbb{C}^2$ is the two dimensional representation. The singularity $\mathbb{C}^{2}/\Gamma$ can be resolved to $X_{\Gamma}=\widetilde{\mathbb{C}^{2}/\Gamma}$. The resolution $X_{\Gamma}$ is such that $H_{2}(X_{\Gamma},\mathbb{Z})$ generated by 2-cycles, which are topologically $\mathbb{P}^{1}$, can be identified with the root lattice of $ADE$ Lie algebra corresponding to $\Gamma$ such that the intersection number of the 2-cycles is given by the inner product on the root lattice which is determined by the Cartan matrix $A_{ij}$, i.e. there exists a basis $\{C_{1},C_{2},\cdots, C_{r}\}$ of $H_{2}(X_{\Gamma},\mathbb{Z})$ such that
\bea
C_{i}\cdot C_{j}=-A_{ij}\,.
\eea
As we blow down these 2-cycles to zero size we get back the singular space $\mathbb{C}^{2}/\Gamma$.

Consider type IIB strings propagating on $\mathbb{R}^{1,5}\times X_{\Gamma}$ and let
 $t_i=\mu_i/g_s$, where $\mu_i$ is the size of blown up 2-cycles.  The conformal limit is achieved by taking
$t_i\rightarrow 0$.  For $\Gamma_{N}=\left\{\left(
                                   \begin{array}{cc}
                                     e^{\frac{2\pi i}{N}} & 0 \\
                                     0 & e^{-\frac{2\pi i}{N}} \\
                                   \end{array}
                                 \right)\,\left |\,i=1,\cdots,N-1\right.\right\}$ the corresponding Dynkin diagram is that of $A_{N-1}$ and the corresponding type IIB theory in the conformal limit gives the $(2,0)$ superconformal theory of $N$ coincident M5 branes. Moving away from the conformal point by turning on $t_{i}$ corresponds to the separation of the M5 branes along a linear direction as discussed before.

The emergence of conformal theory is signalled by the appearance of tensionless strings. In the M-theory setup this arises by M2 branes ending on M5 branes, and the tension of the
resulting string is proportional to the separation of the corresponding M5 branes.
Thus each pair of M5 branes leads to a string which become tensionless in the conformal limit.
Similarly in the type IIB the strings arise by wrapping D3-branes over holomorphic 2-cycles $C$ of the blown-up
geometry $X_{\Gamma}$. Since holomorphic curves satisfy $C^2=2g-2$, where $g$ is the genus of the curve $C$, and the inner product $C^2$ is given by minus the Cartan matrix it follows that the only holomorphic curves in the geometry are 2-spheres $C$ with $C^2=-2$, i.e. they are in one to one correspondence with the positive roots of $A_{N-1}$. The tension of a string coming from  D3 brane wrapped on $\sum_{i}n_{i}C_{i}$ is given by $\sum_{i}n_{i}t_{i}$ hence giving rise to strings with tensions $t_i$ for each 2-sphere $C_{i}$.   Unwrapped D1 branes can also be considered and they would correspond to M2 branes winding along a compactified circle
transverse to the M5 brane.

The theory describing M-strings, when they have finite tension, can be deduced using
the quiver description \cite{Douglas:1996sw}.  This in particular leads to the
affine $ADE$ quiver.  If we are interested in the local behaviour of the 6 dimensional CFT, we will be mostly interested in the limit where the transverse circle
to the M5 brane is infinitely large where we would be ignoring the D1 brane.   One could also consider
the opposite limit where the transverse circle shrinks and consider the little string theory \cite{ Aharony:1998ub,Aharony:1999ks}, where the considerations of this paper will still apply.  If we ignore the D1 brane charge,
this corresponds to deleting the affine root from the quiver and gives rise to the ordinary $ADE$ quiver.   This theory is equivalent
to the reduction to two dimensions of the familiar ${\cal N}=2$ $ADE$ quiver theory in four dimensions.  In two dimensions this leads to a $(4,4)$ supersymmetric quiver theory.\\

Note however that, as already noted, not all the symmetries of the M5 branes are realized in this setup.
This also impacts the symmetries that the M-string sees.  In particular the symmetries of the 2d
quiver theory (i.e. that of the $(4,4)$ quiver theory) are given by
$$Spin(4)\times SU(2)$$
where $SU(2)=Spin(3) \subset Spin(4)_R =SU(2)_L\times SU(2)_R$ where $SU(2)$ is diagonally embedded
in the two $SU(2)$'s. The Cartan of this $SU(2)$ can be identified with the rotation of the normal line
bundle on the blown up ${\mathbb P}^1$'s.  As already noted
we cannot realize the twisting by $m$ in this setup. This particular Cartan
can be viewed as being in the $e_3$(or $e_4$) direction of $Spin(8)$ holonomy.  Thus in the setup of the most
general twisting discussed in the previous section, we see that we are in the limit where $m=(\epsilon_1+\epsilon_2)/2$.
Thus a 2-parameter subspace of the 3-parameter elliptic genus should be computable using the elliptic
genus of ${\cal N}=4$ $ADE$ quiver theories.  Of course, as noted before, we would need to get rid of $U(1)$ zero modes.  In the $D$ and $E$ cases this gives a new way to compute the BPS degeneracies, which is
not so simple in the geometric setup.

For concreteness let us focus on the $A_{N-1}$ case.  Let
$N_i$ D3 branes wrap the cycle $C_{i}$, which correspond to the simple roots forming a basis of positive root lattice
of $A_{N-1}$.  Then this theory has gauge group
\bea
G=\prod_{i=1}^{N-1} U(N_i),
\eea
with bi-fundamental matter between adjacent gauge factors.  From the perspective of M-theory, this
should be identified with the theory living on a collection of $N_i$ M-strings.  For simplicity let us consider the case
of the $A_1$ theory.  This corresponds to having two M5 branes with $N_{1}$ M2 branes between them.  The $(4,4)$ theory in this
case corresponds to the pure $U(N_{1})$ gauge theory \cite{Witten:1997yu}.
This theory has a Coulomb branch which at least far away from the origin of the Coulomb branch gives rise to the sigma model on $\mbox{Sym}^{N_{1}}{\mathbb{R}^4}$, i.e.
the $N_{1}$-fold symmetric product of $\mathbb{R}^4$. These $N_1$ points in ${\mathbb{R}^4}$
can be identified as the end points
of the transverse $\mathbb{R}^4$ to the M2 brane in the M5 brane. This also follows from the fact that broken supercharges $Q^{\dot{\alpha}}_{+\frac{1}{2}\,\dot{a}}$ and $Q^{\alpha}_{-\frac{1}{2}\,a}$ give rise to four left moving bosons $\partial_{+} x^{a\dot{a}}$ and four right moving bosons $\partial_{-}x^{a\dot{a}}$ where $x^{a\dot{a}}=X^{\mu}\gamma^{a\dot{a}}_{\mu}$ such that $\mu=2,3,4,5$ and $\gamma_{
\mu}$ are the $Spin(4)$ gamma matrices.  Modulo the resolution of the singularities when
the points coincide, this can also be viewed as
the Hilbert scheme of $N_{1}$ points on $\mathbb{R}^4$.  What is the status of the theory when
the points coincide is of course critical to the formation of BPS bound states, and therefore
the above heuristic picture for $N_1>1$ is not guaranteed to be correct.  In fact we will find
later that our computation suggests that this picture is not accurate.

\section{Topological partition function of M5 branes}

\par
\begin{figure}[h]
  \centering
  \includegraphics[width=3in]{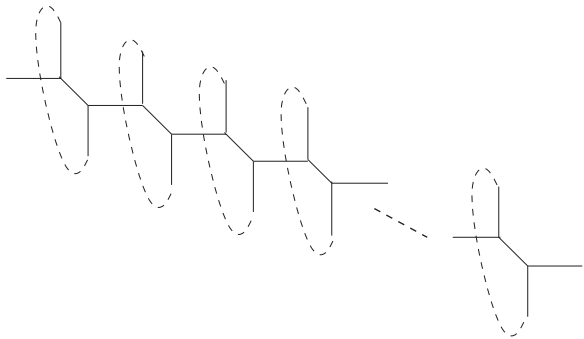}\\
  \caption{The toric diagram of elliptic Calabi-Yau threefold $X_{N}$.}\label{figure1}
\end{figure}

{The ${\cal N}=2^{*}$ $SU(N)$ gauge theories can be geometrically engineered using elliptic Calabi-Yau threefolds. These elliptic Calabi-Yau threefolds, which we will denote by $X_{N}$, are given by a deformation of the $A_{N-1}$ fibration over $T^{2}$. The geometry of $X_{N}$ is captured by the toric diagram shown in \figref{figure1}. }

\par{{The gauge theory partition function can be obtained either from Nekrasov's instanton calculus or by calculating the topological string partition function of $X_{N}$. The topological string partition function can be calculated using the refined topological vertex formalism \cite{Iqbal:2007ii}. The refined topological vertex has a preferred direction which breaks the cyclic symmetry of the topological vertex. For a given toric diagram that engineers a gauge theory we need to pick an orientation for the preferred direction. It was argued that the total amplitude is independent of the choice \cite{Aganagic:2012hs} although the form of the amplitude could have a significantly different looking form\footnote{In a number of non-trivial examples this invariance is shown for high degrees of the curves and is used as a new way to derive identities involving Macdonald polynomials \cite{Iqbal:2008ra}.}. The choice is not necessarily arbitrary and has, as we will see later, important physical meaning. The preferred direction usually determines the instanton directions. In other words, according to the gluing algorithm of the topological vertex we perform sums of Young diagrams along each internal edge in the toric diagram. All of these sums can be explicitly performed except the ones along the preferred directions. The preferred direction is generally chosen such that the left-over sums match with the instanton expansion of the corresponding gauge theory.}

From \figref{figure1} it is clear that there are two choices for the preferred direction. One choice of preferred direction is along the vertical which is compactified on a circle and the other choice is along the horizontal. We will calculate the partition function for both these cases.

Before we begin calculating the partition functions we would like to explain our notation which will appear in later sections. We will denote by Greek letters $\lambda,\mu,\nu$ partitions of natural numbers. An empty partition will be denoted by $\emptyset$. A non-empty partition $\lambda$ is a set of non-negative integers such that $\lambda_{1}\geq \lambda_{2}\geq \lambda_{3}\geq \mathellipsis\geq \lambda_{\ell(\lambda)}>0$. The number of parts of the partition $\lambda$ will be denoted by $\ell(\lambda)$. We will denote by $\lambda^t$ the transpose of the partition $\lambda$. $\lambda^t$ is also a partition such that $\lambda^t_{i}=\mbox{number of parts of $\lambda$ which are greater than $i-1$}$. For example, if $\lambda=\{4,2,2,1\}$ then $\lambda^t=\{4,3,1,1\}$. The following are few functions on the set of partitions which will be of use later,
\bea
|\lambda|\coloneqq\sum_{i=1}^{\ell(\lambda)}\lambda_{i}\,,\,\,\Arrowvert\lambda\Arrowvert^2\coloneqq\sum_{i=1}^{\ell(\lambda)}\lambda_{i}^{2}\,,\,\,\Arrowvert\lambda^{t}\Arrowvert^2=\sum_{i=1}^{\ell(\lambda^t)}(\lambda^{t}_{i})^2\,.
\eea
As is well known the partition $\lambda$ has a two dimensional representation called the Young diagram. A Young diagram corresponding to the partition $\lambda$ is obtained by placing a box in the first quadrant with upper left hand coordinate $(i,j)$ for each $(i,j)\in \{(i,j)\,|\,i=1,\mathellipsis,\ell(\lambda);j=1,\mathellipsis, \lambda_{i}\}$. Thus the number of boxes in the $i^{th}$ column of the Young diagram give $\lambda_i$. We will not distinguish between the partition and its Young diagram so that $(i,j)\in \lambda$ makes sense and means the box in the Young diagram with coordinates $(i,j)$.

To calculate the topological string partition function we will use the refined topological vertex which is given by
\begin{align}
C_{\lambda\,\mu\,\nu}(t,q)=t^{-\frac{\Arrowvert\mu^t\Arrowvert^2}{2}}\,q^{\frac{\Arrowvert\mu\Arrowvert^2+\Arrowvert\nu\Arrowvert^2}{2}}\,\widetilde{Z}_{\nu}(t,q)
\sum_{\eta}\Big(\frac{q}{t}\Big)^{\frac{|\eta|+|\lambda|-|\mu|}{2}}\,s_{\lambda^{t}/\eta}(t^{-\rho}\,q^{-\nu})\,s_{\mu/\eta}(t^{-\nu^t}\,q^{-\rho})\,,
\end{align}
where $\rho=\{-\frac{1}{2},-\frac{3}{2},-\frac{5}{2},\cdots\}$, $s_{\nu}(x_{1},x_{2},\cdots)$ is the Schur function labelled by a partition $\nu$, and
\begin{align}
\widetilde{Z}_{\nu}(t,q)=\prod_{(i,j)\in \nu}\Big(1-q^{\nu_{i}-j}\,t^{\nu^{t}_{j}-i+1}\Big)^{-1}.
\end{align}

We will also calculate gauge theory partition functions using equivariant instanton calculus where the torus action on $\mathbb{C}^2$ is given by $(z_{1},z_{2})\mapsto (e^{2\pi i \epsilon_{1}}\,z_{1},e^{2\pi i\epsilon_{2}}\,z_{2})$. The topological string parameters $q$ and $t$ are related to the gauge theory parameters $\epsilon_{1}$ and $\epsilon_{2}$ as
\bea
q=e^{2\pi i \epsilon_{1}}\,,\,\,\,\,t=e^{-2\pi i \epsilon_{2}}\,.
\eea
\subsection{Case 1: Preferred direction along the compactified circle}
Let us consider the case of $SU(2)$ in detail and then we will generalize this to $SU(N)$. The toric diagram for the $SU(2)$ case is shown in \figref{su2} below. The vertical lines in the toric diagram are glued and the preferred direction is along the vertical.
\begin{figure}[h]
  \centering
  \includegraphics[width=2in]{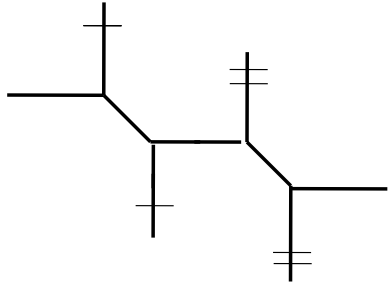}\\
  \caption{Toric diagram of the geometry giving rise to $SU(2)$ ${\cal N}=2^{*}$ theory. The preferred direction is taken to be vertical.}\label{su2}
\end{figure}

The refined topological string partition function in terms of the refined vertex $C_{\lambda\,\mu\,\nu}(t,q)$ is given by,
\begin{align}\nn
Z^{(2)} =  \sum_{\lambda\,\mu\,\sigma\,\nu_{1}\,\nu_{2}}&(-\hat{Q})^{|\nu_{1}|+|\nu_{2}|} (-Q_{m})^{|\sigma|+|\mu|}(-Q)^{|\lambda|}
C_{\mu\,\emptyset\,\nu_{1}}(t^{-1},q^{-1})\,C_{\mu^{t}\lambda\,\nu_{1}^{t}}(q^{-1},t^{-1}) \nn \\
& \times C_{\sigma\,\lambda^{t}\,\nu_{2}}(t^{-1},q^{-1})\,
C_{\sigma^{t}\,\emptyset\,\nu_{2}^{t}}(q^{-1},t^{-1}),
\end{align}
where the superscript refers to the number of M5 branes in the construction. Using standard techniques of summing up the Schur symmetric function given in Appendix B we get ($Q_{\tau}=\hat{Q}Q_{m}$)
\bea
Z^{(2)}=Z^{(2)}_{pert} \,Z^{(2)}_{inst}\,,
\eea
where
\bea
Z^{(2)}_{pert} = \prod_{i,j=1}^{\infty}\frac{(1-Q_{m}\,q^{i-\frac{1}{2}}t^{j-\frac{1}{2}})^2(1-Q_{f}Q_{m}\,q^{i-\frac{1}{2}}t^{j-\frac{1}{2}})(1-Q_{f}Q_{m}^{-1}\,q^{i-\frac{1}{2}}t^{j-\frac{1}{2}})}
{(1-Q_{f}\,q^{i-1}t^{j})(1-Q_{f}\,q^{i}t^{j-1})}
\eea
and
\begin{align}\label{su2pf}\nn
Z^{(2)}_{inst} = \sum_{\nu_{1}\,\nu_{2}}Q_{\tau}^{|\nu_{1}|+|\nu_{2}|}&\\ \nn
\times\prod_{(i,j)\in\nu_{1}}&\frac{(1-Q_{m}\,q^{\nu_{1,j}^{t}-i+\frac{1}{2}}t^{\nu_{1,i}-j+\frac{1}{2}})(1-Q_{m}^{-1}\,q^{\nu_{1,j}^{t}-i+\frac{1}{2}}t^{\nu_{1,i}-j+\frac{1}{2}})}{(1-\,q^{\nu_{1,j}^{t}-i}t^{\nu_{1,i}-j+1})(1-\,t^{\nu_{1,i}-j}q^{\nu_{1,j}^{t}-i+1})}\\\nn
\times&\frac{(1-Q_{m}\,Q_{f}\,q^{\nu_{2,j}^{t}-i+\frac{1}{2}}t^{\nu_{1,i}-j+\frac{1}{2}})
(1-Q_{m}^{-1}\,Q_{f}\,q^{\nu_{2,j}^{t}-i+\frac{1}{2}}t^{\nu_{1,i}-j+\frac{1}{2}})}{(1-Q_{f}\,q^{\nu_{2,j}^{t}-i+1}t^{\nu_{1,i}-j})
(1-Q_{f}\,q^{\nu_{2,j}^{t}-i}t^{\nu_{1,i}-j+1})}\\\nn
\times\prod_{(i,j)\in \nu_{2}}&\frac{(1-Q_{m}\,q^{\nu_{2,j}^{t}-i+\frac{1}{2}}t^{\nu_{2,i}-j+\frac{1}{2}})
(1-Q_{m}^{-1}\,q^{\nu_{2,j}^{t}-i+\frac{1}{2}}t^{\nu_{2,i}-j+\frac{1}{2}})}
{(1-\,q^{\nu_{2,j}^{t}-i}t^{\nu_{2,i}-j+1})(1-\,t^{\nu_{2,j}-i}q^{\nu_{2,i}^{t}-j+1})}\\
\times&\frac{(1-Q_{m}Q_{f}\,q^{-\nu_{1,j}^{t}+i-\frac{1}{2}}t^{-\nu_{2,i}+j-\frac{1}{2}})
(1-Q_{m}^{-1}Q_{f}\,q^{-\nu_{1,j}^{t}+i-\frac{1}{2}}t^{-\nu_{2,i}+j-\frac{1}{2}})}{(1-Q_{f}\,q^{-\nu_{1,j}^{t}+i}t^{-\nu_{2,i}+j-1})
(1-Q_{f}\,q^{-\nu_{1,j}^{t}+i-1}t^{-\nu_{2,i}+j})}.
\end{align}
In the limit $Q_{\tau} \to 0$ the partition function reduces to $Z^{(2)}_{pert}$ which is the contribution of the holomorphic curves that do not wrap the elliptic curve.

\subsubsection*{$Q_{m}=\left(q\,t\right)^{\pm \frac{1}{2}}$ limit}An important property of the above partition function is that for $Q_{m}=\left(q\,t\right)^{\pm \frac{1}{2}}$ the $Q_{\tau}$ dependence goes away, i.e. the sum over $Q_{\tau}$ in Eq.(\ref{su2pf}) only gets contribution from trivial partitions and becomes $1$. To see this consider the following factor which occurs in the sum over $(\nu_{1},\nu_{2})$ in Eq.(\ref{su2pf}),
\bea\label{factor}
\prod_{(i,j)\in \nu_{1}}
(1-Q_{m}\,q^{\nu_{1,j}^{t}-i+\frac{1}{2}}t^{\nu_{1,i}-j+\frac{1}{2}})
(1-Q_{m}^{-1}\,q^{\nu_{1,j}^{t}-i+\frac{1}{2}}t^{\nu_{1,i}-j+\frac{1}{2}})\,.
\eea
If we consider a box in the Young diagram $\nu_{1}$ which is an outer corner then its arm length and the leg length is zero, i.e. if $(i_0,j_0)\in \nu_{1}$ are the coordinates of such a box then $\nu_{1,i_0}-j_0=0$ and $\nu^{t}_{1,j_0}-i_0=0$. Such a box always exists in a nonempty Young diagram and the contribution of such a box to the factor in Eq.(\ref{factor}) is given by,
\bea
(1-Q_{m}\,t^{\frac{1}{2}}\,q^{\frac{1}{2}})(1-Q_{m}^{-1}\,t^{\frac{1}{2}}\,q^{\frac{1}{2}})=0\,\,\,\,\mbox{for}\,\,\,Q_{m}=\left(q\,t\right)^{\pm \frac{1}{2}}\,.
\eea

\subsubsection{Partition function from instanton calculus}
\label{ins}

The partition function of the $U(N)$ ${\cal N}=2^{*}$ theory, calculated in the last section using geometric engineering and the topological vertex formalism, can also be determined using Nekrasov's instanton calculus \cite{Nekrasov:2002qd}. In the case of the massive adjoint hypermultiplet the supersymmetric partition function of the gauge theory reduces to the partition function of supersymmetric quantum mechanics on the instanton moduli space \cite{Dorey:2002ik}. This quantum mechanical model is the reduction to one dimension of the ${\cal N}=(2,2)$ two dimensional sigma model with instanton moduli space as the target space. Recall that the ${\cal N}=(2,2)$ sigma model with target space $M$, a K\"ahler manifold, has Lagrangian given by
\bea
{\cal L}=\int d^{4}\theta\,K(\Phi_i,\overline{\Phi_{j}})\,,
\eea
where $K(\phi^{i},\phi^{\bar{j}})$ is the K\"ahler potential of $M$ and $\Phi_i$ are the chiral superfields. In terms of the component fields the Lagrangian is given by\footnote{We are considering a Lorentzian worldsheet so that $(\theta^{+}, \bar{\theta}^{+})$ are positive chirality spinors and $(\theta^{-},\bar{\theta}^{-})$ are negative chirality spinors. The fermions $\psi^{i}_{+}$ and $\psi^{j}_{-}$ are of negative and positive chirality respectively. Also $\partial_{\pm}=\frac{\partial_{0}\pm\partial_{1}}{2}$.}
\bea\nn
{\cal L}=g_{i\bar{j}}\partial_{+}\phi^{i}\partial_{-}\phi^{\bar{j}}+g_{i\bar{j}}\partial_{-}\phi^{i}\partial_{+}
\phi^{\bar{j}}-2ig_{i\bar{j}}\psi_{+}^{i}D_{-}\psi^{\bar{j}}_{+}
-2ig_{i\bar{j}}\psi^{i}_{-}D_{+}\psi^{\bar{j}}_{-}+R_{\bar{k}i\bar{l}j}\psi^{i}_{+}\psi^{j}_{-}\psi^{\bar{k}}_{-}\psi^{\bar{l}}_{+}\,,
\eea
where the covariant derivatives are given by
\bea
D_{-}\psi^{\bar{j}}_{+}=(\partial_{-}\delta^{\bar{j}}_{\bar{l}}+\Gamma^{\bar{j}}_{\bar{l}\bar{k}}\partial_{-}\phi^{\bar{k}})\psi^{\bar{l}}_{+}\,,
D_{+}\psi^{\bar{j}}_{-}=(\partial_{+}\delta^{\bar{j}}_{\bar{l}}+\Gamma^{\bar{j}}_{\bar{l}\bar{k}}\partial_{+}\phi^{\bar{k}})\psi^{\bar{l}}_{-}\,.
\eea
Reducing to one dimension we get the Lagrangian
\bea\nn
L=\frac{1}{2}g_{i\bar{j}}\dot{\phi}^{i}\dot{\phi}^{\bar{j}}-i\,g_{i\bar{j}}\psi^{i}_{+}\dot{\psi}^{\bar{j}}_{+}
-i\,g_{i\bar{j}}\psi^{i}_{-}\dot{\psi}^{\bar{j}}_{-}-i
g_{i\bar{j}}\Gamma^{\bar{j}}_{\bar{l}\bar{k}}\dot{\phi}^{\bar{k}}(\psi^{i}_{+}\psi^{\bar{l}}_{+}+\psi^{i}_{-}\psi^{\bar{l}}_{-})+
R_{\bar{k}i\bar{l}j}\psi^{i}_{+}\psi^{j}_{-}\psi^{\bar{k}}_{-}\psi^{\bar{l}}_{+}\,.
\eea
This Lagrangian is invariant under 4 supersymmetries since the target space is K\"ahler manifold given by,
\bea
\delta\,\phi^{i}&=&\epsilon^{+}\psi^{i}_{+}+\epsilon^{-}\psi^{i}_{-}\,,\\\nn
\delta\psi^{i}_{+}&=&i\overline{\epsilon}^{+}\dot{\phi}^{i}-\epsilon^{-}\Gamma^{i}_{jl}\psi^{j}_{-}\psi^{l}_{+}\,,\\\nn
\delta\psi^{i}_{-}&=&i\overline{\epsilon}^{-}\dot{\phi}^{i}+\epsilon^{+}\Gamma^{i}_{jl}\psi^{j}_{-}\psi^{l}_{+}\,.
\eea
The Witten index of this quantum mechanical system is the Euler characteristic of the target space,
\bea
\chi(M)=\mbox{Tr}\,(-1)^{F}e^{-\beta\,H}.
\eea
One can define a more general partition function
\cite{Hollowood:2003vt},
\bea
Z\coloneqq\mbox{Tr}\,(-1)^{F_{-}}\,(-y)^{F_{+}}\,e^{-\beta\,H},
\eea
which is invariant under only two of the supersymmetries $\epsilon^{-}$ and $\overline{\epsilon}^{-}$. Inserting $y^{F_{+}}$ has the effect of changing the fermion boundary conditions so that $\psi^{i}_{-}$ and $\psi^{\bar{i}}_{-}$ remain periodic but $\psi^{i}_{+}(\beta)=y\,\psi^{i}_{+}(0)$ and $\psi^{\bar{i}}_{+}(\beta)=y\,\psi^{\bar{i}}_{+}(0)$. This twisted partition function can be calculated using the supersymmetric localization with respect to the remaining supercharges and gives
\bea
Z\coloneqq\int_{{\cal M}}\prod_{i}\frac{x_{i}(1-y\,e^{-x_{i}})}{1-e^{-x_{i}}}\,=\sum_{i,j}(-1)^{i+j}y^{j}\,b_{i,j}({\cal M})\,,
\eea
where $x_{i}$ are the roots of the curvature two form. This is the $\chi_y$ genus of the manifold ${\cal M}$ which for $y=1$ gives the Euler characteristic. Thus the partition function of the $U(N)$ ${\cal N}=2^{*}$ theory is the $\chi_y$ genus of the rank $N$ instanton moduli spaces,
\bea\label{chi}
Z_{U(N)}=\sum_{k\geq 0}\varphi^{k}\,\chi_{y}({\cal M}(N,k))\,,
\eea
where ${\cal M}(N,k)$ is the moduli space of rank $N$ instantons with charge $k$. We will determine Eq.(\ref{chi}) using equivariant localization and see that it exactly reproduces $Z^{(2)}/\left(Z^{(1)}\right)^{2}$ of Eq.(\ref{su2pf}) for $N=2$.

The instanton moduli space of charge $k$ for the $U(1)$ case is the Hilbert scheme of $k$ points in $\mathbb{C}^2$, $\mbox{Hilb}^{k}[\mathbb{C}^2]$. $\mbox{Hilb}^{k}[\mathbb{C}^2]$ is a $4k$ dimensional hyperk\"ahler manifold which is obtained by a resolution of singularities of the $k^{th}$ symmetric product of $\mathbb{C}^2$ \cite{Nakajimabook}. It is also the space of polynomial ideals in $\mathbb{C}[z_{1},z_{2}]$ of codimension $k$,
\bea
\mbox{Hilb}^{k}[\mathbb{C}^2]=\{I\subset \mathbb{C}[z_{1},z_2]~|~ \mbox{dim}(\mathbb{C}[z_{1},z_{2}]/I)=k\}\,.
\eea
The tangent space at $I\in \mbox{Hilb}^{k}[\mathbb{C}^{2}]$ is given by
\bea
T_{I}(\mbox{Hilb}^{k}[\mathbb{C}^2])\simeq \mbox{Hom}(I,\mathbb{C}[z_1,z_2]/I)\,.
\eea
The torus $T=U(1)_{\epsilon_{1}}\times U(1)_{\epsilon_{2}}$ action on $\mathbb{C}^2$,
\bea
(z_{1},z_{2})\mapsto (e^{2\pi i\epsilon_{1}}\,z_{1},e^{2\pi i \epsilon_{2}}\,z_{2})\,,
\eea
induces an action on $\mbox{Hilb}^{k}[\mathbb{C}^2]$ with a finite number of isolated fixed points labelled by the partitions of $k$. The fixed point corresponding to the partition $\lambda$ will be denoted by $I_{\lambda}$. The torus $T$ maps $I_{\lambda}$ to $I_{\lambda}$ and hence maps $T_{I_{\lambda}}(\mbox{Hilb}^{k}[\mathbb{C}^2])$ to itself. The weights of the $T$ action on the tangent space at the fixed point $I_{\lambda}$ are given by \cite{Nakajimabook}
\bea
\{q^{\lambda^{t}_{j}-i}\,t^{\lambda_{i}-j+1},q^{-\lambda^{t}_{j}+i-1}
t^{-\lambda_{i}+j}\,|(i,j)\in \lambda\}\,,\,\,q=e^{2\pi i\epsilon_{1}}\,,\,t=e^{-2\pi i\epsilon_{2}}\,.
\label{weights}
\eea
The $U(1)$ partition function can now be calculated,
\bea\nn
Z^{U(1)}&=&\sum_{k\geq 0}\widetilde{Q}^{k}\,\chi_{y}\left(\mbox{Hilb}^{k}[\mathbb{C}^{2}]\right)\\
&=&\sum_{k\geq 0}\widetilde{Q}^{k}\int_{\mbox{Hilb}^{k}[\mathbb{C}^{2}]}\,\,\prod_{i=1}^{2k}\frac{(1-y\,e^{-x_{i}})x_{i}}{1-e^{-x_{i}}},
\eea
where $x_{i}$ are the Chern roots of the tangent bundle. The integral above can be calculated using equivariant localization \cite{Nakajimabook} and is given by
\bea
Z^{U(1)}&=&\sum_{k\geq 0}\widetilde{Q}^{k}\,\sum_{p\in\{ \mbox{\small fixed points}\}}\prod_{i=1}^{2k}\frac{1-y\,e^{-x_{p,i}}}{1-e^{-x_{p,i}}},
\eea
where for the fixed point $p$ labelled by the partition $\lambda$:
\bea
e^{-x_{p,i}}\in \{q^{\lambda^{t}_{j}-i}\,t^{\lambda_{i}-j+1}, q^{-\lambda^{t}_{j}+i-1}
t^{-\lambda_{i}+j}\,|\,(i,j)\in \lambda\}\,.
\eea
Thus we get
\bea
Z^{U(1)}&=&\sum_{k\geq 0}\,\widetilde{Q}^{k}\,\sum_{|\lambda|=k}\prod_{(i,j)\in \lambda}\frac{(1-y\,q^{\lambda^{t}_{j}-i}t^{\lambda_{i}-j+1})(1-y\,q^{-\lambda^{t}_{j}+i-1}t^{-\lambda_{i}+j})}{(1-q^{\lambda^{t}_{j}-i}
t^{\lambda_{i}-j+1})(1-q^{-\lambda^{t}_{j}+i-1}\,t^{-\lambda_{i}+j})}\,.
\eea

For the case of $U(N)$ the instanton moduli space of charge $k$ ${\cal M}(N,k)$ has a $T=U(1)_{\epsilon_{1}}\times U(1)_{\epsilon_{2}}\times U(1)^{N}$ action with a finite number of fixed points labelled by a set $(\nu_1,\nu_2,\mathellipsis, \nu_N)$ of $N$ partitions such that $|\nu_1|+|\nu_2|+\cdots +|\nu_N|=k$. The weights of the $T$ action on the tangent space above the fixed point labelled by $(\nu_1,\nu_2,\mathellipsis,\nu_N)$ is given by \cite{Nakajima},
\bea\nn
\sum_{i}e^{-x_{p,i}}=\sum_{\alpha,\beta=1}^{N}e^{2\pi i(a_{\alpha}-a_{\beta})}\Big( \sum_{(i,j)\in \nu_{\alpha}}\,q^{-\nu_{\beta,j}^{t}+i}t^{-\nu_{\alpha,i}+j-1}+\sum_{(i,j)\in \nu_{\beta}}\,q^{\nu_{\alpha,j}^{t}-i+1}
t^{\nu_{\beta,i}-j}\Big)~.
\eea
Using these weights we can write down the $U(N)$ partition function,
\bea\nn
Z^{U(N)}&=&\sum_{k\geq 0}\widetilde{Q}^{k}\,\chi_{y}({\cal M}(N,k))\\\label{eq:ZUNN}
&=&\sum_{k\geq 0}\widetilde{Q}^{k}\sum_{|\nu_{1}|+\cdots +|\nu_{N}|=k}\prod_{\alpha,\beta=1}^{N}Z_{\nu_{\alpha},\nu_{\beta}}\,,
\eea
where ($Q_{\alpha\beta}=e^{2\pi i(a_{\alpha}-a_{\beta})}$),
\bea\nn
Z_{\nu_{\alpha},\nu_{\beta}}:=\prod_{(i,j)\in \nu_{\alpha}}\frac{(1-y\,Q_{\alpha\beta}\,q^{-\nu_{\beta,j}^{t}+i}\,t^{-\nu_{\alpha,i}+j-1})}
{(1-\,Q_{\alpha\beta}\,q^{-\nu_{\beta,j}^{t}+i}\,t^{-\nu_{\alpha,i}+j-1})}
\prod_{(i,j)\in \nu_{\beta}}
\frac{(1-y\,Q_{\alpha\beta}\,q^{\nu_{\alpha,j}^{t}-i+1}\,t^{\nu_{\beta,i}-j})}{
(1-Q_{\alpha\beta}\,q^{\nu_{\alpha,j}^{t}-i+1}\,t^{\nu_{\beta,i}-j})}\,.
\eea
For the case of $U(2)$ gauge group the above partition function becomes ($Q_{12}=Q_{f}^{-1}$),
\bea \label{eq:chiy}
Z_{U(2)}=\sum_{\nu_1,\nu_2}\widetilde{Q}^{|\nu_1|+|\nu_2|}\,W(\nu_1,\nu_2)\,W(\nu_2,\nu_1)\,,
\eea
where
\begin{align}\nn
W(\nu_1,\nu_2)=\prod_{(i,j)\in \nu_{1}}&\frac{(1-y\,q^{-\nu_{1,j}^{t}+i}\,t^{-\nu_{1,i}+j-1})(1-y\,q^{\nu_{1,j}^{t}-i+1}\,t^{\nu_{1,i}-j})}
{(1-\,q^{-\nu_{1,j}^{t}+i}\,t^{-\nu_{1,i}+j-1})(1-\,q^{\nu_{1,j}^{t}-i+1}\,t^{\nu_{1,i}-j})}\\
\times&\frac{(1-y\,Q_{f}^{-1}\,q^{-\nu_{2,j}^{t}+i}\,t^{-\nu_{1,i}+j-1})}
{(1-\,Q_{f}^{-1}\,q^{-\nu_{2,j}^{t}+i}\,t^{-\nu_{1,i}+j-1})}\frac{(1-y\,Q_{f}\,q^{\nu_{2,j}^{t}-i+1}\,t^{\nu_{1,i}-j})}{(1-Q_{f}\,q^{\nu_{2,j}^{t}-i+1}\,t^{\nu_{1,i}-j})}.
\end{align}
The above partition function is precisely $\widehat{Z}^{(2)}$ of Eq.(\ref{su2pf}) with,
\bea
Q_{\tau}=\widetilde{Q}\,y^2\,,\,\,\,Q_{m}=y\,\sqrt{\frac{q}{t}}\,.
\eea
For $y=1$ we get the generating function for the Euler characteristic of ${\cal M}(N,k)$,
\bea
Z_{U(N)}|_{y=1}=\Big(\prod_{k=1}^{\infty}(1-Q_{\tau}^k)^{-1}\Big)^{N}
\eea
which is also the partition function of the $U(N)$ ${\cal N}=4$ SYM.

\subsection{Case II: Preferred direction along the horizontal}
\label{sec:horizontal}
\par{This choice of the preferred direction leads to a very interesting representation of the partition function. First, it makes the modular properties with respect to the elliptic fiber manifest. We can perform all the sums along this direction to get infinite products. These products can be recast in terms of Jacobi $\theta$-functions. Second, as we will discuss later this choice is the natural one to understand the world volume theory of M2 branes. For a related discussion see also \cite{Kim:2011mv}.}

For the $SU(N)$ theory the geometry is made of $N$ building blocks depicted in \figref{sunblock}.

\begin{figure}[h]
  \centering
  \includegraphics[width=0.5\textwidth]{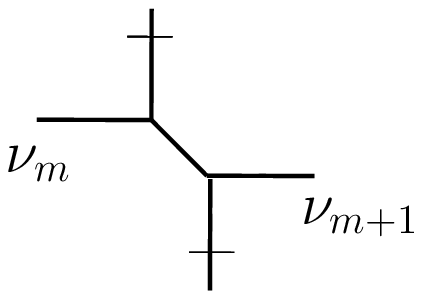}\\
  \caption{The toric diagram of the building block of the $X_N$ geometry with the preferred direction along the horizontal direction.}\label{sunblock}
\end{figure}

\par{We want to outline the computation of these blocks using the refined topological vertex. Further details can be found in Appendix B. According to the gluing rules this open topological string amplitude is given by\footnote{Since we are considering an open topological string amplitude we can consider the branes with some framing. In this case we have taken the branes to be with framing $1$ which is the reason for the prefactor in Eq.(\ref{bsum}) which is given in terms of the framing factor $f_{\mu}(t,q)=q^{\frac{\Arrowvert\mu\Arrowvert^2}{2}}\,t^{-\frac{\Arrowvert\mu^t\Arrowvert^2}{2}}$. This framing will not affect the calculation of the closed topological string partition function since the framing factor will cancel when the two open string amplitudes are glued together because of the identity $f_{\mu}(t,q)f_{\mu^t}(q,t)=1$.},}
\begin{align}\nn\label{bsum}
W_{\nu_{m}^{t}\nu_{m+1}}(\tau,m,\epsilon_{1},\epsilon_{2})&=f_{\nu_{m+1}}(t,q)\,f_{\nu_{m}}(q,t)Q_{m}^{-\frac{|\nu_m|+|\nu_{m+1}|}{2}}\\
&\times\sum_{\lambda\,\mu}(-Q_{m})^{|\lambda|}(-Q_{\tau}Q_{m}^{-1})^{|\mu|}C_{\lambda^{t}\,\mu^{t}\,\nu_{m}^{t}}(q^{-1},t^{-1})C_{\lambda\,\mu\,\nu_{m+1}}(t^{-1},q^{-1})\,.
\end{align}
After some algebra in Schur functions, the building blocks take the form
\begin{align}\nn
&W_{\nu_{m}^{t}\nu_{m+1}}(Q_{\tau},Q,t,q)=t^{-\frac{\Arrowvert\nu^{t}_{m+1}\Arrowvert^2}{2}}\,q^{-\frac{\Arrowvert\nu_{m}\Arrowvert^2}{2}}\,\widetilde{Z}_{\nu_{m}^{t}}(q^{-1},t^{-1})\widetilde{Z}_{\nu_{m+1}}(t^{-1},q^{-1})\,Q_{m}^{-\frac{|\nu_m|+|\nu_{m+1}|}{2}}\\
&\prod_{k=1}^{\infty} \left(1-Q_{\tau}^k\right)^{-1} \prod_{i,j=1}^{\infty}\frac{\left(1-Q_{\tau}^{k}Q^{-1}\,q^{\nu_{m+1,i}-j+\frac{1}{2}}t^{\nu^{t}_{m,j}-i+\frac{1}{2}}\right)
\left(1-Q_{\tau}^{k-1}Q\,q^{\nu_{m,i}-j+\frac{1}{2}}t^{\nu^{t}_{m+1,j}-i+\frac{1}{2}}\right)}
{\left(1-Q_{\tau}^{k}\,q^{\nu_{m+1,i}-j+1}t^{\nu^{t}_{m+1,j}-i}\right)\left(1-Q_{\tau}^{k}\,q^{\nu_{m,i}-j}t^{\nu_{m,j}^{t}-i+1}\right)}.
\end{align}

Let us define the normalized building block
\bea\label{normalize}
D_{\nu_{m}^t\,\nu_{m+1}}(\tau,m,\epsilon_1,\epsilon_2)=\frac{W_{\nu_{m}^t\,\nu_{m+1}}(\tau,m,\epsilon_{1},\epsilon_{2})}{W_{\emptyset\,\emptyset}(\tau,m,\epsilon_{1},\epsilon_{2})}\,.
\eea
The factor in the denominator is the closed topological string partition function of the geometry shown in \figref{sunblock}. Simplifying Eq.(\ref{normalize}) using the identities given in Appendix A we get,
\begin{align}\nn
&D_{\nu_{m}^{t}\nu_{m+1}}(\tau,m,\epsilon_1,\epsilon_2)=t^{-\frac{\Arrowvert\nu^{t}_{m+1}\Arrowvert^2}{2}}\,q^{-\frac{\Arrowvert\nu_{m}\Arrowvert^2}{2}}Q_{m}^{-\frac{|\nu_m|+|\nu_{m+1}|}{2}}\\ \nn &\times\prod_{k=1}^{\infty}\prod_{(i,j)\in\nu_{m}}\frac{(1-Q_{\tau}^{k}Q_{m}^{-1}\,q^{-\nu_{m,i}+j-\frac{1}{2}}\,t^{-\nu_{m+1,j}^{t}+i-\frac{1}{2}})
(1-Q_{\tau}^{k-1}Q_{m}\,q^{\nu_{m,i}-j+\frac{1}{2}}\,t^{\nu_{m+1,j}^{t}-i+\frac{1}{2}})}{(1-Q_{\tau}^{k}\,q^{\nu_{m,i}-j}\, t^{\nu_{m,j}^{t}-i+1})(1-Q_{\tau}^{k-1}\,q^{-\nu_{m,i}+j-1}\,t^{-\nu_{m,j}^{t}+i})}\\
&\times\prod_{(i,j)\in\nu_{m+1}}\frac{(1-Q_{\tau}^{k}Q_{m}^{-1}\,q^{\nu_{m+1,i}-j+\frac{1}{2}}t^{\nu_{m,j}^{t}-i+\frac{1}{2}})(1-Q_{\tau}^{k-1}Q_{m} \,q^{-\nu_{m+1,i}+j-\frac{1}{2}}t^{-\nu_{m,j}^{t}+i-\frac{1}{2}})}{(1-Q_{\tau}^{k}\,q^{\nu_{m+1,i}-j+1}t^{\nu_{m+1,j}^{t}-i})(1-Q_{\tau}^{k-1}\,q^{-\nu_{m+1,i}+j}t^{-\nu_{m+1,j}^{t}+i-1})} \label{eq:domainwall}
\end{align}

In the unrefined case $\epsilon_{2}=-\epsilon_1=-\epsilon$ we have $q=t$ and the above open string amplitude can be written as
\bea\label{bb}
D_{\nu_m^t\,\nu_{m+1}}(\tau,m,\epsilon,-\epsilon)&=&\,\prod_{(i,j)\in \nu_{m}}\frac{\theta_{1}(\tau;\alpha_{ij})}{\theta_{1}(\tau;\beta_{ij})}\,\prod_{(i,j)\in \nu_{m+1}}\frac{\theta_{1}(\tau;\gamma_{ij})}{\theta_{1}(\tau;\delta_{ij})},
\eea
with
\begin{equation}
\begin{array}{ll}
e^{2\pi i\alpha_{ij}}=Q_{m}^{-1}\,q^{-\nu_{m,i}+j-\nu^{t}_{m+1,j}+i-1},\qquad&e^{2\pi i \gamma_{ij}}=Q_{m}^{-1}q^{\nu_{m+1,i}-j+\nu^{t}_{m,j}-i+1},\\
e^{2\pi i \beta_{ij}}=q^{\nu_{m,i}-j+\nu^{t}_{m,j}-i+1}, & e^{2\pi i \delta_{ij}}=q^{\nu_{m+1,i}-j+\nu^{t}_{m+1,j}-i+1},
\end{array}
\end{equation}
where
\bea\nn
\theta_{1}(\tau;z)=-i\,e^{i\pi\,\tau/4}e^{i\pi\,z}\prod_{k=1}^{\infty}(1-\,e^{2\pi i\,k\,\tau})(1-\,e^{2\pi iz}\,e^{2\pi i\,k\,\tau})(1-\,e^{-2\pi i z}\,e^{-2\pi i\,(k-1)\,\tau})\,.
\eea
Recall that the theta function can be written in terms of the Eisenstein series as
\bea\label{Eisenstein}
\theta_{1}(\tau;z)&=&\eta^{3}(\tau)\,(2\pi i\,z)\,\mbox{exp}\left(\sum_{k\geq 1}\frac{B_{2k}}{(2k)(2k)!}\,E_{2k}(\tau)\,(2\pi i z)^{2k}\right)\,.
\eea
In the above equation $E_{2k}(\tau)$ is the weight $2k$ Eisenstein series defined as
\bea
E_{2k}(\tau)&=&1-\frac{4k}{B_{2k}}\sum_{n=1}^{\infty}\sigma_{2k-1}(n)\,q^{n}\,,
\eea
where $B_{2k}$ are the Bernoulli numbers and $\sigma_{k}(n)=\sum_{d|n}d^k$ is the divisor sum function\footnote{$\sum_{k=0}^{\infty}B_{k}\frac{x^{k}}{k!}=\frac{x}{e^{x}-1}$}. Under modular transformation $ \left(
                                                                                                                                                   \begin{array}{cc}
                                                                                                                                                     a & b \\
                                                                                                                                                     c & d \\
                                                                                                                                                   \end{array}
                                                                                                                                                 \right)\in SL(2,\mathbb{Z})
$ the Eisenstein series $E_{2k}(\tau)$ transforms in the following way,

\begin{equation}
E_{2k}\left(\frac{a\tau+b}{c\tau+d}\right)=\left\{\begin{array}{l}(c\tau+d)^2\,E_{2}(\tau)-i\pi c(c\tau+d),\,k=1,\\(c\tau+d)^{2k}\,E_{2k}(\tau)\,,\,k>1\,. \end{array}\right.
\end{equation}
Thus the Eisenstein series $E_{2}(\tau)$ is not a modular form. $E_{2}(\tau)$ can be made modular form by adding a non-holomorphic term to it. Define
\bea
\widehat{E}_{2}(\tau,\bar{\tau})=E_{2}(\tau)-\frac{3}{\pi \mbox{Im}(\tau)},
\eea
then $\widehat{E}_{2}(\tau)$ transforms as modular form,
\bea
\widehat{E}_{2}\left(\frac{a\tau+b}{c\tau+d}\right)=(c\tau+d)^{2}E_{2}(\tau)\,.
\eea
Thus if we replace $E_{2}(\tau)$ with $\widehat{E}_{2}(\tau,\bar{\tau})$ in the $\theta_{1}(\tau;z)$ of Eq.(\ref{bb}) then under the transformation\,,
\bea
(\tau,m,\epsilon_{1},\epsilon_{2})\mapsto \left(\frac{a\tau+b}{c\tau+d},\frac{m}{c\tau+d},\frac{\epsilon_{1}}{c\tau+d},\frac{\epsilon_{2}}{c\tau+d}\right)\,,
\eea
the open topological string amplitude $W_{\nu^{t}_m\nu_{m+1}}(\tau,\bar{\tau},m,\epsilon,-\epsilon)$ is invariant but no longer holomorphic. The open string amplitude satisfies a holomorphic anomaly equation\footnote{It would be interesting to analyze
this anomaly from the perspective of the worldsheet theory.}
\bea\nn
\frac{\partial D_{\nu_{m}^{t}\nu_{m+1}}(\tau,\bar{\tau},m,\epsilon)}{\partial \widehat{E}_{2}(\tau,\bar{\tau})}=\frac{1}{24}\left(\sum_{(i,j)\in \nu_m} (\beta_{ij}^2-\alpha_{ij}^2)+\sum_{(i,j)\in \nu_{m+1}}(\delta_{ij}^2-\gamma_{ij}^2)\right)D_{\nu_{m}^{t}\,\nu_{m+1}}(\tau,\bar{\tau},m,\epsilon),
\eea
where,
\begin{align}\nn
&\sum_{(i,j)\in \nu_m}(\beta_{ij}^2-\alpha_{ij}^2)+\sum_{(i,j)\in \nu_{m+1}}(\delta_{ij}^2-\gamma_{ij}^2)=\\\nn&\sum_{(i,j)\in \nu_m}\left((m+(\nu_{m,i}-j+\nu^{t}_{m+1,j}-i+1)\epsilon\right)^2-\left((\nu_{m,i}-j+\nu^{t}_{m,j}-i+1)\epsilon\right)^2\\+&\sum_{(i,j)\in \nu_{m+1}}\left((m-(\nu_{m+1,i}-j+\nu^{t}_{m,j}-i+1)\epsilon\right)^2-\left((\nu_{m+1,i}-j+\nu^{t}_{m+1,j}-i+1)\epsilon\right)^2\,.
\end{align}

\subsubsection{$U(2)$ partition function}

\par{Using the open topological string amplitude we can now determine the closed topological string partition function for $X_N$ taking the preferred direction to be horizontal.
\begin{figure}[h]
  \centering
  \includegraphics[width=5in]{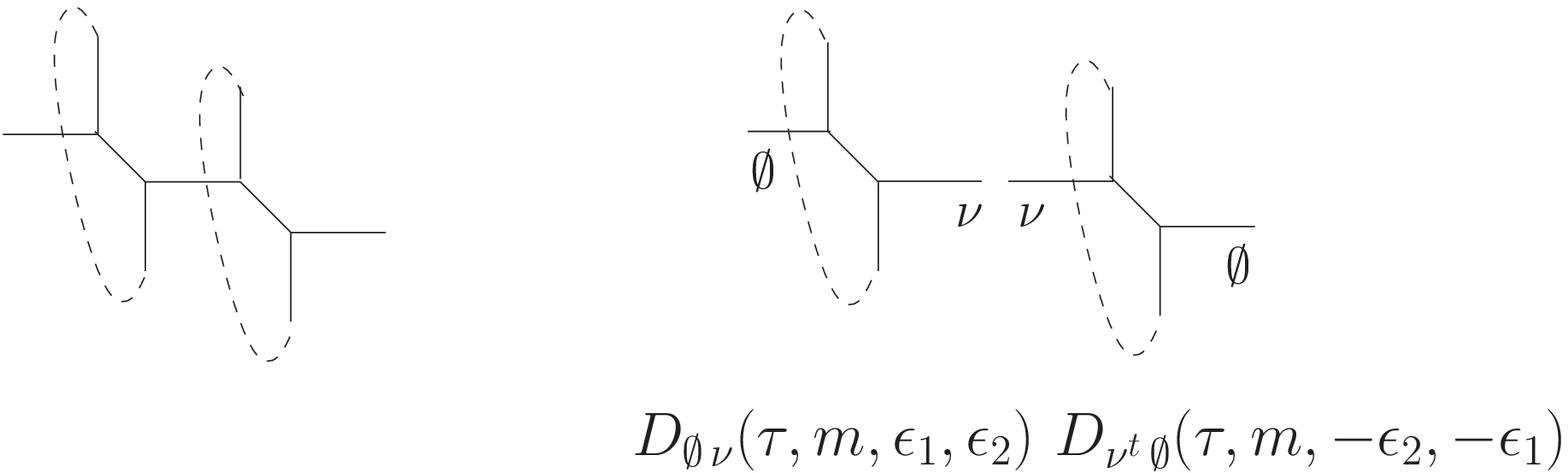}\\
  \caption{(a) Toric diagram of the geometry giving rise to $U(2)$ $N=2^{*}$ theory. The preferred direction is taken to be horizontal. (b) Partition function can be obtained by gluing open topological string amplitudes. }\label{u2}
\end{figure}

As shown in \figref{u2} we can glue two copies of $D_{\emptyset\,\nu}$ to obtain the $U(2)$ partition function,}
\bea
\widehat{Z}^{(2)}(\tau,m,t_{f},\epsilon_1,\epsilon_2)=\sum_{\nu}(-Q_f)^{|\nu|}\,D_{\emptyset\nu}(\tau,m,\epsilon_1,\epsilon_2)\,
D_{\nu^t\emptyset}(\tau,m,-\epsilon_2,-\epsilon_1),
\eea
where $Q_{f}=e^{2\pi i\,t_{f}}$ is the parameter of the fiber $\mathbb{P}^1$, the compact part of the geometry is a $\mathbb{P}^1$ bundle over $T^2$, whose local geometry is ${\cal O}(0)\oplus{\cal O}(-2)$. Using Eq.(\ref{normalize}) and Eq.(\ref{eq:domainwall}) the $U(2)$ partition function is given by,
\bea\nn
Z^{(2)}(\tau,m,t_{f},\epsilon_1,\epsilon_2)&=&\left(Z^{(1)}(\tau,m,\epsilon_1,\epsilon_2)\right)^{2}\\\label{u2pf1}
&\times&\underbrace{\sum_{\nu}(-Q_{f})^{|\nu|}\,\prod_{(i,j)\in \nu}\frac{\theta_{1}(\tau;z_{ij})\,\theta_{1}(\tau;v_{ij})}{\theta_{1}(\tau;w_{ij})\theta_{1}(\tau;u_{ij})}}_{\widehat{Z}^{(2)}(\tau,m,t_{f},\epsilon_1,\epsilon_2)}
\eea
with the following definitions for the arguments of the $\theta$-functions
\begin{equation}\label{var1}
\begin{array}{ll}
e^{2\pi i \,z_{ij}}=Q_{m}^{-1}\,q^{\nu_{i}-j+1/2}\,t^{-i+1/2},\qquad& e^{2\pi i\,v_{ij}}=Q_{m}^{-1}\,t^{i-1/2}\,q^{-\nu_{i}+j-1/2},\\
e^{2\pi i \,w_{ij}}=q^{\nu_{i}-j+1}\,t^{\nu_{j}^{t}-i},&e^{2\pi i \,u_{ij}}=q^{\nu_{i}-j}\,t^{\nu_{j}^{t}-i+1}.
\end{array}
\end{equation}
Notice that for $Q_{m}=\left(\frac{q}{t}\right)^{\pm\frac{1}{2}}$, i.e. $m=\pm\frac{\epsilon_{1}+\epsilon_{2}}{2}$ we have either $z_{11}=0$ or $v_{11}=0$ and since the $(1,1)$ box is present in every non-trivial Young diagram the sum over $\nu$ in $\widehat{Z}^{(2)}$ will only get a non-trivial contribution for $\nu=\emptyset$, therefore it reduces to
\bea
\widehat{Z}^{(2)}\left(\tau,m=\pm\frac{\epsilon_1+\epsilon_2}{2},t_{f},\epsilon_1,\epsilon_2\right)=1.
\eea
However, notice that if,
\bea
\widehat{Z}_{U(2)}(\tau,m,t_{f},\epsilon_1,\epsilon_2)=\sum_{k\geq 0}(-Q_{f})^{k}\,\widehat{Z}_{k}(\tau,m,\epsilon_1,\epsilon_2)\,,
\eea
then
\bea
\lim_{m\mapsto \pm\frac{\epsilon_1+\epsilon_2}{2}}\frac{\widehat{Z}_{k}(\tau,m,\epsilon_1,\epsilon_2)}{\widehat{Z}_{1}(\tau,m,\epsilon_1,\epsilon_2)}\neq 0\,.
\eea
Thus the vanishing of $\widehat{Z}_{k}(\tau,m,\epsilon_1,\epsilon_2)$ for $m=\pm\frac{\epsilon_1+\epsilon_2}{2}$ is entirely due to the fact that $\widehat{Z}_{k}(\tau,m,\epsilon_1,\epsilon_2)$ has $\widehat{Z}_{1}(\tau,m,\epsilon_1,\epsilon_2)$ as a factor. From Eq.(\ref{u2pf1}) and Eq.(\ref{var1}) it is easy to see that,
\bea \label{prediction}
\frac{\widehat{Z}_{k}(\tau,m,\epsilon_1,\epsilon_2)}{\widehat{Z}_{1}(\tau,m,\epsilon_1,\epsilon_2)}=\sum_{|\nu|=k}\frac{\prod_{(i,j)\in \nu,(i,j)\neq (1,\nu_1)}\,\theta_{1}(\tau;z_{ij})\,\theta_{1}(\tau;v_{ij})}{\prod_{(i,j)\in \nu,(i,j)\neq (\ell(\nu),\nu_{\ell(\nu)})}\theta_{1}(\tau;w_{ij})\theta_{1}(\tau;u_{ij})}\,.
\eea
From Eq.(\ref{var1}) it also is clear that the above expression does not vanish. Below we list $\frac{\widehat{Z}_{k}(\tau,m,\epsilon_1,\epsilon_2)}{\widehat{Z}_{1}(\tau,m,\epsilon_1,\epsilon_2)}$ for $k=2$:
\bea
\frac{\widehat{Z}_{2}(\tau,m,\epsilon_1,\epsilon_2)}{\widehat{Z}_{1}(\tau,m,\epsilon_1,\epsilon_2)}&=&
\frac{\theta_{1}(\tau;m-\frac{3}{2}\epsilon_{1}-\frac{1}{2}\epsilon_{2})
\theta_{1}(\tau;m+\frac{3}{2}\epsilon_{1}+\frac{1}{2}\epsilon_{2})}{\theta_{1}(\tau;2\epsilon_{1})\theta_{1}(\tau;\epsilon_{1}-\epsilon_{2})}\\\nn
&+&
\frac{\theta_{1}(\tau;m-\frac{1}{2}\epsilon_{1}-\frac{3}{2}\epsilon_{2})
\theta_{1}(\tau;m+\frac{1}{2}\epsilon_{1}+\frac{3}{2}\epsilon_{2})}{\theta_{1}(\tau;\epsilon_{1}-\epsilon_{2})\theta_{1}(\tau;-2\epsilon_{2})}\\\nn
&\xrightarrow{m=\pm\,\frac{\epsilon_1+\epsilon_2}{2}}& \frac{\theta_{1}(\tau;-\epsilon_1)
\theta_{1}(\tau;2\epsilon_{1}+\epsilon_{2})}{\theta_{1}(\tau;2\epsilon_{1})\theta_{1}(\tau;\epsilon_{1}-\epsilon_{2})}+
\frac{\theta_{1}(\tau;-\epsilon_{2})
\theta_{1}(\tau;\epsilon_{1}+2\epsilon_{2})}{\theta_{1}(\tau;\epsilon_{1}-\epsilon_{2})\theta_{1}(\tau;-2\epsilon_{2})}\\\nn
\eea

\subsubsection{Modular properties of the partition function}

The $\theta$-function factor in the $U(2)$ partition function Eq.(\ref{u2pf1}) can be written as,
\begin{align}
\prod_{(i,j)\in \lambda}\frac{\theta_{1}(\tau;z_{ij})\theta_{1}(\tau;v_{ij})}{\theta_{1}(\tau;w_{ij})\,\theta_{1}(\tau;u_{ij})}=
\left(\prod_{(i,j)\in \lambda}\frac{z_{ij}\,v_{ij}}{w_{ij}\,u_{ij}}\right)\mbox{exp}\left(\sum_{k\geq 1}\frac{B_{2k}}{(2k)(2k)!}E_{2k}(\tau)\,f^{(2k)}_{\nu}(m,\epsilon_{1},\epsilon_{2})\right),
\end{align}
where
\bea\nn
f^{(2k)}_{\nu}(m,\epsilon_{1},\epsilon_{2})&=&\sum_{(i,j)\in \nu}(2\pi i\,z_{ij})^{2k}+(2\pi i\,v_{ij})^{2k}-(2\pi i\,w_{ij})^{2k}-(2\pi i u_{ij})^{2k}\,.
\eea
Thus we get,
\begin{align}\nn
\widehat{Z}^{(2)}(\tau,m,t_{F},\epsilon_1,\epsilon_2)&=\sum_{\nu}(-Q_{f})^{|\nu|}\left(\prod_{(i,j)\in \nu}\frac{z_{ij}\,v_{ij}}{w_{ij}\,u_{ij}}\right)
\\ &\times\mbox{exp}\left(\sum_{k\geq 1}\frac{B_{2k}}{(2k)(2k)!}E_{2k}(\tau)\,f^{(2k)}_{\nu}(m,\epsilon_{1},\epsilon_{2})\right).
\end{align}
Because of the presence of (holomorphic) $E_{2}(\tau)$ the partition function is not invariant under the modular transformation,
\bea\nn
(\tau,m,t_{f},\epsilon_{1},\epsilon_{2})\mapsto \left(\frac{a\tau+b}{c\tau+d},\frac{m}{c\tau+d},t_{f},\frac{\epsilon_{1}}{c\tau+d},\frac{\epsilon_{2}}{c\tau+d}\right),\,\,\,\left(
                                                                                                                                                        \begin{array}{cc}
                                                                                                                                                          a & b \\
                                                                                                                                                           c& d \\
                                                                                                                                                        \end{array}
                                                                                                                                                      \right)\in SL(2,\mathbb{Z})\,.
\eea
As before if we replace $E_{2}(\tau)$ with $\widehat{E}_{2}(\tau,\bar{\tau})=E_{2}(\tau)-\frac{3}{\pi\,Im(\tau)}$ then since $f^{(2)}_{\nu}(m,\epsilon_1,\epsilon_2)\,E_{2}(\tau,\bar{\tau})$ is modular invariant (but not holomorphic) $\widehat{Z}^{(2)}(\tau,m,t_{f},\epsilon_{1},\epsilon_{2})$ is modular invariant as well (but not holomorphic). The partition function $\widehat{Z}^{(2)}(\tau,\bar{\tau},m,t_{f},\epsilon_{1},\epsilon_{2})$ now satisfies a holomorphic anomaly equation
\bea \label{eq:refholan}
\frac{\partial \widehat{Z}^{(2)}(\tau,\bar{\tau},m,t_{f},\epsilon_1,\epsilon_2)}{\partial \widehat{E}_{2}(\tau,\bar{\tau})}&=&\frac{1}{12}\,D_{m,\epsilon_{1},\epsilon_{2},t_{f}}\,\widehat{Z}^{(2)}(\tau,\bar{\tau},t_{f},m,\epsilon_1,\epsilon_2),\\\nn
D_{m,\epsilon_{1},\epsilon_{2},t_{f}}&\coloneqq&\epsilon_{1}\epsilon_{2}\,\frac{\partial^{2}}{\partial T_{f}^{2}}+(m^2-(\epsilon_{+}/2)^2)\frac{\partial}{\partial T_{f}},
\eea
where $\epsilon_{+}=\epsilon_{1}+\epsilon_{2}$ and $T_f = 2\pi i t_f$. This can be interpreted, as has been discussed in the context of elliptic Calabi-Yau threefolds \cite{Hosono:1999qc,Klemm:2012sx,Hosono:2002xj}
as the holomorphic anomaly of topological strings \cite{Bershadsky:1993cx}.  Here we are in the unusual situation to have
also fixed the `holomorphic ambiguity' to all orders in the genus expansion, as we have the full expansion
of the topological string amplitude. Equation (\ref{eq:refholan}) is a refined version of the holomorphic anomaly equations and it would be interesting to relate it to the results of \cite{Huang:2010kf,Krefl:2010fm}.

\subsubsection{$U(N)$ partition function}

As shown in \figref{figure2} we can calculate the partition function of the $U(N)$ theory by gluing $N$ open string amplitudes $W_{\nu_{a}\nu_{a+1}}$.

\begin{figure}[h]
  \centering
  \includegraphics[width=5in]{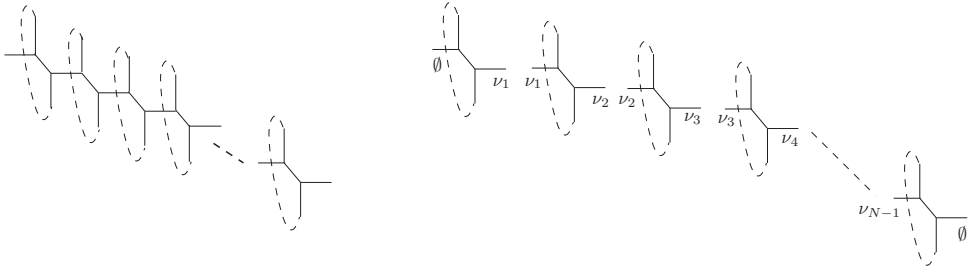}\\
  \caption{(a) The toric diagram of elliptic CY3fold $X_{N}$. (b) Partition function from gluing $W_{\nu_a\nu_{a+1}}$.}\label{figure2}
\end{figure}

The $U(N)$ partition function in terms of the open string amplitude is given by\,,
\bea\nn
Z^{(N)}(\tau,m,t_{f_a},\epsilon_1,\epsilon_2)&=&\sum_{\nu_{1},\mathellipsis,\nu_{N-1}}\left(\prod_{a=1}^{N-1}(-Q_{f_{a}})^{|\nu_{a}|}\right)\\\nn&\times&W_{\emptyset\nu_1}(\epsilon_1,\epsilon_2)W_{\nu_{1}\nu_{2}}(-\epsilon_2,-\epsilon_1)W_{\nu_{2}\nu_{3}}(\epsilon_1,\epsilon_2)\cdots W_{\nu_{N-1}\emptyset}\,.
\eea
If we separate the $U(1)$ piece by defining,
\bea\label{eq:UN}
Z^{(N)}(\tau,m,t_{f_a},\epsilon_1,\epsilon_2)&=&
\left(Z^{(1)}(\tau,m,\epsilon_1,\epsilon_2)\right)^{N}\,\widehat{Z}^{(N)}(\tau,m,t_{f_{a}},\epsilon_1,\epsilon_2)\,,
\eea
then,
\begin{align}\label{unpf}
\widehat{Z}^{(N)}(\tau,m,t_{f_{a}},\epsilon_1,\epsilon_2)=\sum_{\nu_{1},\mathellipsis,\nu_{N-1}}\left(\prod_{a=1}^{N-1}(-Q_{f_{a}})^{|\nu_{a}|}\right)
\prod_{a=1}^{N-1}\prod_{(i,j)\in \nu_{a}}\frac{\theta_{1}(\tau;z^{a}_{ij})\,\theta_{1}(\tau;v^{a}_{ij})}{\theta(\tau;w^{a}_{ij})\theta(\tau;u^{a}_{ij})}
\end{align}
where ($\nu_{0}=\nu_{N}=\emptyset$),
\begin{equation}
\begin{array}{ll}
e^{2\pi i \,z^{a}_{ij}}=Q_{m}^{-1}\,q^{\nu_{a,i}-j+\frac{1}{2}}\,t^{\nu^{t}_{a+1,j}-i+\frac{1}{2}},\qquad &
e^{2\pi i\,v^{a}_{ij}}=Q_{m}^{-1}\,t^{-\nu^{t}_{a-1,j}+i-\frac{1}{2}}\,q^{-\nu_{a,i}+j-\frac{1}{2}},\\
e^{2\pi i \,w^{a}_{ij}}=q^{\nu_{a,i}-j+1}\,t^{\nu_{a,j}^{t}-i},& e^{2\pi i \,u^{a}_{ij}}=q^{\nu_{a,i}-j}\,t^{\nu_{a,j}^{t}-i+1}.
\end{array}
\end{equation}
Notice that for $m=\pm \frac{\epsilon_{1}+\epsilon_{2}}{2}$ we have $z^{(N-1)}_{1,\nu_{a,1}}=-(\nu^{t}_{N,\nu_{N-1,1}})\epsilon_{2}=0$ since $\nu_{N}=\emptyset$ and therefore
\bea
\widehat{Z}^{(N)}\left(\tau,m=\pm \frac{\epsilon_{1}+\epsilon_{2}}{2},t_{f_a},\epsilon_1,\epsilon_2\right)=1\,.
\eea
Using the relation between the theta function and Eisenstein series given in Eq.(\ref{Eisenstein}) we can write $\widehat{Z}^{(N)}(\tau,m,t_{f_{a}},\epsilon_1,\epsilon_2)$ as,
\begin{align}\nn
\widehat{Z}^{(N)}(\tau,m,t_{f_{a}},\epsilon_1,\epsilon_2)&=\sum_{\nu_{1},\mathellipsis, \nu_{N-1}}(-Q_{f_{1}})^{|\nu_{1}|}\cdots(-Q_{f_{N-1}})^{|\nu_{N-1}|} \prod_{a=1}^{N-1}\left(\prod_{(i,j)\in \nu_{a}}\frac{z^{a}_{ij}\,v^{a}_{ij}}{w^{a}_{ij}\,u^{a}_{ij}}\right)\\\nn &\times
\mbox{exp}\left(\sum_{k\geq 1}\frac{B_{2k}}{(2k)(2k)!}E_{2k}(\tau)\,f^{(2k)}_{\nu_{1},\mathellipsis,\nu_{N-1}}(m,\epsilon_{1},\epsilon_{2})\right)\\
f^{(2k)}_{\nu_{1},\mathellipsis,\nu_{N-1}}(m,\epsilon_{1},\epsilon_{2})&=\sum_{a=1}^{N-1}\sum_{(i,j)\in \nu_{a}}\left((2\pi i\,z^{a}_{ij})^{2k}+(2\pi i\,v^{a}_{ij})^{2k}-(2\pi i\,w^{a}_{ij})^{2k}-(2\pi i\,u^{a}_{ij})^{2k}\right)
\end{align}
which shows that if we replace $E_{2}(\tau)$ with the non-holomorphic modular form $E_{2}(\tau,\bar{\tau})$ the partition function becomes modular invariant under the transformation,
\bea\nn
(\tau,m,t_{f_a},\epsilon_{1},\epsilon_{2})\mapsto \Big(\frac{a\tau+b}{c\tau+d},\frac{m}{c\tau+d},t_{f_a},\frac{\epsilon_{1}}{c\tau+d},\frac{\epsilon_{2}}{c\tau+d}\Big),\,\,\,\left(
                                                                                                                                                        \begin{array}{cc}
                                                                                                                                                          a & b \\
                                                                                                                                                           c& d \\
                                                                                                                                                        \end{array}
                                                                                                                                                      \right)\in SL(2,\mathbb{Z})\,.
\eea

\subsubsection{Partition function from instanton calculus}\label{ins2}

In section \ref{ins} we calculated the partition function of the ${\cal N}=2^{*}$ theory using Nekrasov's instanton calculus and observed that it agreed completely with the refined topological string partition function of the elliptic threefold $X_{N}$. A change of the preferred direction gave a different representation of the same refined partition function. In this section we will see that this different representation of the same partition function can also be calculated using instanton calculus for a different gauge theory. This fact reflects the fiber-base duality between the ${\cal N}=2^{*}$ theory and a quiver theory \cite{Katz:1997eq}.

To see let us first consider the case of $U(2)$ ${\cal N}=2^{*}$ theory. The web diagram is shown in \figref{u2de}(a) below.

\begin{figure}[h]
  \centering
  \includegraphics[width=3in]{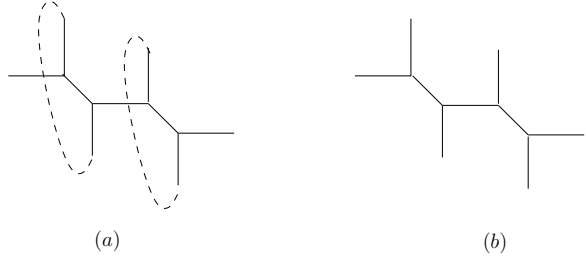}\\
  \caption{(a) The theory on the compactified vertical branes is $U(2)\mapsto U(1)^2$ ${\cal N}=2^{*}$ theory and the theory on the horizontal brane is a $U(1)$ theory. (b) In the limit the circle is decompactified the $U(1)$ theory on the horizontal brane becomes $U(1)$ with $N_{f}=2$.}\label{u2de}
\end{figure}

If we decompactify the circle we get the usual kind of web diagram as shown in \figref{u2de}(b). If we consider the theory on the horizontal branes then the web diagram of \figref{u2de}(b) corresponds to a $U(1)$ theory with two fundamental hypermultiplets. The partition function of the $U(1)$ theory with $N_{f}=2$ can be derived using equivariant instanton calculus \cite{Nekrasov:2002qd},
\begin{align}
\label{eqnf2}\nn
Z&=\sum_{\nu}\varphi^{|\nu|}\,q^{\Arrowvert\nu\Arrowvert^2}\,\prod_{(i,j)\in \nu}\frac{(1-e^{2\pi i m_{1}}\,q^{-j+1}\,t^{i-1})(1-e^{2\pi i m_{2}}\,q^{-j+1}\,t^{i-1})}{(1-q^{\nu_{i}-j+1}\,t^{\nu^{t}_{j}-i})(1-q^{-\nu_{i}+j}\,t^{-\nu^{t}_{j}+i-1})}\\\nn
&=\sum_{\nu}\left(-\varphi\,e^{2\pi i m_{1}}\sqrt{\frac{q}{t}}\right)^{|\nu|}q^{\frac{\Arrowvert\nu\Arrowvert^2}{2}}t^{\frac{\Arrowvert\nu^t\Arrowvert^2}{2}}\,\prod_{(i,j)\in \nu}\frac{(1-e^{-2\pi i m_{1}}\,q^{j-1}\,t^{-i+1})(1-e^{2\pi i m_{2}}\,q^{-j+1}\,t^{i-1})}{(1-q^{\nu_{i}-j+1}\,t^{\nu^{t}_{j}-i})(1-q^{\nu_{i}-j}\,t^{\nu^{t}_{j}-i+1})}\\
&\begin{array}{r}
=\sum_{\nu}(-\widetilde{\varphi})^{|\nu|}\prod_{(i,j)\in \nu} \frac{(e^{-i\pi m_{1}}\,q^{\frac{j-1}{2}}\,t^{-\frac{i-1}{2}}-e^{i\pi m_{1}}\,q^{-\frac{j-1}{2}}\,t^{\frac{i-1}{2}})}{(q^{\frac{\nu_{i}-j+1}{2}}\,t^{\frac{\nu^{t}_{j}-i}{2}}-q^{-\frac{\nu_{i}-j+1}{2}}\,t^{-\frac{\nu^{t}_{j}-i}{2}})}\\\\
\times \frac{(e^{i\pi m_{2}}\,q^{-\frac{j-1}{2}}\,t^{\frac{i-1}{2}}-e^{-i\pi m_{2}}\,q^{\frac{j-1}{2}}\,t^{-\frac{i-1}{2}})}{(q^{\frac{\nu_{i}-j}{2}}\,t^{\frac{\nu^{t}_{j}-i+1}{2}}-q^{-\frac{\nu_{i}-j}{2}}\,t^{-\frac{\nu^{t}_{j}-i+1}{2}})}
\end{array}
\end{align}

where $\widetilde{\varphi}=\varphi\,e^{i\pi (m_{1}+m_{2})}\sqrt{\frac{q}{t}}$. The above partition function is precisely $\widehat{Z}^{(2)}$ given in Eq.(\ref{u2pf1}) in the limit $Q_{\tau}\mapsto 0$ if we make the following identification of parameters\footnote{To see that the two partition functions are the same, apart from parameter identification, one needs the fact that for a fixed $i$ the set $\{j-1\,|\,(i,j)\in \nu\}$ is the same as the set $\{\nu_{i}-j\,|\,(i,j)\in \nu\}$.}:
\bea\label{id}
\widetilde{\varphi}=Q_{f},\,\,\,\,m_{1}=m-\frac{\epsilon_{1}+\epsilon_{2}}{2}\,,\,m_{2}=-m-\frac{\epsilon_{1}+\epsilon_{2}}{2}\,.
\eea
Notice that the above partition function becomes trivial for $m_1=0$ or $m_2=0$ which corresponds to precisely $m=\pm \frac{\epsilon_1+\epsilon_2}{2}$ as expected from the results of section 2.

Eq.(\ref{eqnf2}) is the topological string partition function of the geometry shown \figref{nf2}(a).
\begin{figure}[h]
  \centering
  \includegraphics[width=3in]{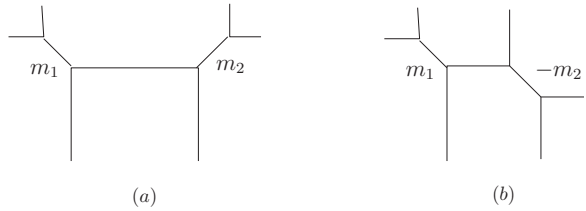}\\
  \caption{Toric realization of the flop transition with two mass parameters.} \label{nf2}
\end{figure}
In Eq.(\ref{id}) taking $m_{1}$ and $m_2$ to be related to $m$ with opposite sign essentially means that in the geometry one of the exceptional curves giving the fundamental hypermultiplet has undergone flop transition and the geometry has become the one shown in \figref{nf2}(b).  Note however that in order for it to come from 6 dimensions, which corresponds to making the
toric geometry periodic, places a restriction on the parameters of the $U(1)$ gauge theory.  In particular
we are at the origin of Coulomb branch where the vev of the scalar field is set to zero.  The only
left over parameters are the coupling constant of the gauge theory which is related to $t_f=1/g^2$ which in turn
is proportional to the separation of M5 branes, as well as the masses which have the same value
(up to sign discussed above).

The instanton partition function for supersymmetric gauge theories can be calculated using an appropriate topological index on the moduli space of instantons, this is essentially due to supersymmetric localization of the partition function with respect to one of the preserved supercharges. The details of the index computation depend on the details of the gauge theory such as the gauge group and the matter content. The matter fields of the theory are sections of a vector bundle on the instanton moduli space. In the case of a fundamental hypermultiplet the vector bundle on the instanton moduli space is the tautological bundle $E$. The tautological bundle $E$ over $\mbox{Hilb}^{k}[\mathbb{C}^2]$ and has rank $k$ . The fiber of $E$ over the point $I$ (which is a codimension $k$ ideal in $\mathbb{C}[z_{1},z_{2}]$) is given by $\mathbb{C}[z_{1},z_{2}]/I$. If we consider $N_f$ hupermultiplets we have $U(N_f)$ global symmetry and $N_f$ copies of the bundle $E$, i.e. the appropriate bundle is
\bea\nn
E\otimes \mathbb{C}^{N_{f}}\cong \underbrace{E\oplus E\oplus \cdots \oplus E}_{N_{f}-\mbox{\tiny copies}}\,,
\eea
where $\mathbb{C}^{N_f}$ is the fundamental representation of $U(N_f)$. $U(N_f)$ acts on the above bundle  on $\mathbb{C}^{N_f}$ and, therefore, the Cartan of $U(N_f)$ acts on the i-th copy of $E$ by scaling by $e^{2\pi i m_i}$ where $m_i\,(i=1,\cdots,N_f)$ is the mass of the i-th hypermultiplet.

The case we are considering is $N_f=2$ in which case we have $E\oplus E$ and the action of $U(1)_{m_1}\times U(1)_{m_2}\subset U(2)$ is given by $(e^{2\pi i m_{1}},e^{2\pi im_{2}})$ on the two factors. However,  due to flop transition on the curve giving the hypermultiplet of mass $m_2$ the relevant bundle for us is $E\oplus E^{*}$ on which the $U(1)_{m_1}\times U(1)_{m_2}$ action is given by $(e^{2\pi i m_{1}},e^{-2\pi im_{2}})$. Apart from this we also have a $U(1)_{\epsilon_1}\times U(1)_{\epsilon_2}$ action on $E\oplus E^{*}$ which comes from lifting the action on $\mathbb{C}^2$. As mentioned before the fiber of $E$ over a point $I\in \mbox{Hilb}^{k}[\mathbb{C}^2]$ is given by $\mathbb{C}[z_1,z_2]/I$ which is a $k$ dimensional vector space. The fixed points of $\mbox{Hilb}^{k}[\mathbb{C}^2]$ under the $U(1)_{m_1}\times U(1)_{m_2}\times U(1)_{\epsilon_1}\times U(1)_{\epsilon_2}$ are labelled by partitions of $k$ and correspond to monomial ideals. \footnote{The fixed points under $U(1)_{m_1}\times U(1)_{m_2}\times U(1)_{\epsilon_1}\times U(1)_{\epsilon_2}$ are the same as fixed points under $U(1)_{\epsilon_1}\times U(1)_{\epsilon_2}$ since $U(1)_{m_1}\times U(1)_{m_2}$ does not act on $\mbox{Hilb}^{k}[\mathbb{C}^2]$.} The ideal corresponding to the partition $\nu$ of $k$ is given by,
\bea
I_{\nu}=\bigoplus_{(i,j)\notin \nu} \mathbb{C}\,z_{1}^{i-1}\,z_{2}^{j-1}\,.
\eea
Thus the fiber of $E$ over the fixed point $I_{\nu}$ is give by
\bea\label{tauw}
E|_{I_{\nu}}=\mathbb{C}[z_1,z_2]/I_{\nu}=\bigoplus_{(i,j)\in \nu}\mathbb{C}\,z_{1}^{i-1}z_{2}^{j-1}\,,
\eea
which shows that the fiber $E$ above the fixed point $I_{\nu}$ decomposes in terms of one dimensional vector spaces spanned by basis vectors $z_{1}^{i-1}z_{2}^{j-1}$. Eq.(\ref{tauw}) also gives us the weights of the $U(1)_{m_1}\times U(1)_{\epsilon_1}\times U(1)_{\epsilon_2}$ action on $E|_{I_{\nu}}$ which are,
\bea
\mbox{weights of $E|_{I_{\nu}}$}=\{\,e^{2\pi i m_1}\,q^{-i+1}\,t^{j-1}\,|\,(i,j)\in \nu\}\,.
\eea
Similarly the weights of $E^*$ at $I_{\nu}$ under $U(1)_{m_2}\times U(1)_{\epsilon_1}\times U(1)_{\epsilon_2}$ are given by
\bea
\mbox{weights of $E^{*}|_{I_{\nu}}$}=\{\,e^{-2\pi i m_2}\,q^{i-1}\,t^{-j+1}\,|\,(i,j)\in \nu\}\,.
\eea
Using the relation between $m_{1},m_{2}$ and $m$ given in Eq.(\ref{id}) we see that under $U(1)_{m}\times U(1)_{\epsilon_1}\times U(1)_{\epsilon_2}$,
\bea\label{weightsE}
\mbox{weights of $E|_{I_{\nu}}$}=\{\,e^{2\pi i m}\,q^{-i+\frac{1}{2}}\,t^{j-\frac{1}{2}}\,|\,(i,j)\in \nu\}\,,\\\nn
\mbox{weights of $E^{*}|_{I_{\nu}}$}=\{\,e^{2\pi i m}\,q^{i-\frac{1}{2}}\,t^{-j+\frac{1}{2}}\,|\,(i,j)\in \nu\}\,.
\eea

Eq.(\ref{eqnf2}) can now be expressed in terms of the equivariant holomorphic Euler characteristic of $E\oplus E^{*}$ as follows,
\bea\label{sum}
Z=\sum_{k\geq 0}\widehat{\varphi}^{k}\,\chi_{y}(\mbox{Hilb}^{k}[\mathbb{C}^2],E\oplus E^{*})\,,
\eea
where $\chi_{y}({\cal M},V)=\sum_{p,q\geq 0}(-1)^{p+q}\,y^{p}\,h^{p,q}({\cal M},V)$ and $h^{p,q}({\cal M},V)=h^{0,q}({\cal M},\wedge^{p}T^{*}{\cal M}^{1,0}\otimes V)$. As $\chi_{y}({\cal M},V)$ is the index of a twisted Dirac operator, it can be given in terms of the Chern classes of $V$ and the ones of $T_{{\cal M}} $\footnote{This can also be written as $\mbox{Tr}(-1)^{F}y^{J}e^{-\beta H}$ for a suspersymmetric quantum mechanical system which is the dimensional reduction of $(2,0)$ $d=2$ theory.},
\bea
\chi_{y}({\cal M},V)=\int_{{\cal M}}\,\mbox{ch}(V_{y})\,\mbox{Td}({\cal M})\,,\,\,\,\,\,V_{y}=\sum_{n\geq 0}\,(-y)^{n}\,\wedge^{n}V\,.
\eea
Using equivariant localization,
\bea
\chi_{y}(\mbox{Hilb}^{k}[\mathbb{C}^2],E\oplus E^{*})&=&\int_{\mbox{Hilb}^{k}[\mathbb{C}^2]}\,\mbox{ch}((E\oplus E^{*})_{y})\,\mbox{Td}(\mbox{Hilb}^{k}[\mathbb{C}^2])\\\nn
&=&\sum_{p\in \{\mbox{\tiny fixed points}\}}\frac{\prod_{j=1}^{2k}(1-y\,e^{-\tilde{x}_{p,i}})}{\prod_{i=1}^{2k}(1-e^{-x_{p,i}})}\,,
\eea
where $\tilde{x}_{p,i}$ and $x_{p,i}$ are weights at the fixed point $p$ of the $U(1)_{m}\times U(1)_{\epsilon_1}\times U(1)_{\epsilon_2}$ action for $E\oplus E^{*}$ and $T_{\mbox{Hilb}^{k}[\mathbb{C}^2]}$ respectively.
 The Chern character of the tautological bundle $E$ was calculated in \cite{tautological} and it was shown that
\bea
\mbox{ch}(T_{\mbox{Hilb}^{k}[\mathbb{C}^2]})=
\mbox{ch}(E\oplus E^{*})\,.
\eea
However as equivariant bundles $T_{\mbox{Hilb}^{k}[\mathbb{C}^2]}$ has different weights than $E\oplus E^{*}$. Using the weights given in Eq.(\ref{weightsE}) we get
\bea\nn
\chi_{y}(\mbox{Hilb}^{k}[\mathbb{C}^2],E\oplus E^{*})&=&\sum_{|\nu|=k}\prod_{(i,j)\in \nu}\frac{(1-y\,Q_{m}^{-1}\,q^{i-\frac{1}{2}}t^{-j+\frac{1}{2}})(1-y\,Q_{m}^{-1}\,q^{-i+\frac{1}{2}}t^{j-\frac{1}{2}})}
{(1-q^{\nu^{t}_{j}-i+1}t^{\nu_{i}-j})(1-q^{-\nu^{t}_{j}+i}t^{-\nu_{i}+j-1})}\,.
\eea
After some simplification Eq.(\ref{sum}) then becomes
\begin{align}\label{inspf}
Z=\sum_{\nu}(-\widehat{\varphi}\,y\,Q_{m}^{-1}\sqrt{\frac{t}{q}})^{|\nu|}\prod_{(i,j)\in \nu}\frac{y^{\frac{1}{2}}Q_{m}^{-\frac{1}{2}}q^{\frac{i-\frac{1}{2}}{2}}t^{\frac{-j+\frac{1}{2}}{2}}-y^{-\frac{1}{2}}Q_{m}^{\frac{1}{2}}
q^{-\frac{i-\frac{1}{2}}{2}}t^{-\frac{-j+\frac{1}{2}}{2}}}
{q^{\frac{\nu^{t}_{j}-i+1}{2}}\,t^{\frac{\nu_{i}-j}{2}}-q^{-\frac{\nu^{t}_{j}+i-1}{2}}\,t^{-\frac{\nu_{i}+j}{2}}}\\\nn
\prod_{(i,j)\in \nu}\frac{y^{\frac{1}{2}}Q_{m}^{-\frac{1}{2}}q^{-\frac{i-\frac{1}{2}}{2}}t^{-\frac{-j+\frac{1}{2}}{2}}-y^{-\frac{1}{2}}Q_{m}^{\frac{1}{2}}
q^{\frac{i-\frac{1}{2}}{2}}t^{\frac{-j+\frac{1}{2}}{2}}}
{q^{\frac{\nu^{t}_{j}-i}{2}}t^{\frac{\nu_{i}-j+1}{2}}-q^{\frac{-\nu^{t}_{j}+i}{2}}t^{\frac{-\nu_{i}+j}{2}}}
\end{align}
This is precisely the partition function Eq.(\ref{u2pf1}) with $y=1$, $Q_{f}=\widehat{\varphi}\,y\,Q_{m}^{-1}\sqrt{\frac{t}{q}}$ and $Q_{\tau}=0$. Thus the choice of the horizontal preferred direction gives the twisted $\chi_y$ genus of the Hilbert scheme of points on $\mathbb{C}^2$. It is easy to generalize this to the case $Q_{\tau}\neq 0$. From Eq.(\ref{u2pf1}) it is clear that we need to replace each factor in the product in Eq.(\ref{inspf}) with $\theta_{1}(\tau;z)$. This is achieved by considering the elliptic genus rather than the $\chi_{y}$ genus.

Now consider the following formal combination of bundles \cite{gritsenko}:
\bea\label{fc}
V_{Q_{\tau,y}}=\bigotimes_{k=0}^{\infty} \bigwedge_{-y\,Q_{\tau}^{k-1}}\,V\otimes \bigotimes_{k=1}^{\infty}\bigwedge_{-y^{-1}\,Q_{\tau}^{k}}V^{*}\otimes \bigotimes_{k=1}^{\infty}S_{Q_{\tau}^{k}}T^{*}_{{\cal M}}\otimes \bigotimes_{k=1}^{\infty}S_{Q_{\tau}^{k}}T_{{\cal M}}
\eea
Then
\bea\label{ellipticgenus}
\mbox{ch}(V_{Q_{\tau},y})\mbox{Td}(T_{{\cal M}})=\prod_{k=1}^{\infty}\frac{\prod_{i=1}^{r}(1-Q_{\tau}^{k-1}y\,e^{-\tilde{x}_{i}})
(1-Q_{\tau}^{k}\,y^{-1}\,e^{\tilde{x}_{i}})}{\prod_{j=1}^{d}x_{j}^{-1}(1-Q_{\tau}^{k-1}e^{-x_{j}})
(1-Q_{\tau}^{k}e^{x_{j}})}
\eea
where $r$ is the rank of $V$ which in this case is equal to the dimension of ${\cal M}$, $x_{j}$ are the Chern roots of the tangent bundle and $\tilde{x}_{i}$ are Chern root of $V$. In terms of theta functions we have $(y=e^{2\pi i z})$:
\bea
\mbox{ch}(V_{Q_{\tau},y})\mbox{Td}(T_{{\cal M}})=y^{d/2}\,\,e^{\sum_{i=1}^{d}\frac{1}{2}(x_{i}-\tilde{x}_{i})}\,\prod_{i=1}^{d}\frac{\theta_{1}(\tau;-z+\frac{\tilde{x}_{i}}{2\pi i})}{\theta_{1}(\tau;\frac{x_{i}}{2\pi i})}
\eea
Taking $V=E\oplus E^{*}$ and ${\cal M}=\mbox{Hilb}^{k}(\mathbb{C}^{2})$ we get
\bea\nn
&&\sum_{k\geq 0}\widetilde{Q}^{k}\int_{{\cal M}}\mbox{ch}((E\oplus E^{*})_{Q_{\tau},y})\mbox{Td}(T_{{\cal M}})=
\sum_{k\geq 0}(\widetilde{Q}y)^{k}\int_{{\cal M}}e^{\sum_{i=1}^{d}\frac{1}{2}(x_{i}-\tilde{x}_{i})}\prod_{i=1}^{d}\frac{\theta_{1}(\tau;z-\frac{\tilde{x}_{i}}{2\pi i})}{\theta_{1}(\tau;-\frac{x_{i}}{2\pi i})}\\\nn
&=&\sum_{\nu}(\widetilde{Q}y)^{|\nu|}\prod_{(i,j)\in \nu}\frac{\theta_{1}(\tau;z-m+i\,\epsilon_{1}-j\,\epsilon_{2}-\frac{\epsilon_{+}}{2})\,
\theta_{1}(\tau;z-m-i\,\epsilon_{1}+j\,\epsilon_{2}+\frac{\epsilon_{+}}{2})}
{\theta_{1}(\tau;-(\nu^{t}_{j}-i+1)\epsilon_{1}+(\nu_{i}-j)\epsilon_{2})
\theta_{1}(\tau;(\nu^{t}_{j}-i)\epsilon_{1}-(\nu_{i}-j+1)\epsilon_{2})}
\eea
Comparing with Eq.(\ref{u2pf1}) we see that the above precisely agrees with $\widehat{Z}^{(2)}$, the topological vertex result for $z=0$.\\

The bundle $V=E\oplus E^{*}$ on $\mbox{Hilb}^{k}[\mathbb{C}^2]$ played a crucial role in the calculation above which corresponds to the case of two M5-branes. If $I\in \mbox{Hilb}^{k}[\mathbb{C}^2]$ then $I$ is an ideal such that $\mbox{dim}_{\mathbb{C}}(\mathbb{C}[x,y]/I)=k$. Given such an ideal the fiber of $V$ above $I$ is given by \cite{Nakajimabook,CO,smirnov},
\bea\label{bundlev}
V|_{I}&=&\underbrace{\mbox{Ext}^{1}({\cal O},I)\otimes L^{-\frac{1}{2}}}_{E}\oplus \underbrace{\mbox{Ext}^{1}(I,{\cal O})\otimes L^{-\frac{1}{2}}}_{E^{*}}\,,
\eea
where $L$ is a trivial line bundle and ${\cal O}=\mathbb{C}[x,y]$. $L$ is in fact the canonical line bundle on $\mathbb{C}^2$ such that the weight of $L^{-\frac{1}{2}}$ is $e^{-2\pi i\frac{\epsilon_{1}+\epsilon_{2}}{2}}$. The appearance of $\mbox{Ext}$-groups is not unexpected since it has been shown that Ext-groups count the open string states between D-branes wrapped on holomorphic submanifolds \cite{Katz:2002gh}. Notice that $\mbox{Ext}^{1}({\cal O},I)$ and $\mbox{Ext}^{1}(I,{\cal O})$ are not dual of each other but are such that \cite{CO,smirnov},
\bea
\mbox{Ext}^{1}(I,{\cal O})=\mbox{Ext}^{1}({\cal O},I)^{*}\otimes L|_{I}\,,
\eea
this implies that the two factors of $V$ in Eq.(\ref{bundlev}) are dual to each other.

Now consider $(I,J)\in \mbox{Hilb}^{k_1}[\mathbb{C}^2]\times \mbox{Hilb}^{k_2}[\mathbb{C}^2]$ and a bundle $V$ on $\mbox{Hilb}^{k_1}[\mathbb{C}^2]\times \mbox{Hilb}^{k_2}[\mathbb{C}^2]$ such that its fiber over the point $(I,J)$ is given by,
\bea
V|_{(I,J)}=\Big(\mbox{Ext}^{1}({\cal O},I)\oplus \mbox{Ext}^{1}(I,J)\oplus \mbox{Ext}^{1}(J,{\cal O})\Big)\otimes L^{-\frac{1}{2}}\,.
\eea
If we take $k_1=0$ or $k_2=0$ then the above bundle becomes the bundle of Eq.(\ref{bundlev}) on $\mbox{Hilb}^{\bullet}[\mathbb{C}^2]$. At the fixed point $(I_{\lambda},J_{\mu})$ labelled by two partitions $\lambda$ and $\mu$ of $k_1$ and $k_2$ respectively, the weights of the bundle $V$ are given by \cite{CO},\footnote{The Young diagram convention of \cite{CO} is different from ours so we have to take the transpose of their partitions.}
\bea
\mbox{weights of $V$}&=&\{Q_{m}\,q^{-i+\frac{1}{2}}\,t^{j-\frac{1}{2}}\,,\,Q_{m}\,q^{\lambda^{t}_{j}-i+\frac{1}{2}}\,t^{\mu_{i}-j+\frac{1}{2}}\,|\,(i,j)\in \lambda\}\cup \\\nn &&\{Q_{m}\,q^{-\mu^{t}_{j}+i-\frac{1}{2}}\,t^{-\lambda_{i}+j-\frac{1}{2}}\,,\,Q_{m}\,q^{i-\frac{1}{2}}\,t^{-j+\frac{1}{2}}|\,(i,j)\in \mu\}\,,
\eea
where we have included the $U(1)_{m}$ weight as well in the above. Thus for each M5-brane with M2-branes ending on the left and the right we have a factor $\mbox{Ext}^{1}(I,J)\otimes L^{-\frac{1}{2}}$ in the corresponding bundle where $I$ is a point on $\mbox{Hilb}^{k1}[\mathbb{C}^2]$ corresponding to the M2-brane on the left and $J$ is a point on $\mbox{Hilb}^{k_2}[\mathbb{C}^2]$ corresponding to the M2-brane on the right as shown in \figref{figbundle}.

\begin{figure}[h]
  \centering
  \includegraphics[width=1.5in]{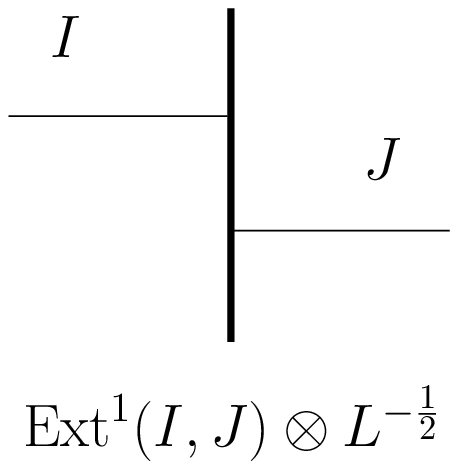}\\
  \caption{Each pair of M2 branes ending on the M5 brane from opposite sides gives rise to a factor $\mbox{Ext}^{1}(I,J)\otimes L^{-\frac{1}{2}}$ in the corresponding bundle.}\label{figbundle}
\end{figure}

We can now generalize this construction for ${\cal M}_{k_1,\mathellipsis,k_{N-1}}=\mbox{Hilb}^{[k_1]}[\mathbb{C}^2]\times \cdots \times \mbox{Hilb}^{[k_{N-1}]}[\mathbb{C}^2]$. Let $V$ be the bundle on ${\cal M}_{k_{1},\mathellipsis,k_{N-1}}$ such that its fiber above $(I_{1},I_2,\mathellipsis,I_{N-1})\in {\cal M}_{k_{1},\mathellipsis,k_{N-1}}$ is given by,
\bea
V|_{(I_{1},I_2,\cdots,I_{N-1})}=\Big(\oplus_{a=0}^{N-1}\,\mbox{Ext}^{1}(I_{a},I_{a+1})\Big)\otimes L^{-\frac{1}{2}}\,,
\eea
where $I_{0}=I_{N}={\cal O}$. At a fixed point labelled by $(\nu_1,\nu_2,\cdots,\nu_{N-1})$ the weights of $V$ are given by \cite{CO},
\bea\label{weights3}
&&\{Q_{m}\,q^{-i+\frac{1}{2}}\,t^{j-\frac{1}{2}}\,|\,(i,j)\in \nu_{1}\}\cup\{Q_{m}\,q^{i-\frac{1}{2}}\,t^{-j+\frac{1}{2}}\,|\,(i,j)\in \nu_{N-1}\}\\\nn
&&\Big(\cup_{a=1}^{N-2}\{Q_{m}\,q^{\nu^{t}_{a,j}-i+\frac{1}{2}}\,t^{\nu_{a+1,i}-j+\frac{1}{2}}\,|\,(i,j)\in \nu_{a}\}\cup\{Q_{m}\,q^{-\nu^{t}_{a+1,j}+i-\frac{1}{2}}\,t^{-\nu_{a,i}+j-\frac{1}{2}}|\,(i,j)\in \nu_{a+1}\}\Big)\,.
\eea
We can now write down the partition function,
\bea\label{calc1}
Z_{U(1)^{N}}=\sum_{k_{1},\mathellipsis, k_{N-1}\geq 0}\left(\prod_{a=1}^{N-1}\widetilde{Q}_{a}^{k_{a}}\right)\,\int_{{\cal M}_{k_1,\mathellipsis,k_{N-1}}}\mbox{ch}(V_{Q_{\tau},y})\mbox{Td}({\cal M}_{k_1,\mathellipsis,k_{N-1}})
\eea
using Eq.(\ref{weights3}) to obtain,
\bea
Z_{U(1)^{N}}=\sum_{\nu_1,\mathellipsis,\nu_{N-1}}\left(\prod_{a=1}^{N-1}\widetilde{Q}_{a}^{|\nu_a|}\right)\prod_{a=1}^{N-1}\prod_{(i,j)\in \nu_a}\frac{\theta_{1}(\tau;a_{ij})\theta_{1}(\tau;b_{ij})}{\theta_{1}(\tau;c_{ij})\theta_{1}(\tau;d_{ij})}\,,
\eea
with
\begin{equation}
\begin{array}{ll}
e^{2\pi i\,a_{ij}}=y\,Q_{m}^{-1}q^{-\nu^{t}_{a-1,j}+i-\frac{1}{2}}\,t^{-\nu_{a,i}+j-\frac{1}{2}},\qquad & e^{2\pi i b_{ij}}=y\,Q_{m}^{-1}\,q^{\nu^{t}_{a+1,j}-i+\frac{1}{2}}\,t^{\nu_{a,i}-j+\frac{1}{2}},\\
e^{2\pi i c_{ij}}=q^{\nu^{t}_{a,j}-i}\,t^{\nu_{a,i}-j+1},& e^{2\pi i d_{ij}}=q^{-\nu^{t}_{a,j}+i-1}\,t^{-\nu_{a,i}+j}\,.
\end{array}
\end{equation}
The above partition function is precisely the partition function of Eq.(\ref{unpf}) if we interchange $t$ and $q$, take $y=1$ and $\widetilde{Q}_{a}=Q_{f_{a}}$.

\section{Elliptic Genus of M-strings}

In this section we interpret the results of the computations done in section 3,
using the fact that topological string partition function is the same as the BPS partition function.
We start by reviewing the types of
BPS states which we obtain by compactifying the M5 brane theory on $S^1$.  The BPS states can be divided
to two classes:  Those that are coming from tensor multiplets in 6$d$ and their tower of KK modes, and
those that arise due to wrapping of the strings around the circle.  The latter is the main focus for our work,
but we also review how the KK tower shows up in the topological string computation, and
how we can normalize the topological string partition function to be left entirely with the
partition function of the suspended M2 branes and the corresponding wrapped M-strings.
Turning our attention to wrapped string states and their elliptic genus, we first study the case with
two M5 branes and a single M2 brane suspended between them.  In this case we find
that the theory on the resulting string is as expected the sigma model on ${\mathbb R}^4$, and show how our
BPS formula agrees with this result.  We then consider the case with two M2 branes suspended
between two M5 branes.  One may naively expect the theory on the string to be $\mbox{Sym}^2({\mathbb R}^4)$, but
it turns out not to be the case.  We explain in detail the similarity and the difference of our
result with this expectation.  Finally we explain how the elliptic genus for $n$ M2 branes suspended
between 2 M5 branes can be related to elliptic genus of a $(4,0)$ theory on
$\mbox{Hilb}^n(\mathbb{R}^4)$ but where the right-moving fermions are coupled to a bundle distinct from, but with the same
Chern character as, the tangent bundle.
Finally we discuss the situation where we have more than two M5 branes.  In this case we find that
new bound states of wrapped strings can arise due to compactification on the circle, and the BPS
degeneracies do not factorize as product of pairs of bound states (this is related to the fact
that $m \not= 0$).
We then discuss the interpretation of our result in terms of domain walls formed between different
numbers of M2 branes
ending on the same M5 branes.  Our computations lead to the elliptic genus of these $2d$ domain walls.
Finally we consider special values of $m$ and in particular when $m=\pm\frac{1}{2} (\epsilon_1+\epsilon_2)$
and confirm that our partition function agrees with the elliptic genus of $(4,4)$ $A_n$ quiver theories
(at least for some of the $A_1$ cases we checked).

\subsection{Sigma model, BPS states and elliptic genus}
Beginning with the case of the single M5 brane let us try to understand in detail the physical meaning of the partition functions calculated in the last section.
We are interested in identifying the BPS particles and understanding their multiplicities. These have two sources: BPS states which come from particles in six dimensions, and BPS states which come from BPS strings in six dimensions wrapped around the $S^1$.
For a single M5 brane the only 5$d$ BPS states come from particles in 6 dimensions \cite{ Lockhart:2012vp} as we
review next.

\subsubsection{Single M5 brane and $Z^{(1)}$} Consider the case of a single M5 brane wrapped on the circle. As mentioned in the last section the six dimensional theory on the M5 brane has a tensor multiplet which consists of a self-dual two form field, four symplectic Majorana-Weyl spinors and five real scalar fields. Compactification of the M5 brane on the $S^1$ then gives the Kaluza-Klein modes of the tensor multiplet in five dimensions. There will be two kinds of multiplets in five dimensions, the massless and the massive. The little group of massless particles in five dimensions is $Spin(3)=SU(2)$ and the little group of massive particle in five dimensions is $Spin(4)=SU(2)_{L}\times SU(2)_{R}$. If the radius of the circle is $R$ then there will be a massive multiplet, $\Phi_{k}$, for each $k\in \mathbb{Z},\,k\neq 0$ with mass $\frac{|k|}{R}$. For each $k$ the fields in the massive multiplet (which is actually $(2,0)$ multiplet of the ${\cal N}=2$ supersymmetry in five dimensions) are in the following representation
\bea\label{massive}
SU(2)_{L}\times SU(2)_{R}:\,\,\,\Phi_{k}=\left(1,0\right)\oplus 4\left(\frac{1}{2},0\right)\oplus 5\left(0,0\right).
\eea
The massless multiplet, $\Phi_{0}$, (which is the five dimensional vector multiplet containing the massless vector field) is given by
\bea\label{massless}
Spin(3):\,\,\,\Phi_{0}=\left(1\right)\oplus 4\left(\frac{1}{2}\right)\oplus 5\left(0\right).
\eea
Now consider the topological string partition function corresponding to the geometry $X_{1}$ with toric diagram given by \figref{figure1} for $N=1$. This geometry is dual to the brane web in which we have a single M5 brane wrapped on a circle with a mass deformation $m$. In the limit $m\mapsto 0$ we get the ${\cal N}=2$ SYM in five dimensions. The topological string partition function for this geometry is given by \cite{Iqbal:2008ra}\,,
\bea\nn
Z^{(1)}=M(t,q) \prod_{k=1}^{\infty}(1-Q_{\tau}^{k})^{-1}\prod_{i,j=1}^{\infty}\frac{(1-Q_{\tau}^{k-1}\,Q_{m}\,q^{i-\frac{1}{2}}\,t^{j-\frac{1}{2}})
(1-Q_{\tau}^{k-1}Q_{1}q^{i-\frac{1}{2}}t^{j-\frac{1}{2}})}{
(1-Q_{\tau}^{k}q^{i}t^{j-1})(1-Q_{\tau}^{k}q^{i-1}t^{j})}\,,\\ \label{u1pf}
\eea
where $M(t,q)=\prod_{i,j=1}^{\infty}(1-q^{i}t^{j-1})^{-1}$, $Q_{m}=e^{2\pi i m}$ and $Q_{1}Q_{m}=Q_{\tau}=e^{2\pi i \tau}$ with $\tau=\frac{1}{R}$.

Recall that in M-theory compactification on a Calabi-Yau threefold the massive particles in five dimensions, classified by the little group $SU(2)_{L}\times SU(2)_{R}$, come from M2 branes wrapping the holomorphic curves. The topological string partition function of a Calabi-Yau threefold contains the information about the spin content of these BPS particles \cite{Gopakumar:1998ii, Gopakumar:1998jq}. The refined topological string partition function can be written in terms of BPS multiplicities as follows \cite{Gopakumar:1998ii, Gopakumar:1998jq,Hollowood:2003cv}:
\bea\nn
Z(\omega,q,t) &=& \\\nn
&&\hspace{-.88in}\prod_{\Sigma\in H_{2}(X,\mathbb{Z})}\prod_{j_{L},j_{R}}\prod_{k_{L}=-j_{L}}^{+j_{L}}\prod_{k_R=-j_R}^{+j_R} \prod_{i,j=1}^{\infty}\Big(1-q^{k_{L}-k_{R}+i-\frac{1}{2}}\,t^{k_{L}+k_{R}+j-\frac{1}{2}}\,e^{-\int_{\Sigma}\omega}\Big)^{(-1)^{2j_{L}+2j_{R}}\,N_{j_{L},j_{R}}(\Sigma)}\,\\\nn
&=&\mbox{exp}\left(-\sum_{n=1}^{\infty}\frac{F(n\omega,q^{n},t^{n})}{n(q^{\frac{n}{2}}-q^{-\frac{n}{2}})(t^{\frac{n}{2}}-t^{-\frac{n}{2}})}\right),
\eea
where
\begin{align}
F(\omega,q,t) = \sum_{\Sigma\in H_{2}(X,\mathbb{Z})}\sum_{j_{L},j_{R}}\sum_{k_{L},k_{R}}\,e^{-\int_{\Sigma}\omega}\,(-1)^{2j_{L}+2j_{R}}\,N_{j_{L},j_{R}}(\Sigma)
\,\mbox{Tr}_{j_{L}}\left(q\,t\right)^{\frac{j_{L,3}}{2}}\,\mbox{Tr}_{j_{R}}\left(\frac{q}{t}\right)^{\frac{j_{R,3}}{2}} \label{tspf}
\end{align}
and $N_{j_{L},j_{R}}(\Sigma)$ is the number of particles with charge $\Sigma$ and $SU(2)_{L}\times SU(2)_{R}$ representation $(j_{L},j_{R})$ and $\omega$ is the complexified K\"ahler class of $X$. Comparing Eq.(\ref{u1pf}) and Eq.(\ref{tspf}) we see that,
\bea\nn
F_{U(1)}+\overline{F_{U(1)}}=\underbrace{Q_{m}+Q_{m}^{-1}-\sqrt{\frac{q}{t}}-\sqrt{\frac{t}{q}}}_{massless}+\sum_{k\in \mathbb{Z},k\neq 0}^{\infty}\underbrace{Q_{\tau}^{k}\Big[(Q_{m}+Q_{m}^{-1})-(\sqrt{q\,t}+\frac{1}{\sqrt{q\,t}})\Big]}_{massive}\,.
\eea
In the limit $m\mapsto 0$ we see that the massless and the massive multiplets are in the following representation of the $SU(2)_{L}\times SU(2)_{R}$,
\bea
\mbox{\it massless}:&&\, \left(0,\frac{1}{2}\right)\oplus 2\left(0,0\right)\,,\\\nn
\mbox{\it massive}:&&\,\,\left(\frac{1}{2},0\right)\oplus 2\left(0,0\right)\,.
\eea
This, however, is not the complete story. We need to take into account the universal half-hypermultiplet associated with the position of the particle to get the full $Spin(4)$ content \cite{Gopakumar:1998jq}. Tensoring the above with the half hypermultiplet $(\frac{1}{2},0)\oplus 2(0,0)$ we get
\bea\nn
\mbox{\it massless}:&&\,\,\left(\left(\frac{1}{2},0\right)\oplus 2\left(0,0\right)\right)\otimes\left( \left(0,\frac{1}{2}\right)\oplus 2\left(0,0\right)\right)\\\nn
&=&\left(\frac{1}{2},\frac{1}{2}\right)\oplus 2\left(\frac{1}{2},0\right)\oplus 2
\left(0,\frac{1}{2}\right)\oplus 4\left(0,0\right)\mapsto \left(1\right)\oplus 4\left(\frac{1}{2}\right)\oplus 5\left(0\right),\\\nn
\mbox{\it massive}:&&\,\,\left(\left(\frac{1}{2},0\right)\oplus 2\left(0,0\right)\right)\otimes\left( \left(\frac{1}{2},0\right)\oplus 2\left(0,0\right)\right)\\\nn
&=&\left(1,0\right)\oplus 4\left(\frac{1}{2},0\right)\oplus 5\left(0,0\right).
\eea
This is precisely the spin content of the massless and the massive modes of the tensor multiplet on the circle given in Eq.(\ref{massive}) and Eq.(\ref{massless}).

\subsubsection{One M2 brane suspended between two M5 branes}
Recall the partition function we obtained for two M5 branes:

\bea\nn
Z^{(2)}(\tau,m,t_{f},\epsilon_1,\epsilon_2)&=&\left(Z^{(1)}(\tau,m,\epsilon_1,\epsilon_2)\right)^{2}\\\label{utwopf}
&\times&\underbrace{\sum_{\nu}(-Q_{f})^{|\nu|}\,\prod_{(i,j)\in \nu}\frac{\theta_{1}(\tau;z_{ij})\,\theta_{1}(\tau;v_{ij})}{\theta_{1}(\tau;w_{ij})\theta_{1}(\tau;u_{ij})}}_{\widehat{Z}^{(2)}(\tau,m,t_{f},\epsilon_1,\epsilon_2)}
\eea

with the following definitions for the arguments of the $\theta$-functions
\begin{equation}\label{varii}
\begin{array}{ll}
e^{2\pi i \,z_{ij}}=Q_{m}^{-1}\,q^{\nu_{i}-j+1/2}\,t^{-i+1/2},\qquad &
e^{2\pi i\,v_{ij}}=Q_{m}^{-1}\,t^{i-1/2}\,q^{-\nu_{i}+j-1/2},\\
e^{2\pi i \,w_{ij}}=q^{\nu_{i}-j+1}\,t^{\nu_{j}^{t}-i},& e^{2\pi i \,u_{ij}}=q^{\nu_{i}-j}\,t^{\nu_{j}^{t}-i+1}.
\end{array}
\end{equation}
where $q=e^{2\pi i \epsilon_1}, t=e^{-2\pi i \epsilon_2}, Q_m=e^{2\pi im}$.
We can interpret these contributions as follows:  The $(Z^{(1)})^2$ term captures the
KK tower of states of the two tensor multiplets we have in 6$d$, extending our discussion for
the case of single M5 brane in the previous section.  The term ${\widehat{Z}^{(2)}}$ we interpret
as the contribution of BPS states coming from M-strings which are wrapped around the
circle.  Moreover the term $|\nu|$ counts the number of M-strings branes wrapping the circle.
In particular we write
$$ {\widehat{Z}^{(2)}}(\tau,m,t_{f},\epsilon_1,\epsilon_2)=\sum_n (-Q_f)^n \widehat{Z}_n^{(2)}(\tau,m,\epsilon_1,\epsilon_2)$$
where $\widehat{Z}_n^{(2)}$ corresponds to the partition function of $n$ suspended and wrapped M2 branes between two M5 branes:
$$\widehat{Z}_n^{(2)}=\sum_{|\nu|=n}\prod_{(i,j)\in \nu}\frac{\theta_{1}(\tau;z_{ij})\,\theta_{1}(\tau;v_{ij})}{\theta_{1}(\tau;w_{ij})\theta_{1}(\tau;u_{ij})}$$
Considering the case of a single M2 brane, corresponding to $n=1$ we have
\begin{align}\nn
\widehat{Z}_1^{(2)}&=\frac{\theta_{1}(\tau;-m+(\epsilon_1+\epsilon_2)/2)\,\theta_{1}(\tau;m+(\epsilon_1+\epsilon_2)/2))}{\theta_{1}(\tau;\epsilon_1)\theta_{1}(\tau;\epsilon_2)}\\
&=\prod_{k=1}^{\infty} \frac{(1-Q_\tau^k Q_m^{\pm 1}\,q^{1/2}t^{-1/2})(1-Q_\tau^{k-1}Q_m^{\pm 1}\, q^{-1/2}t^{1/2})}{(1-Q_\tau^k \,q)(1-Q_\tau^{k-1}\,q^{-1})(1-Q_\tau^n\, t^{-1})(1-Q_\tau^{n-1}\,t)}
\end{align}
where by $\pm 1$ we mean to include one factor with each of the two signs.   This expression can be
manifestly identified with the partition function of a sigma model on ${\mathbb{R}}^4$ where the four fermionic
oscillators in the numerator are twisted according to the $Spin(4)_R$ twist parameters and the four
bosonic oscillators in the denominator are twisted according to the $Spin(4)$ parameter, and represent
the 4 directions parallel to the M5 brane but perpendicular to the M2 brane.  In particular the twists
in the Cartan of
$Spin(4)\times Spin(4)_R$ are
$$(\epsilon_1,\epsilon_2, -m+({\epsilon_1}+\epsilon_2)/2,m+({\epsilon_1}+\epsilon_2)/2)$$
This is also what we would have expected for the light-cone partition function of the chiral Green-Schwarz
string in 6 dimensions, where $\epsilon_i$ are identified with the Cartans of $SO(4)$ rotation group.

\subsubsection{Two M2 branes suspended between two M5 branes}
To extract the contribution from two M2 branes suspended between two M5 branes we simply take
the term $|\nu|=2$ in Eq.(\ref{utwopf}).  The most naive expectation, as already discussed is
that this should correspond to sigma model on $\mbox{Sym}^2{\mathbb{R}^4}$.  Indeed this term has some similarities
to this expression, namely the ratio of the four theta functions can be interpreted as that of eight fermionic oscillators and eight bosonic oscillators.  However, the structure of the sum of the partitions and the corresponding
charges of the fermions are not what one may expect based on the symmetric product structure.
Namely, if one studies the elliptic genus of the 2-fold symmetric product of $\mathbb{R}^4$ one finds
$$\hspace{-.5in}Z_{\mbox{Sym}^2\left(\mathbb{R}^4\right)}= \frac{1}{2}\bigg(\bigg(\sq{1}{1}+\sq{g}{1}\bigg)+\bigg(\sq{1}{g}+\sq{g}{g}\bigg)\bigg)$$
$$={1\over 2}\bigg[\widehat{Z}_1^{(2)}(\tau,\epsilon_1,\epsilon_2,m)^2 +\widehat{Z}^{(2)}_1(2\tau,2\epsilon_1,2\epsilon_2,2m)$$
$$  \hspace{.75in} +\, \widehat{Z}^{(2)}_1(\tau /2 ,\epsilon_1,\epsilon_2,m)+\widehat{Z}^{(2)}_1((\tau+1)/2,\epsilon_1,\epsilon_2,m)\bigg].$$
where $g$ is the order two twist given by exchanging the two $R^4$'s.
However, surprisingly enough we find
$$\widehat{Z}^{(2)}_2\not= Z_{\mbox{Sym}^2\left(\mathbb{R}^4\right)}!$$
The closest we can make this to look like the symmetric product partition function is

\begin{eqnarray}
	\widehat{Z}^{(2)}_2 & =& \frac{1}{2}\bigg(\bigg(\sq{1}{1}-\sq{g}{1}\bigg)+\bigg(\sq{1}{g}-\sq{g}{g}\bigg)\cdot Y\bigg)\nonumber \\
	~    & = & {1\over 2}\bigg((\widehat{Z}^{(2)}_1(\tau,\epsilon_1,\epsilon_2,m)^2 -\widehat{Z}^{(2)}_1(2\tau,2\epsilon_1,2\epsilon_2,2m)) \nonumber \\
  ~     &  ~ &+(\widehat{Z}^{(2)}_1(\tau /2 ,\epsilon_1,\epsilon_2,m)
 -\widehat{Z}^{(2)}_1((\tau+1)/2,\epsilon_1,\epsilon_2,m))Y(\tau, \epsilon_1,\epsilon_2)\bigg),
\end{eqnarray}
with
\[Y(Q_\tau,q,t) =\frac{\theta_2(\tau;0)\theta_2(\tau;\epsilon_1+\epsilon_2)}{\theta_2(\tau;\epsilon_1)\theta_2(\tau;\epsilon_2)}
\frac{\frac{\theta_3(\tau;0)\theta_3(\tau;\epsilon_1+\epsilon_2)}{\theta_3(\tau;\epsilon_1)\theta_3(\tau;\epsilon_2)}\frac{\theta_4(\tau;0)\theta_4(\tau;\epsilon_1+\epsilon_2)}{\theta_4(\tau;\epsilon_1)\theta_4(\tau;\epsilon_2)}-1}{\frac{\theta_3(\tau;0)\theta_3(\tau;\epsilon_1+\epsilon_2)}{\theta_3(\tau;\epsilon_1)\theta_3(\tau;\epsilon_2)}-\frac{\theta_4(\tau;0)\theta_4(\tau;\epsilon_1+\epsilon_2)}{\theta_4(\tau;\epsilon_1)\theta_4(\tau;\epsilon_2)}},\]
where
\bea\nn
\theta_2(\tau; z) &=& -\theta_1(\tau; z-1/2),\\\nn
\theta_3(\tau; z) &=& -\exp(\pi i (-z+\tau/4))\theta_1(\tau; z-1/2-\tau/2),\\\nn
\theta_4(\tau; z) &=& \,\,i\,\exp(\pi i(-z+\tau/4))\theta_1(\tau; z-\tau/2).
\eea
Note that this structure is not entirely a trivial rewriting of $\widehat{Z}^{(2)}_2$ because $Y$ does not depend on $m$.
So in particular the $m$ dependence of $\widehat{Z}^{(2)}_2$ is captured by the symmetric product Hilbert space.  It would be interesting
to see if this structure can be better understood.

Despite the fact that we could not write the $\widehat{Z}^{(2)}_2$ in terms of the $\mbox{Hilb}^2(\mathbb{R}^4)$ it turns out there is a way to
relate it to the sigma model on the same space but a bundle different from the tangent bundle.  We explain this in the next section, for the
general $\widehat{Z}^{(2)}_n$ case.

\subsubsection{Many M2 branes suspended between two M5 branes and $E\oplus E^*$ (4,0) sigma model}
\label{sec:sigmamodel}
As we have seen the description of the theory on $n$ suspended M2 branes between two M5 branes is close to a sigma model
on $\mbox{Hilb}^n(\mathbb{R}^4)$.  Whatever this theory is, it should have $(4,4)$ supersymmetry.  However, if one is only considering
the $(2,0)$ elliptic genus of this theory, any deformation of this theory which preserves $(2,0)$ supersymmetry will
yield the same elliptic genus.  In this section we show that there is a sigma model with $(2,0)$ supersymmetry whose
elliptic genus exactly reproduces the result we have found.

The sigma model we propose is on $\mbox{Hilb}^n(\mathbb{R}^4)$, but it is a $(4,0)$ model.  As is familiar in the context of heterotic
string vacua this means that even though the left-moving fermions are coupled to the tangent bundle, the right-moving fermions are coupled to a different bundle $V$ than the tangent bundle.  In fact the bundle that we find is $V=E\oplus E^*$ where
$E$ is the $n$-dimensional complex tautological bundle over the $\mbox{Hilb}^n(\mathbb{R}^4)$ already discussed.  Note that $V$ has the same dimension
as the tangent bundle.  In fact more is true:  It turns out that they are equivalent K-theoretically \cite{tautological}\footnote{This observation should be relevant in verifying the anomaly cancellation of the 2d theory on the M-string (see \cite{Berman:2004ew} for a related discussion).}.
Therefore it is conceivable that $V$ is continuously deformable to the tangent bundle to $\mbox{Hilb}^n(\mathbb{R}^4)$.
Before explaining the potential implication of this statement let us see why the elliptic genus for this sigma model
gives the same answer as we have found.  As already discussed in section 3, if we view the case of two M5 branes
as an elliptic version of the $5d$ supersymmetric $U(1)$  gauge theory with two fundamental matter multiplets of mass $m$, we
saw that the contribution to elliptic genus is obtained by counting the index of the bundle

\begin{align}
E_{Q_{\tau,m}}=\bigotimes_{k=0}^{\infty} \bigwedge_{Q_{m}^{-1}Q_{\tau}^k}\,(E\oplus E^*)^{*}\otimes \bigotimes_{k=1}^{\infty}\bigwedge_{Q_{m}\,Q_{\tau}^{k}}(E\oplus E^*)\otimes \bigotimes_{k=1}^{\infty}S_{Q_{\tau}^{k}}T^{*}_{{\cal M}}\otimes \bigotimes_{k=1}^{\infty}S_{Q_{\tau}^{k}}T_{{\cal M}},
\end{align}
but this is exactly the elliptic genus of the $(4,0)$ theory coupled to $E\oplus E^*$ bundle (where
the twisting by $Q_m$ is identified with right-moving fermion number, and the actions of $\epsilon_{1,2}$
is inherited from its action on the $\mathbb{R}^4$).

Now we attempt to demystify the appearance of a $(4,0)$ theory.  The first question is why does it have 1/2 the supersymmetry
expected?  This is explained by noting that counting the BPS states for a 5$d$ ${\cal N}=1$ supersymmetric gauge theory
will preserve only 4 supercharges, which is consistent with a $(4,0)$ theory. Indeed in this context the above description
of the index computation directly follows from the instanton calculus.  This is related to the fact that even if we turn off
$m=\epsilon_i=0$ the theory is not really the compactification of the M5 brane theory on untwisted $S^1$ due to the fact that
the geometry is not quite the product structure.  To make it the product structure we need to change the geometry
which in the brane description corresponds to lifting the horizontal brane off the two vertical branes (see \figref{fig:ldeformation}).

\begin{figure}[here!]
  \centering
  \includegraphics[width=0.8\textwidth]{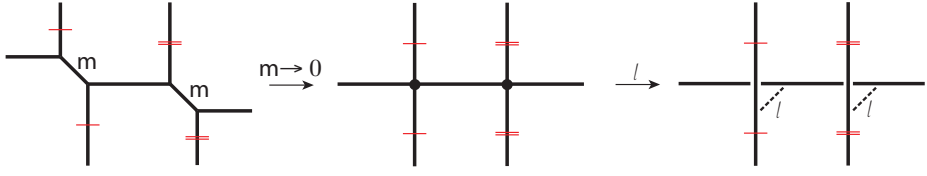}\\
  \caption{The $l$-deformation of the brane system.  In the first step the mass of the adjoint hypermultiplet is sent to zero. Then the NS5 brane is removed from the D4 branes leaving a system with $16$ preserved supercharges as
$l\rightarrow \infty$.}\label{fig:ldeformation}
\end{figure}

It is conceivable that the separation of the horizontal brane from the vertical branes is a deformation which
deforms the $E\oplus E^*$
 bundle to $T(\mbox{Hilb}^n(\mathbb{R}^4))$, in the limit of infinite separation.  One may then
ask if this is indeed the case, why we were not able to compute the elliptic genus using directly the tangent bundle
of the Hilbert scheme?  The natural answer is that the only way to get a non-trivial answer is to turn on $m,\epsilon_i$ in
which case the action of these is not the same between the tangent bundle and the $E\oplus E^*$ bundle.
Of course we already knew that the actions have to be different because even for a single M2 brane, the
fermions transform in the spinor representation of the $Spin(4)$ which is different from the boson.  So even then
we do not expect the action of these twistings to be the canonical action on the tangent bundle.  The surprise is that
the answer is not the same as one would obtain by considering the symmetric product theory which would have
had a $(4,4)$ supersymmetry.

\subsubsection{Multiple M2 branes suspended between multiple M5 branes}

For the case of $N$ M5 branes the degeneracy of BPS states is captured by the partition function Eq.(\ref{unpf}),
\begin{align}
\widehat{Z}^{(N)}(\tau,m,t_{f_{a}},\epsilon_1,\epsilon_2)=\sum_{\nu_{1},\mathellipsis, \nu_{N-1}}\left(\prod_{a=1}^{N-1}(-Q_{f_{a}})^{|\nu_{a}|}\right)
\prod_{a=1}^{N-1}\prod_{(i,j)\in \nu_{a}}\frac{\theta_{1}(\tau;z^{a}_{ij})\,\theta_{1}(\tau;v^{a}_{ij})}{\theta(\tau;w^{a}_{ij})\theta(\tau;u^{a}_{ij})}
\end{align}
with
\begin{equation}
\begin{array}{ll}
e^{2\pi i \,z^{a}_{ij}}=Q_{m}^{-1}\,q^{\nu_{a,i}-j+\frac{1}{2}}\,t^{\nu^{t}_{a+1,j}-i+\frac{1}{2}},\qquad &
e^{2\pi i\,v^{a}_{ij}}=Q_{m}^{-1}\,t^{-\nu^{t}_{a-1,j}+i-\frac{1}{2}}\,q^{-\nu_{a,i}+j-\frac{1}{2}},\\
e^{2\pi i \,w^{a}_{ij}}=q^{\nu_{a,i}-j+1}\,t^{\nu_{a,j}^{t}-i},& e^{2\pi i \,u^{a}_{ij}}=q^{\nu_{a,i}-j}\,t^{\nu_{a,j}^{t}-i+1}.
\end{array}
\end{equation}

This is the refined topological string partition function of the Calabi-Yau threefold $X_{N}$. In section \ref{ins2} we showed that this partition function is also given by the index of a twisted Dirac operator coupled to the bundle $V_{Q_{\tau},y}$. If $V$ was the holomorphic tangent bundle of ${\cal M}$ then we would be calculating the index of the Dirac operator on the loop space of ${\cal M}$, i.e. the $(2,2)$ elliptic genus of ${\cal M}$ \cite{Witten:1986bf}. However, as we have seen in the last section $V=E\oplus E^*$ is not the holomorphic tangent bundle and therefore the above partition function does not give the $(2,2)$ elliptic genus. To put the calculation of the section \ref{ins2} in a physical perspective recall that the theory on the string which is the intersection of M2 branes and the M5 branes is a $(2,0)$ two dimensional theory \cite{Witten:1993yc} with target space the Hilbert scheme of points on $\mathbb{C}^2$ or product of such spaces for $N>2$. The partition function of this theory on $T^2$ is the $(2,0)$ elliptic genus where left handed fermions are sections of the tangent bundle and the right handed fermions are sections of the bundle $E\oplus E^*$ for $N=2$. The $(2,0)$ elliptic genus is given by,
\bea
Z(\tau,y)=\mbox{Tr}\,(-1)^{F}\,y^{J_{R}}\,q^{L_{0}}\,\overline{q}^{\overline{L_{0}}}\,,
\eea
where $J_{R}$ is the conserved charge associated with the right $U(1)$ symmetry. As in section \ref{ins2} if we denote by $\widetilde{x}_{i}$ and $x_{j}$ the roots of the Chern polynomial of $E\oplus E^*$ and $T{\cal M}$ respectively then \cite{Kawai:1994np}\footnote{Assuming $\mbox{rank}\,E\oplus E^*=\mbox{rank}\,(T{\cal M})$ which will be the case for us.}\,,
\bea
Z(\tau,y)&=&\int_{{\cal M}}\,\prod_{j=1}^{d}\frac{x_{j}\theta_{1}(\tau;-m+\widetilde{x}_{i})}{\theta_{1}(\tau,x_{j})}\,,\\\nn
&=&y^{-\frac{d}{2}}\int_{{\cal M}}\mbox{ch}(E_{Q_{\tau},Q_{m}})\,\mbox{Td}({\cal M})\\\nn
&=&\sum_{k,n}Q_{\tau}^{k}\,Q_{m}^{n-\frac{d}{2}}\,\chi(\mathcal{M},V_{k,n})
\eea
where $d=\mbox{dim}_{\mathbb{C}}\mathcal{M}$ and
\bea\nn
V_{Q_{\tau,y}}&=&\bigotimes_{k=0}^{\infty} \bigwedge_{-Q_m\,Q_{\tau}^{k-1}}\,(E\oplus E^{*})\otimes \bigotimes_{k=1}^{\infty}\bigwedge_{-Q_m^{-1}\,Q_{\tau}^{k}}(E\oplus E^{*})^{*}\otimes \bigotimes_{k=1}^{\infty}S_{Q_{\tau}^{k}}T^{*}_{{\cal M}}\otimes \bigotimes_{k=1}^{\infty}S_{Q_{\tau}^{k}}T_{{\cal M}},\\\nn
&=&\oplus_{k,n}\,Q_{\tau}^{k}\,Q_{m}^{n}\,V_{k,n}\,.
\eea

This is precisely the calculation carried out in Eq.(\ref{calc1}) for the target space $\mbox{Hilb}^{k_{1}}[\mathbb{C}^2]\times \mbox{Hilb}^{k_{2}}[\mathbb{C}^2]\times \cdots \times \mbox{Hilb}^{k_{N-1}}[\mathbb{C}^2]$ and bundle $V$ described in section \ref{ins2} generalizing the holomorphic Euler characteristic $\chi_{y}({\cal M},E\oplus E^*)$ which appears in the calculation of the partition function of the $U(1)$ gauge theory with fundamental hypermultiplets.

\subsubsection{Factorisation from the topological string theory and KK surprise}

\par{The BPS states for more general $A_{N-1}$ for $N>2$ corresponding to $N$ M5 branes are
interesting and teach us about the dynamics of these strings: If the M2 branes act independently one would expect that answer for $A_{N-1}$ to be obtainable from the $A_1$ theory
by considering arbitrary pairs of M5 branes. This is certainly the case before compactification on $S^1$. However, as we will see this is not the case after compactification on $S^1$ and there are new bound states of M2 branes stretched between different pairs
of M5 branes.
In this section, we want to review the factorisation property of the topological string partition function of the so-called strip geometry. We show that the partition function ceases to factorize upon the partial compactification of this geometry. The failure of such a factorisation is argued to indicate the existence of new bound states of M-strings `glued' by non-trivial
momentum around the circle.}
\begin{figure}[h]
  \centering
  \includegraphics[width=5in]{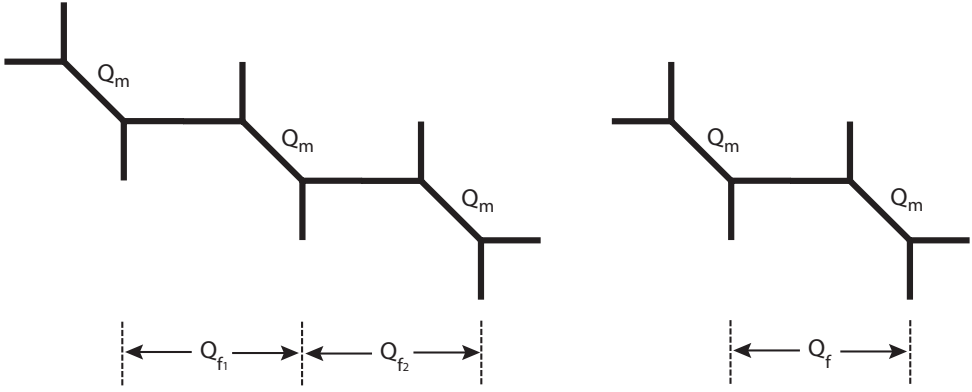}\\
  \caption{The strip geometry for the $SU(3)$ (left) and the geometry for the $SU(2)$ (right) theories.}\label{strip}
\end{figure}
\par{The strip geometry in our setup is the half of the toric geometry that engineers $SU(N)$ gauge theory with $N=2^*$.
 flavours\footnote{We will refer to the strip geometries referring to the gauge group only.}, it is depicted in \figref{strip}. It is also the same geometry as when we decompactify the sixth direction to zero size by taking the limit $Q_{\tau}\rightarrow 0$. Let us review the factorisation on a specific example, for the $SU(3)$ theory. The partition function can be computed using the refined topological vertex,  }
\begin{align}\nn
Z_{SU(3)}&(Q_{f_{1}},Q_{f_{2}},Q_{m})= \\\nn
&\prod_{i,j=1}^{\infty}\frac{(1-Q_{m}\,t^{-i+\frac{1}{2}}q^{-j+\frac{1}{2}})^{3}(1-Q_{f_{1}}Q_{m}\,t^{-i+\frac{1}{2}}q^{-j+\frac{1}{2}})(1-Q_{f_{1}}Q_{m}^{-1}\,t^{-i+\frac{1}{2}}q^{-j+\frac{1}{2}})}{(1-Q_{f_{1}}t^{-i+1}q^{-j})(1-Q_{f_{1}}t^{-i}q^{-j+1})}\\\nn
\times&\prod_{i,j=1}^{\infty}\frac{(1-Q_{f_{2}}Q_{m}\,t^{-i+\frac{1}{2}}q^{-j+\frac{1}{2}})(1-Q_{f_{2}}Q_{m}^{-1}\,t^{-i+\frac{1}{2}}q^{-j+\frac{1}{2}})}{(1-Q_{f_{2}}t^{-i+1}q^{-j})(1-Q_{f_{2}}t^{-i}q^{-j+1})}\\
\times&\prod_{i,j=1}^{\infty}\frac{(1-Q_{f_{1}}Q_{f_{2}}Q_{m}\,t^{-i+\frac{1}{2}}q^{-j+\frac{1}{2}})(1-Q_{f_{1}}Q_{f_{2}}Q_{m}^{-1}\,t^{-i+\frac{1}{2}}q^{-j+\frac{1}{2}})}{(1-Q_{f_{1}}Q_{f_{2}}t^{-i+1}q^{-j})(1-Q_{f_{1}}Q_{f_{2}}t^{-i}q^{-j+1})}.
\end{align}
\par{We previously pointed out that the partition function is invariant under the different choices of the preferred direction although the functional form may vary. The above factorized form is the result of our choice for the preferred direction along the external legs of the toric diagram. The curve counting arguments suggest that the factors appearing in the partition function can be grouped in such a way that the partition function can be written in terms of $SU(2)$ partition functions. The $SU(2)$ partition function is given by}
\begin{align}
Z_{SU(2)}(Q_{f_{1}},Q_{m})=\prod_{i,j=1}^{\infty}\frac{(1-Q_{m}\,t^{-i+\frac{1}{2}}q^{-j+\frac{1}{2}})^{2}(1-Q_{f_{1}}Q_{m}\,t^{-i+\frac{1}{2}}q^{-j+\frac{1}{2}})(1-Q_{f_{1}}Q_{m}^{-1}\,t^{-i+\frac{1}{2}}q^{-j+\frac{1}{2}})}{(1-Q_{f_{1}}t^{-i+1}q^{-j})(1-Q_{f_{1}}t^{-i}q^{-j+1})}.
\end{align}
\par{Comparing the two partition functions above, it is easy to see that they satisfy}
\begin{align}
Z_{SU(3)}(Q_{f_{1}},Q_{f_{2}},Q_{m})=\frac{1}{(1-Q_{m})^{3}}Z_{SU(2)}(Q_{f_{1}},Q_{m})Z_{SU(2)}(Q_{f_{2}},Q_{m})Z_{SU(2)}(Q_{f_{1}}Q_{f_{2}},Q_{m}),
\end{align}
where we have used the short-hand notation
\bea
\frac{1}{(1-Q_{m})^{3}}\equiv \prod_{i,j=1}^{\infty}\frac{1}{(1-Q_{m}\,t^{-i+\frac{1}{2}}q^{-j+\frac{1}{2}})^{3}}.
\eea
\par{A priori we do not have any reason to expect that the factorisation would be lost when we partially compactify the geometry along the vertical external lines to geometrically engineer ${\cal N}=2^{*}$ theory with the gauge group $SU(3)$. However, there is no choice of the preferred direction that would allow us to express the partition function in terms of products as in the corresponding strip geometry. Therefore, we are forced to compare the expansions in K\"{a}hler parameters to see whether the factorisation still holds. We observe that the ${\cal N}=2^{*}$ with $SU(3)$ partition function can be written (up to the pre-factor including only $Q_{m}$)} as
\begin{align}\nn
Z_{SU(3)}(Q_{f_{1}},Q_{f_{2}},Q_{\tau},Q_{m})&=Z_{SU(2)}(Q_{f_{1}},Q_{\tau},Q_{m})Z_{SU(2)}(Q_{f_{2}},Q_{\tau},Q_{m})\\ &\times Z_{SU(2)}(Q_{f_{1}}Q_{f_{2}},Q_{\tau},Q_{m})(1+c(t,q)\, Q_{f_{1}}Q_{f_{2}}Q_{\tau}+\mathellipsis),
\end{align}
where $c(t,q)$ is a non-vanishing and the $(\mathellipsis)$ involve higher powers in $Q_{f_{1}}$, $Q_{f_{2}}$ and $Q_{\tau}$. Obviously, in the limit when $Q_{\tau}\rightarrow 0$, the expansion reduces to $1$ and we recover the factorisation.
\par{We can actually shed more light into this observation by isolating the same curves from the $SU(3)$ partition function and the products of the $SU(2)$ partition functions. We determine the BPS content that these curves give rise to. Let us start with the curve $Q_{f_{1}}Q_{f_{2}}Q_{\tau}$. From the $SU(3)$ partition function we obtain the following states }
\bea
\left(\frac{1}{2},0\right)\oplus3\left(0,\frac{1}{2} \right).
\eea
On the other hand, we obtain from the products of the $SU(2)$ partition functions
\bea
\left(0,\frac{1}{2} \right).
\eea
 for the same curve $Q_{f_{1}}Q_{f_{2}}Q_{\tau}$. Clearly, the $SU(3)$ partition function includes more states than the product of $SU(2)$'s. Let us work out another curve: $(Q_{f_{1}}Q_{f_{2}}Q_{\tau})^{2}$. This curve is particularly interesting since there are no new states appearing for the product of $SU(2)$'s. This curve does not have any other contribution than the multi-covering contributions. However, the BPS content for the $SU(3)$ partition consists of the following states
\bea
\left(\frac{1}{2},2\right)\oplus5\left(\frac{1}{2},1\right)\oplus3\left(\frac{1}{2},0\right)\oplus\left(0,\frac{5}{2}\right)\oplus6\left(0,\frac{3}{2}\right)\oplus10\left(0,\frac{1}{2}\right)
\eea
The difference in BPS spectrum originating from the same curves continues to hold for higher degree curves. The above discussion can be extended for $SU(N)$ partition functions without any complications. We always have a smaller Hilbert space of states originating from the factor partition functions.
\par{The lack of factorisation is rather surprising. Let us try to understand the implication of this observation from the point of view of the M-theory construction. In the Coulomb branch of the $SU(3)$ theory the three M5 branes are separated. There are M2 branes stretched between them. In the case of the uncompactified $x_{6}$ direction, the only bound states consist of the $(13)$ strings, stretching between the first and the third M5 branes, in addition to the $(12)$ and the $(23)$ strings. However, once we compactify $x_{6}$, the $(13)$ strings are not the only bound states the topological string partition function counts, there are additional states that we have found. The momentum along the circle could account for the additional bound states and the $Q_{\tau}$ dependence of the partition function. Using the momentum along the circle we have new junction states.    }

\subsection{M5 branes as domain walls between M2 branes}

The computation of the BPS partition function using the topological vertex (in the second method discussed in section 3)
can be reformulated as introducing a Hilbert space whose basis is formed from arbitrary Young diagrams $\nu$, with the identity operator $I=\sum_{\nu} |\nu\rangle \langle \nu^t|$, and whose
`Hamiltonian' is $H=|\nu|$ and an operator $D$ whose matrix elements are given by $D_{\nu^{t}\mu}(\tau,m,\epsilon_1,\epsilon_2)$,
\bea\nn
&&\langle \nu^t |D|\mu\rangle =D_{\nu^{t}\mu}(\tau,m,\epsilon_1,\epsilon_2)=t^{-\frac{\Arrowvert\mu^{t}\Arrowvert^2}{2}}\,q^{-\frac{\Arrowvert\nu\Arrowvert^2}{2}}Q_{m}^{-\frac{|\nu|+|\mu|}{2}}\\ \nn &&\times\prod_{k=1}^{\infty}\prod_{(i,j)\in\nu}\frac{(1-Q_{\tau}^{k}Q_m^{-1}\,q^{-\nu_{i}+j-\frac{1}{2}}\,t^{-\mu_{j}^{t}+i-\frac{1}{2}})(1-Q_{\tau}^{k-1}Q_m\,q^{\nu_{i}-j+\frac{1}{2}}\,t^{\mu_{j}^{t}-i+\frac{1}{2}})}{(1-Q_{\tau}^{k}\,q^{\nu_{i}-j}\, t^{\nu_{j}^{t}-i+1})(1-Q_{\tau}^{k-1}\,q^{-\nu_{i}+j-1}\,t^{-\nu_{j}^{t}+i})}\\
&&\times\prod_{(i,j)\in\mu}\frac{(1-Q_{\tau}^{k}Q_m^{-1}\,q^{\mu_{i}-j+\frac{1}{2}}t^{\nu_{j}^{t}-i+\frac{1}{2}})(1-Q_{\tau}^{k-1}Q_m \,q^{-\mu_{i}+j-\frac{1}{2}}t^{-\nu_{j}^{t}+i-\frac{1}{2}})}{(1-Q_{\tau}^{k}\,q^{\mu_{i}-j+1}t^{\mu_{j}^{t}-i})(1-Q_{\tau}^{k-1}\,q^{-\mu_{i}+j}t^{-\mu_{j}^{t}+i-1})}
\eea

  Letting
the $\beta_a=2\pi i\, t_{f_{a}}$ where $t_{f_{a}}$ label the fiber sizes are equivalently the separation of the M5 branes.
In this language the partition function can be written as
$${\widehat Z}^{(N)}={Z^{(N)}\over \left(Z^{(1)}\right)^N }=\langle 0| De^{-\beta_1H}De^{-\beta_2 H}D\cdots e^{-\beta_{N-1} H}D|0\rangle$$
where $|0\rangle$ is identified with the Young diagram of zero size.  We would like to explain the physical meaning of this expression.

The idea is very simple.  As already noted, $Z^{(N)}/(Z^{(1)})^N$ is computing the partition function of
the M2 branes stretched between M5 branes and wrapping a $T^2$ with suitable twists
along the cycles of $T^2$.  So far we viewed the distance between the M5 branes as small compared to the size of the $T^2$.  Since nothing depends
on the relative sizes of $T^2$ and the separation of M5 branes, we can take the opposite
limit in which we view the $T^2$ as small.  This is shown in \figref{fig:M2domainwall}.
\begin{figure}[here!]
	\begin{center}
	\includegraphics[width=\textwidth]{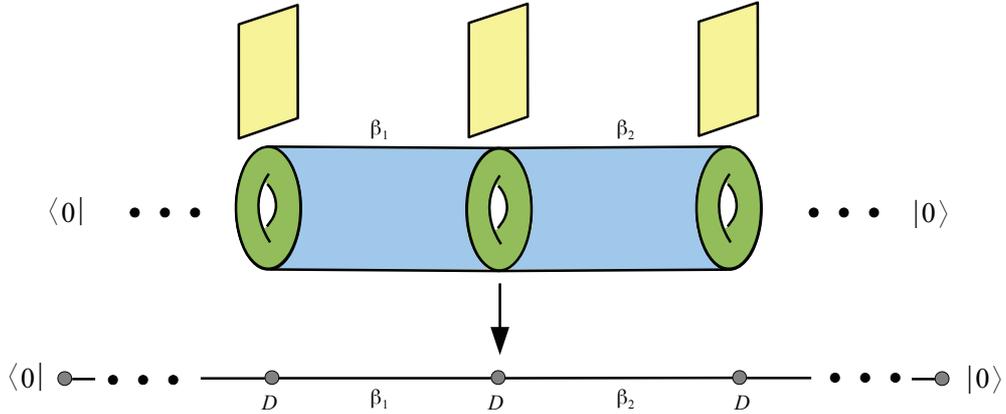}
	\caption{System of M2 branes and M5 brane induced domain walls. If we take the size of the $T^2$ to be much smaller that the distance between the M5 branes then the theory reduces down to dimensions $d=1$. Again the M5 branes are depicted in yellow and M2 branes in blue.}
	\label{fig:M2domainwall}
	\end{center}
\end{figure}
  In this case we get a reduction of the M2 brane theory
down to 1 dimension, where the time dimension is punctuated by M5 branes where the M2 branes
end on.  So it is natural to identify the Hilbert space of this one dimensional theory
with the Hilbert space of the M2 brane on $T^2$.  In fact it has been argued that these
vacua (at least in the mass deformed version of the M2 brane theory, which is effectively
what we have due to twistings along the cycles of $T^2$) should correspond to partitions
of $n$ where $n$ is the number of M2 branes \cite{Lin:2004nb,Gomis:2008vc, Kim:2010mr}.  This is in agreement with what
we have found here.  Moreover here $H$ is the energy of the ground state of M2 brane, which
is 0, up to the addition of the M2 brane mass, given by the tension of the M2 brane times the size of $T^2$
which we have effectively normalized to 1, times the number of the M2 branes.  Thus $H= |\nu|$ as we have
here.
 The effect of the M5 brane domain wall reduced on $T^2$ should be an operator
acting on this 1 dimensional theory.  So on this effective theory, we identify $D$ with this
operator.  Note that we are not fixing the number of M2 branes (i.e. the size
of the Young diagrams) on each interval and summing over all of them.

Another way to say this is that from the viewpoint of the M2 brane theory, ending on M5 brane
is like putting a particular boundary condition on the M2 brane theory, as is familiar
in the context of D-branes in 2d.  More generally we can have a number of M2 branes on
one side of the M5 brane and another number of M2 branes on the other.  In this way we can view
M5 brane as defining a domain wall separating two different theories on the left and right of the
domain wall with different number of M2 brane on either side.  In this set up we can view
$\langle \nu^t |D|\mu\rangle $ as the partition function of this domain wall which connects
a particular vacuum of the M2 brane theory labeled by $\nu^t$ on the left, to the vacuum labeled
by $\mu$ on the right, compactified and twisted on $T^2$ with complex structure $\tau$ and
twist parameters $(m,\epsilon_1,\epsilon_2)$.  We can also view $D$ as an operator taking a vacuum of
of the left $M2$ brane to a vacuum of the right one.

One may expect the partition function $D_{\nu^t\mu}$ to be modular.  This turns out to be only true,
as discussed in Section 3, when we have the unrefined parameters.  Otherwise pairs of adjacent
domain walls $D_{\nu^t\mu}$ need to be included for it to be modular.  This can be interpreted as saying
that the modular transformation acts also on the boundary data at the other end of the M2 brane
and only the combined object should be invariant, which is the case.

\subsection{Special limits of $m$ and comparison with elliptic genus of $A_{N-1}$ quiver theories}

As already noted we expect that the partition function for the elliptic genus simplifies in special limits:
If $m=\pm (\epsilon_1-\epsilon_2)/2$ the supersymmetry enhances from $(2,0) \rightarrow (2,2)$
and the partition function becomes a constant.  This expectation agrees with the results we have,
as already noted.
On the other hand if $m=\pm (\epsilon_1+\epsilon_2)/2$, the partition function vanishes again.
However in this case the vanishing has nothing to do with supersymmetry, but rather it has to do
with the fact that the center of mass of the string (in the absence of M5 brane) has an extra direction to move which leads
to a fermionic zero mode.  Moreover, as already discussed, this is the case where
we have a description of M-strings in terms of $A_{N-1}$ quiver theory.  In particular
for two M5 branes we have the $A_1$ quiver theory which for $N$ suspended
M2 branes is the pure $U(N)$ Yang-Mills theory with $(4,4)$
supersymmetry.  It is equally true that the elliptic genus of this theory also vanishes
due to the $U(1)\subset U(N)$.  Moreover the elliptic genus of the $SU(N)$ theory does not vanish.
 It is therefore natural to factor out
the vanishing contribution of the $U(1)$ piece from our computation and obtain the $SU(N)$ result.
Recall the expression we found for the partition function of two M5 branes:

\bea\nonumber
Z^{(2)}(\tau,m,t_{f},\epsilon_1,\epsilon_2)&=&\left(Z^{(1)}(\tau,m,\epsilon_1,\epsilon_2)\right)^{2}\\\label{u2pf}
&\times&\underbrace{\sum_{\nu}(-Q_{f})^{|\nu|}\,\prod_{(i,j)\in \nu}\frac{\theta_{1}(\tau;z_{ij})\,\theta_{1}(\tau;v_{ij})}{\theta_{1}(\tau;w_{ij})\theta_{1}(\tau;u_{ij})}}_{\widehat{Z}^{(2)}(\tau,m,t_{f},\epsilon_1,\epsilon_2)}
\eea
with the following definitions for the arguments of the $\theta$-functions
\begin{equation}\label{var}
\begin{array}{ll}
e^{2\pi i \,z_{ij}}=Q_{m}^{-1}\,q^{\nu_{i}-j+1/2}\,t^{-i+1/2},\qquad&
e^{2\pi i\,v_{ij}}=Q_{m}^{-1}\,t^{i-1/2}\,q^{-\nu_{i}+j-1/2},\\
e^{2\pi i \,w_{ij}}=q^{\nu_{i}-j+1}\,t^{\nu_{j}^{t}-i},& e^{2\pi i \,u_{ij}}=q^{\nu_{i}-j}\,t^{\nu_{j}^{t}-i+1}.
\end{array}
\end{equation}
We can extract the contribution for two suspended M2 branes by considering the $|\nu|=2$ term and we take the limit where $m=(\epsilon_1+\epsilon_2)/2$, i.e. $Q_{m}=e^{i\pi(\epsilon_{1}+\epsilon_{2})}$ to obtain
\begin{align}
\widehat{Z}^{(2)}_{2}&=\frac{\theta_{1}(\tau;0)\theta_{1}(\tau;-\epsilon_{1}-\epsilon_{2})\theta_{1}(\tau;\epsilon_{2})\theta_{1}(\tau;-\epsilon_{1}-2\epsilon_{2})}{\theta_{1}(\tau;\epsilon_{1}-\epsilon_{2})\theta_{1}(\tau;\epsilon_{1})\theta_{1}(\tau;-\epsilon_{2})\theta_{1}(\tau;-2\epsilon_{2})}\\
&+\frac{\theta_{1}(\tau;0)\theta_{1}(\tau;\epsilon_{1})\theta_{1}(\tau;-2\epsilon_{1}-\epsilon_{2})\theta_{1}(\tau;-\epsilon_{1}-\epsilon_{2})}{\theta_{1}(\tau;\epsilon_{1})\theta_{1}(\tau;2\epsilon_{1})\theta_{1}(\tau;-\epsilon_{2})\theta_{1}(\tau;\epsilon_{1}-\epsilon_{2})},
\end{align}
where the first term is the contribution of $\nu=(1,1)$ and the second term is the one of $\nu=(2)$.
As is manifest, the above expression vanishes due to the fermion zero mode from the $U(1)$ part of the
partition function which in this limit becomes
$$\widehat{Z}^{(2)}_1\rightarrow \frac{\theta_{1}(\tau;0)\theta_{1}(\tau;\epsilon_1+\epsilon_2)}{\theta_{1}(\tau;\epsilon_1)\theta_{1}(\tau;\epsilon_2)}$$
Dividing\footnote{We can also consider instead an insertion of $F_L$ in the elliptic genus
to absorb the extra fermionic zero mode of the $U(1)$ which has the effect of replacing
the vanishing theta function with its derivative.} both terms of the above expression by $\widehat{Z}^{(2)}_1$ we get a theory which should
be the elliptic genus of the $SU(2)$ partition function:
\begin{align}
{\widehat{Z}^{(2)}_{2}\over\widehat{Z}^{(2)}_1}&=\frac{\theta_{1}(\tau;\epsilon_{2})\theta_{1}(\tau;\epsilon_{1}+2\epsilon_{2})}{\theta_{1}(\tau;\epsilon_{2}-\epsilon_{1})\theta_{1}(\tau;2\epsilon_{2})}+\frac{\theta_{1}(\tau;\epsilon_{1})\theta_{1}(\tau;2\epsilon_{1}+\epsilon_{2})}{\theta_{1}(\tau;2\epsilon_{1})\theta_{1}(\tau;\epsilon_{1}-\epsilon_{2})}
\end{align}

Notice that in the limit $\epsilon_1\rightarrow 0$ the above expression
becomes a constant, independent of the modulus.  This is consistent with the fact that in the limit $m=\epsilon_2$,
$\epsilon_1=0$ we have a $(2,2)$ supersymmetric index, and so it should be independent of
the modulus of $T^2$.

Eq.(\ref{prediction}) gives our prediction for the $SU(N)$ elliptic genus ($m=\pm \frac{\epsilon_{1}+\epsilon_{2}}{2}$),
\bea
{\widehat{Z}^{(2)}_{N}\over\widehat{Z}^{(2)}_1}&=&\sum_{|\nu|=N}\frac{\prod_{(i,j)\in \nu,(i,j)\neq (1,\nu_1)}\,\theta_{1}(\tau;z_{ij})\,\theta_{1}(\tau;v_{ij})}{\prod_{(i,j)\in \nu,(i,j)\neq (\ell(\nu),\nu_{\ell(\nu)})}\theta_{1}(\tau;w_{ij})\theta_{1}(\tau;u_{ij})}\,,
\eea
where we identify
\begin{equation}
\begin{array}{ll}
z_{ij}=(\nu_{i}-j)\epsilon_{1}+(i-1)\epsilon_{2}\,,\qquad & v_{ij}=-(\nu_{i}-j+1)\epsilon_{1}-i\,\epsilon_{2}\,,\\
w_{ij}=(\nu_{i}-j+1)\epsilon_{1}-(\nu^{t}_{j}-i)\epsilon_{2}\,,& u_{ij}=(\nu_{i}-j)\epsilon_{1}-(\nu^{t}_{j}-i+1)\epsilon_{2}\,.
\end{array}
\end{equation}
For $N=3$ the above gives,
\bea
{\widehat{Z}^{(2)}_{3}\over\widehat{Z}^{(2)}_1}&=&\frac{\theta(\tau;2\epsilon_{1})\theta(\tau;3\epsilon_{1}+\epsilon_{2})\theta(\tau;\epsilon_{1})\theta(\tau;2\epsilon_{1}+\epsilon_{2})}
{\theta(\tau;3\epsilon_{1})\theta(\tau;2\epsilon_{1}-\epsilon_{2})\theta(\tau;2\epsilon_{1})\theta(\tau;\epsilon_{1}-\epsilon_{2})}+\\\nn
&&\frac{\theta(\tau;\epsilon_{1})\theta(\tau;2\epsilon_{1}+\epsilon_{2})\theta(\tau;\epsilon_{2})\theta(\tau;\epsilon_{1}+2\epsilon_{2})}
{\theta(\tau;2\epsilon_{1}-\epsilon_{2})\theta(\tau;\epsilon_{1}-2\epsilon_{2})\theta(\tau;\epsilon_{1})\theta(\tau;-\epsilon_{2})}+\\\nn
&&\frac{\theta(\tau;\epsilon_{2})\theta(\tau;\epsilon_{1}+2\epsilon_{2})\theta(\tau;2\epsilon_{2})\theta(\tau;\epsilon_{1}+3\epsilon_{2})}
{\theta(\tau;\epsilon_{1}-2\epsilon_{2})\theta(\tau;3\epsilon_{2})\theta(\tau;\epsilon_{1}-\epsilon_{2})\theta(\tau;2\epsilon_{2})}\,.
\eea

These predictions are to be compared with the elliptic genus of $(4,4)$ supersymmetric
Yang-Mills for $SU(N)$ theory as computed using the results \cite{Benini:2013nda,Gadde:2013dda}\footnote{This was computed and communicated to us for the $SU(2)$ case by K. Hori and for the $SU(N)$ case by A. Gadde and S. Gukov.}.  The elliptic genus result for $U(N)$ as computed in \cite{Gadde:2013dda} reads as follows
\begin{eqnarray}
	\mathcal{I}^{(N)} & = & \sum_{|\nu| =N} \prod_{(i_1,j_1)\in \nu, (i_2,j_2)\in \nu}\frac{ \theta_1(\tau;\epsilon_1 (i_2-i_1)+ \epsilon_2 (j_2-j_1))}{\theta_1(\tau;\epsilon_1 (1+i_2-i_1)+ \epsilon_2 (j_2-j_1))} \nonumber \\
	~  & ~ & ~~~~~ \times \prod_{(i_1,j_1)\in \nu, (i_2,j_2)\in \nu} \frac{\theta_1(\epsilon_1 (1+i_2-i_1) + \epsilon_2 (1+j_2-j_1))}{\theta_1(\epsilon_1 (i_2-i_1) + \epsilon_2 (1+j_2-j_1) )}~.
\end{eqnarray}
The expressions
look slightly different, however, we have checked
that they indeed agree at least for $N\leq 10$.  This gives a satisfactory confirmation of the overall
picture we have and the connection between BPS degeneracies computed by the topological strings
and the elliptic genus of M-strings.

\section{M5 brane partition function on $ S^4\times T^2 $ and $ S^5\times S^1 $}
We now turn to the computation of the M5 brane partition functions on $ S^4\times T^2  $ and $ S^1\times S^5 $, along the lines of \cite{Iqbal:2012xm, Lockhart:2012vp}. Before discussing the explicit calculations, let us briefly review the relation between refined topological string theory and five-dimensional gauge theories.  Recall \cite{Dijkgraaf:2006um} that refined topological string theory on a Calabi-Yau threefold $X$ is equivalent to M-theory on
\bea
M_{\epsilon_1,\epsilon_2} = (\mathbb{C}^2\times S^1)_{\epsilon_1,\epsilon_2}\times X.
\eea
The $\mathbb{C}^2$ is fibered non-trivially over $ S^1 $ in a way that as one goes around the circle the coordinates parameterizing $ \mathbb{C}^2 $ are rotated by $ (z_1,z_2)\to (e^{2\pi i \epsilon_1}z_1, e^{2\pi i \epsilon_2}z_2) $, and simultaneously one performs a twist on $ X $ in order to preserve supersymmetry.  Moreover, more precisely $\mathbb{C}^2$ should be
viewed as the Taub-NUT space.
 At the level of partition functions, the following statement holds
\bea
Z_{top}(t_i,m_j;\epsilon_1,\epsilon_2) = Z_{M-theory}(M_{\epsilon_1,\epsilon_2}),
\eea
\hyphenation{pa-ra-me-tri-ze}\noindent where we are denoting by $ t_i $ the normalizable K\"ahler parameters of $ X $ and $ m_j $ are the non-normalizable K\"ahler parameters. The parameters $ \epsilon_1,\epsilon_2 $ play the roles of couplings of the refined topological string theory, and in the limit $ \epsilon_1+\epsilon_2\to 0 $ one recovers the unrefined topological string, whose coupling constant is given by $ g_s = \epsilon_1=-\epsilon_2 $. Upon compactification on $ X $, for choices of internal geometry that geometrically engineer a gauge theory \cite{Katz:1997eq}, the M-theory partition function is identified with the 5$d$ gauge theory partition function with equivariant parameters $ (\epsilon_1,\epsilon_2) $ and one arrives at the following statement:
\bea
Z_{top}(t_i,m_j;\epsilon_1,\epsilon_2) = Z_{1-loop}(t_i,m_j;\epsilon_1,\epsilon_2)\, Z_{Nekrasov}(t_i,m_j;\epsilon_1,\epsilon_2),
\eea
where $ Z_{1-loop} $ and $ Z_{Nekrasov} $ capture, respectively, the perturbative and instanton contributions to the gauge theory partition function. From the gauge theory perspective, the $ t_i $ parameterize the Coulomb branch, while the $ m_j $ correspond to hypermultiplet masses.\newline

\noindent  As we will review in more detail below, the topological string can be used as a building block to compute the partition function of such superconformal theories on certain compact geometries. In particular, the SCFT partition function on $ S^4\times S^1 $ is given simply by
\bea\label{eq:s4s1}
Z_{S^4\times S^1} = \int \bigg(\prod_{i} dt_i\bigg)\; Z_{top} \, \overline{Z_{top}},
\eea
while the partition function on $ S^5 $ is given by the integral of three factors of $ Z_{top} $, with appropriate choices of parameters:
\bea\label{eq:s5} Z_{S^5} = \int \bigg(\prod_{i} dt_i\bigg)\; Z_{top}\, Z'_{top} \, Z''_{top}.
\eea
Since the five-dimensional gauge theories we are investigating in the present context come from compactifying the worldvolume theory of $ N $ M5 branes on a circle, Eq.\eqref{eq:s4s1} and Eq.$\eqref{eq:s5} $ produce respectively the partition function of $ N $ M5 branes on $ S^4\times T^2 $ and $ S^5\times S^1 $.  The latter partition function corresponds to the superconformal
index of M5 branes. \newline

\subsection{Partition function on $S^4 \times T^2$}
\label{sec:t2s4}
We are interested in computing the partition function of a supersymmetric gauge theory on $ S^4\times S^1 $ , where the fields of the theory are twisted in an appropriate way along the $ S^1 $.  For superconformal theories, this is equivalent to computing the five-dimensional superconformal index
\bea I_{5d} = \text{Tr}\,(-1)^F q^{J_{12}-R}t^{-(J_{34}-R)}\prod_{j}z_j^{f_j},\eea
where $ J_{12}, J_{34} $ are the generators of rotations of the two planes in $ S^4 $ and $ f_i $ are generators of flavor symmetries. It was argued in \cite{Iqbal:2012xm} that
\bea
I_{5d} = \int_{0}^1  \bigg(\prod_i dt_i\bigg) \,\big\vert Z_{top}(t_i,m_j, \epsilon_1,\epsilon_2)\big\vert^2.\label{eq:5dindex}
\eea
In the subset of theories corresponding to gauge theories this naturally from compactifying on $ S^1 $ and following Pestun's computation of the gauge theory partition function on $S^4$ \cite{Pestun:2007rz}:
\bea Z^{gauge}_{S^4} = \int_{Cartan}d\vec a \; Z_{S^4}^{1-loop}\, \vert Z_{\mathbb{R}^4}^{gauge}\vert^2.\eea
where appearing in the integrand is the 4$d$ Nekrasov partition function. The form  of this expression follows from the fact that instanton contributions to the partition function localize to the north and south poles of the four-sphere (the fixed points of the $ J_{12} $ and $ J_{34} $ rotations). In fact, the factor of $ \vert Z_{top}\vert^2 $ correctly accounts for both the instanton factors and the perturbative determinant. Furthermore, it is expected that an analog of the AGT relation \cite{Alday:2009aq} holds in the five-dimensional case under consideration (see for example \cite{Awata:2010yy,Nieri:2013yra}) in which the factors appearing in the integrand are related the correlation functions of a $ q $-deformed version of Toda theory. Let us now use this prescription to compute the partition function of $ N $ M5 branes on $ S^4\times T^2 $.\newline

\noindent\textbf{Single M5 brane}: Let us rewrite the partition function for a single M5 brane on $ \mathbb{R}^4\times T^2 $ as
\bea\label{eq:U1modular}
Z_{\mathbb{R}^4\times T^2}^{(1)}(\tau,m,\epsilon_1,\epsilon_2)&=&\frac{1}{\eta(\tau)}\prod_{i,j=0}^{\infty}\frac{\theta_1(\tau;-m+\epsilon_1(i+1/2)-\epsilon_2(j+1/2))}{\theta_1(\tau;i\epsilon_1-(j+1)\epsilon_2)};
\eea
To obtain this expression from Eq.(\ref{u1pf}) we used the identity $\prod_{j=0}^\infty (1-A\,x^p)=\prod_{p=0}^\infty(1-A\,x^{-p-1})^{-1}$ and multiplied Eq.(\ref{u1pf}) by a factor of
\bea\label{eq:gravit}
Q_\tau^{-1/24}\prod_{i,j=0}^\infty e^{-\pi i(m-\epsilon_1/2-\epsilon_2/2)}=\exp\left[\pi i\left(-\frac{\tau}{12}+\frac{1}{2}\left(m-\frac{\epsilon_1+\epsilon_2}{2}\right)\right)\right],
\eea
which we interpret as gravitational contributions to the genus 1 piece of the topological string partition function. As in section \ref{sec:horizontal}, we can turn each theta function into a modular function by replacing $ E_2(\tau) $ by its non-holomorphic counterpart $ \widehat{E}_2(\tau,\bar \tau) $ in its Eisenstein series representation. Thus Eq.\eqref{eq:U1modular} becomes a weight $ -1/2 $ Jacobi form, $ \eta(\tau)^{-1} $ being the only non-constant piece under $ \tau\to-1/\tau $. The partition function on $ S^4\times T^2 $ is obtained simply by taking two copies of $ Z^{(1)}_{\mathbb{R}^4\times S^1} $\footnote{Both factors of $ Z^{(1)} $ contributing to the partition function are convergent for $ \text{Im}(\tau) > 0 $, since both $ \vert Q_\tau\vert <1 $ and $ \vert Q_{\bar\tau}^{-1}\vert = \vert e^{-2\pi i \bar \tau}\vert  < 1 $.}:
\bea\label{eq:U1S4T2}
Z_{S^4\times T^2}^{(1)}(\tau, \bar\tau, m, \bar m, \epsilon_1, \bar \epsilon_1, \epsilon_2, \bar \epsilon_2)=\big| Z^{(1)}(\tau,m,\epsilon_1,\epsilon_2)\big|^2.\;
\eea
We immediately see that Eq.\eqref{eq:U1S4T2} transforms as a weight $(-1/2,-1/2)$ Jacobi form:
\bea Z_{S^4\times T^2}^{(1)}(-1/\tau,-1/\bar\tau,m/\tau,\bar m/\bar \tau,\dots) = \tau^{-1/2}\bar\tau^{-1/2}Z^{(1)}_{S^4\times T^2}(\tau,\bar \tau,m,\bar m,\dots).\eea
Note that this can be made modular invariant by dividing by $\sqrt{\tau_2}$.\newline

\noindent\textbf{Multiple M5 branes:} Recall that the partition function for $ N $ M5 branes on $ \mathbb{R}^4\times T^2 $ is given by Eq.(\ref{eq:UN}), which we repeat here for convenience:
\bea\label{eq:UNR4T2}
Z_{\mathbb{R}^4\times T^2}^{(N)}(\tau,m,,t_{f_{a}},\epsilon_1,\epsilon_2)=\Big(Z_{\mathbb{R}^4\times T^2}^{(1)}(\tau,m,,t_{f_{a}},\epsilon_1,\epsilon_2)\Big)^{N}\widehat Z^{(N)}(\tau,m,t_{f_{a}},\epsilon_1,\epsilon_2),\hspace{.3in}
\eea
where
\bea\nn
\widehat Z^{(N)}(\tau,m,t_{f_{a}},\epsilon_1,\epsilon_2) = \sum_{\nu_{1},\mathellipsis, \nu_{N-1}}(-Q_{f_{1}})^{|\nu_{1}|}\cdots(-Q_{f_{N-1}})^{|\nu_{N-1}|}\,\prod_{a=1}^{N-1}\prod_{(i,j)\in \nu_{a}}\frac{\theta_{1}(\tau;z^{a}_{ij})
\theta_{1}(\tau;v^{a}_{ij})}{\theta_{1}(\tau;w^{a}_{ij})\theta_{1}(\tau;u^{a}_{ij})},
\eea
\vskip -.3in\bea\eea
with $ z_{ij}^a,\,\dots $ defined as in Eq.(\ref{eq:UN}). We can again make this into a modular form by replacing $ E_2(\tau) $ by $ E_2(\tau,\bar \tau) $ for each theta function. Then the partition function on $ S^4\times T^2 $ is simply given by
\bea
Z_{S^4\times T^2}^{(N)}=\left(Z^{(1)}_{S^4\times T^2}\right)^{N}\oint \prod_{a=1}^{N-1}\frac{dQ_{f_{a}}}{2\pi iQ_{f_{a}}}\,\big| \widehat Z^{(N)}(\tau,m,t_{f_{a}},\epsilon_1,\epsilon_2)\big|^2,
\eea
where the integrals are to be performed along the unit circle. The integrand is now invariant under $ (\tau,m,\epsilon_1,\epsilon_2)\to (-1/\tau,m/\tau,\epsilon_1/\tau,\epsilon_2/\tau) $, and $ Z^{U(N)}_{S^4\times T^2}$ transforms as a modular form of weight $(-N/2,-N/2)$, which can again be made modular invariant by dividing by $\tau_2^{N/2}$.

\subsection{The index for multiple M5 branes}
\label{sec:s5s1}

Analogously to the $ S^4\times S^1 $ case, it was argued in \cite{Lockhart:2012vp} that one can use the refined topological string partition function as a building block to compute the partition function of any superconformal theory on a squashed five-sphere, whose geometry is captured by the equation
\bea
\omega_1^2 |z_1|^2+\omega_2^2 |z_2|^2+\omega_3^2 |z_3|^2=1.
\eea
Conformal invariance implies that one can capture the squashing of the five-sphere in terms of just two parameters $ \epsilon_1 = \omega_1/\omega_3, \epsilon_2=\omega_2/\omega_3 $. Explicitly, we define the following non-perturbative version of the refined topological string partition function:
\begin{align}\nn
Z_{np}(t_i,m_j;\epsilon_1,\epsilon_2)&= e^{C(t_i,m_j;\epsilon_1,\epsilon_2)} \, Z_{top}(t_i,m_j;\epsilon_1+1,\epsilon_2)\\&\times Z_{top}(t_i/\epsilon_1,m/\epsilon_1;1/\epsilon_1+1,\epsilon_2/\epsilon_1)\, Z_{top}(t_i/\epsilon_2,m/\epsilon_2;\epsilon_1/\epsilon_2+1,1/\epsilon_2),\label{eq:nonpert}
\end{align}
where
$C(t_i,m_j;\epsilon_1,\epsilon_2)$ is a cubic prefactor receiving contributions from the genus 0 and genus 1 parts of the topological string partition function.

 Then the partition function of the theory is given simply by
\bea\label{eq:np}
Z_{S^5}(m_j;\epsilon_1,\epsilon_2) = \int dt_i \,Z_{np}(t_i,m_j;\epsilon_1,\epsilon_2).
\eea
We can use this fact to compute the superconformal index for $ N $ M5 branes,
\bea
I_{(2,0)}(\mathbf{q}_1,\mathbf{q}_2,\mathbf{q},\mathbf{Q}_m) = \text{Tr }(-1)^F \mathbf{q}_1^{J_{12}-R_1}\mathbf{q}_2^{J_{34}-R_1}\mathbf{q}^{J_{56}-R_1} \mathbf{Q}_m^{R_2-R_1},
\eea
where $ J_{12},J_{34},J_{56} $ are the generators of $ SO(6) $ rotations acting on $ S^5 $, while $ R_1 $ and $ R_2 $ are generators of the $ Sp(4) $ R-symmetry group, and the trace is over the Hilbert space obtained from radial quantization of the (2, 0) theory. Following the discussion in \cite{Lockhart:2012vp}, this index is equal to the squashed five-sphere partition function of the $5d$ theory obtained by compactifying the  (2,0) theory on a circle, and the parameters $ \mathbf{q},\mathbf{q}_1,\mathbf{q}_2,\mathbf{Q}_m $ are related to the $5d$ gauge theory parameters in the following way:
\bea\nn
\mathbf{q} = \exp(-2\pi i/\tau),\quad \mathbf{q}_1 = \exp(2\pi i \epsilon_1/\tau),\quad  \mathbf{q}_2 = \exp(2\pi i \epsilon_2/\tau),\quad \mathbf{Q}_m = \exp(2\pi i m/\tau).
\eea
In the case of a single M5 brane, it was observed that, surprisingly, the $6d$ index is identical to the $5d$ partition function, after performing a modular transformation with respect to $ \tau $ on the $5d$ parameters\footnote{Recall that to obtain the $ S^5 $ partition function we are required to shift one of the equivariant parameters by 1; since fractional powers of $q$ appear, this leads to sign changes in some terms of the partition function.}:
\bea
I^{(1)}_{(2,0)}(\mathbf{q}_1,\mathbf{q}_2,\mathbf{q},\mathbf{Q}_m) =Z_{\mathbb{R}^4\times S^1}^{(1)}(-1/\tau,m/\tau,\epsilon_2/\tau,\epsilon_1/\tau).
\eea
In the case of $N$
 M5 branes, the partition function involves an integral over the Coulomb branch parameterized by $ (t_1,\dots,t_{N-1}) $.

\noindent In addition to the classical prefactor, we need to include the three factors of the topological string partition function Eq.\eqref{eq:UNR4T2}:
\bea\nn
I^{(N)}_{(2,0)} = \int dt_i \, e^{C(t_i,m_j;\epsilon_1,\epsilon_2)}\, &\Bigg[&Z_{\mathbb{R}^4\times S^1}^{(N)}(\tau,m,t_i,\epsilon_1+1,\epsilon_2)\\\nn
&\times& Z_{\mathbb{R}^4\times S^1}^{(N)}(\tau/\epsilon_1,m/\epsilon_1,t_i/\epsilon_1,1/\epsilon_1+1,\epsilon_1/\epsilon_2)\\\nn
&\times& Z_{\mathbb{R}^4\times S^1}^{(N)}(\tau/\epsilon_2,m/\epsilon_2,t_i/\epsilon_2,\epsilon_1/\epsilon_2+1,1/\epsilon_2)\Bigg]\\\nn
&&\hspace{-1.65in} = \Big(Z^{(1)}(-1/\tau,m/\tau,\epsilon_2/\tau,\epsilon_1/\tau+1)\Big)^{N}\\\nn
\times \int dt_i \, e^{C(t_i,m_j;\epsilon_1,\epsilon_2)}\, &\Bigg[&\widehat Z_{U(N)}(\tau,m,t_i,\epsilon_1+1,\epsilon_2)\\\nn
&\times& \widehat Z^{(N)}(\tau/\epsilon_1,m/\epsilon_1,t_i/\epsilon_1,1/\epsilon_1+1,\epsilon_1/\epsilon_2)\\\nn
&\times& \widehat Z^{(N)}(\tau/\epsilon_2,m/\epsilon_2,t_i/\epsilon_2,\epsilon_1/\epsilon_2+1,1/\epsilon_2)\Bigg].
\eea
We are left with a $ (N-1) $-dimensional integral.  It would be interesting to perform this integral explicitly and study the properties of the six-dimensional index in more detail.

\subsubsection{Comparison with the localization computation}

The index of $ N $ M5 branes has also been computed in \cite{Kim:2012qf} by localization of the gauge theory partition function on $ S^5 $ coming from Scherk-Schwarz reduction of the worldvolume theory of the M5 branes. The aim of this section is to show that our computation is equivalent to the one presented there. This is the case despite the fact that the authors of \cite{Kim:2012qf} consider a different squashing of $ S^5 $: while the deformation parameters we consider enter the equation defining the five-sphere,
\bea
\omega_1^2 |z_1|^2+\omega_2^2 |z_2|^2+\omega_3^2 |z_3|^2=1
\eea
(this type of geometry is also frequently denoted as the ellipsoid), the computation of \cite{Kim:2012qf} is for a round sphere
\bea
|z_1|^2+ |z_2|^2+|z_3|^2=1
\eea
with a non-trivial metric obtained by Scherk-Schwarz reduction of the 6$d$ theory on a circle of radius $ r $, which depends on three squashing parameters $ a, b, c $ satisfying the constraint $ a+b+c=0 $. The round sphere limit for them corresponds to setting $ a = b = c = 0 $. The equality of the partition functions for these two geometries is akin to the fact that the 3d partition functions for the three-dimensional squashed sphere \cite{Hama:2011ea} and ellipsoid \cite{Imamura:2011wg} turn out to be identical.\newline

\noindent To compare results, we make the following identification between our parameters and the ones appearing in \cite{Kim:2012qf} (the squashing parameters $(a,b,c)$, Coulomb branch parameters $ (\lambda_{1},\dots,\lambda_{N}) $, hypermultiplet mass $ \mu $, and dimensionless $5d$ gauge coupling $ \beta=g_{YM}^{5d}/2\pi r $):
\bea
\epsilon_1 = \frac{1+a}{1+c}, \quad \epsilon_2 = \frac{1+b}{1+c}, \quad t_\alpha = \frac{i \lambda_\alpha}{1+c}, \quad m = \frac{\mu}{1+c},\quad \tau = \frac{2\pi i}{\beta(1+c)}.
\eea
Then the partition function computed in \cite{Kim:2012qf} is
\bea\label{eq:ZGauge}
\frac{1}{N!}\int_{\mathbb{R}} \big(\prod_{\alpha=1}^N d \lambda_\alpha\big) \; Z_{cl}\, Z_{pert} \, Z_{inst},
\eea
where $ Z_{cl} $ and $ Z_{inst} $ are given, respectively, in equations (2.22) and (2.50) of  \cite{Kim:2012qf}, while $ Z_{pert} $ can be obtained from equations (2.57), (2.59) and was also computed in \cite{Imamura:2012bm}. Written in our variables, the classical piece of the integrand is
\bea
Z_{cl} = e^{-2\pi i \,\frac{\tau \sum t_\alpha^2}{2\epsilon_1\epsilon_2}}.\eea
The perturbative determinant can be expressed in terms of the triple sine function $S_3(z|\omega_1,\omega_2,\omega_3)$ as\footnote{We refer to Appendix A of \cite{Lockhart:2012vp} for the definition and basic properties of the triple sine function.}:
\bea
Z_{pert} &=& \left(\frac{S_3(0|1+a,1+b,1+c)}{S_3(\mu+\frac{3}{2}|1+a,1+b,1+c)}\right)^N\times\\\nn
&&\hspace{-.5in} \prod_{\alpha > \beta}\frac{S_3(i\lambda_\alpha-i\lambda_\beta|1+a,1+b,1+c)\,S_3(i\lambda_\alpha-i\lambda_\beta+3|1+a,1+b,1+c)}{S_3(i\lambda_\alpha-i\lambda_\beta+\mu+\frac{3}{2}|1+a,1+b,1+c)\,S_3(i\lambda_\alpha-i\lambda_\beta-\mu+\frac{3}{2}|1+a,1+b,1+c)},
\eea
where we have used the fact that $ S_3(-z | \omega_1,\omega_2,\omega_3) = S_3(\omega_1+\omega_2+\omega_3+z|\omega_1,\omega_2,\omega_3) $. The $N$ massless vector multiplets contribute $ N $ simple zeros which are removed by replacing $S_3(0 | 1+a,1+b,1+c)^N$ by $\big(\partial_{\lambda_\alpha}S_3(i\lambda_\alpha|1+a,1+b,1+c)\vert_{\lambda_i=0}\big)^N$. Since the triple sine function is invariant under rescaling,
\bea
S_3(z | \omega_1,\omega_2,\omega_3) = S_3(z/\omega_3|\omega_1/\omega_3,\omega_2/\omega_3,1),
\eea
we can rewrite the perturbative determinant in terms of our variables as
\bea
Z_{pert} &=& \left(\frac{S_3(0|\epsilon_1,\epsilon_2,1)}{S_3(m+\frac{1+\epsilon_1+\epsilon_2}{2}|\epsilon_1,\epsilon_2,1)}\right)^N\\\nn
&\times& \prod_{\alpha > \beta}\frac{S_3(t_\alpha-t_\beta|\epsilon_1,\epsilon_2,1)S_3(t_\alpha-t_\beta+\epsilon_1+\epsilon_2+1|\epsilon_1,\epsilon_2,1)}{S_3(t_\alpha-t_\beta+m+\frac{1+\epsilon_1+\epsilon_2}{2}|\epsilon_1,\epsilon_2,1)S_3(t_\alpha-t_\beta-m+\frac{1+\epsilon_1+\epsilon_2}{2}|\epsilon_1,\epsilon_2,1)}.
\eea
Note that, in this normalization, removing the zero mode corresponds to substituting
\bea
\big(S_3(0|\epsilon_1,\epsilon_2,1)\big)^N \to \big(\partial_{t_\alpha} S_3(t_\alpha|\epsilon_1,\epsilon_2,1)\big)^N
\eea
and replacing the integration measure in Eq.\eqref{eq:ZGauge} with $\prod_{\alpha=1}^N dt_\alpha$, in accordance with the prescription of \cite{Lockhart:2012vp}.
The perturbative determinant can be written in factorized form as
\bea
Z_{pert} = {Z_0 \,\hat Z_{pert}(t_\alpha,m,\epsilon_1,\epsilon_2) \over \hat Z_{pert}(\frac{t_\alpha}{\epsilon_1},\frac{m}{\epsilon_1},-\frac{1}{\epsilon_1},\frac{\epsilon_2}{\epsilon_1})\, \hat Z_{pert}(\frac{t_\alpha}{\epsilon_2},\frac{m}{\epsilon_2},\frac{\epsilon_1}{\epsilon_2},-\frac{1}{\epsilon_2})},
\eea
where
\bea\nn
Z_0=\exp\left[\pi i N\left(\frac{m^3}{6\epsilon_1\epsilon_2}-\left(\frac{1}{\epsilon_1\epsilon_2}+\frac{\epsilon_1}{\epsilon_2}+\frac{\epsilon_2}{\epsilon_1}\right)\frac{m}{24}+\frac{(1+\epsilon_1+\epsilon_2)(\epsilon_1\epsilon_2+\epsilon_1+\epsilon_2)}{24\epsilon_1\epsilon_2}\right)\right]\\
\eea
and
\bea\nn
\hat Z_{pert}(t_\alpha,m,\epsilon_1,\epsilon_2) = \left(\frac{M(t,q)}{\prod_{j=1}^\infty(1-t^j)}\right)^N\prod_{j,k=0}^\infty(1+Q_m q^{j+1/2}t^{k+1/2})^N\\
\times\prod_{\alpha>\beta}\prod_{j,k=0}^\infty \frac{(1+e^{2\pi i (t_\alpha-t_\beta)}Q_mq^{j+1/2}t^{k+1/2})(1+e^{2\pi i (t_\alpha-t_\beta)}Q_m^{-1}q^{j+1/2}t^{k+1/2})}{(1-e^{2\pi i (t_\alpha-t_\beta)}q^{j}t^{k+1})(1-e^{2\pi i (t_\alpha-t_\beta)}q^{j+1}t^{k})}.
\eea
Combining $ Z_0 $ and $ Z_{cl} $ we find the prefactors\footnote{This is consistent with what one would expect in the topological string context with the triple intersection coefficients $ C_{\tau t_\alpha t_\alpha} = 1 $ and $ C_{m m m} = -N/2$, as well as $ \int c_2 \wedge m = -N$.}
\bea
\exp\left[-2\pi i\left(\frac{\tau\sum t_\alpha^2}{2\epsilon_1\epsilon_2}-\frac{N m^3}{12 \epsilon_1\epsilon_2}+\frac{Nm}{48}\left(\frac{1}{\epsilon_1\epsilon_2}+\frac{\epsilon_1}{\epsilon_2}+\frac{\epsilon_2}{\epsilon_1}\right) \right)+\text{const}\right]
\eea

\noindent Also, $ \hat Z_{pert} $ is precisely the contribution to the topological string partition function coming from M2 branes that do not wrap the elliptic fiber (including constant maps, which are responsible for the factor of $ (M(t,q)/\prod_{j=1}^\infty(1-t^j))^N $ as reviewed in \cite{Lockhart:2012vp}). Note that, by using analytic continuation, we can write the perturbative piece as
\bea\nn
&& \hat Z_{pert}(t_\alpha,m,\epsilon_1,\epsilon_2) \bigg/\bigg( \hat Z_{pert}(\frac{t_\alpha}{\epsilon_1},\frac{m}{\epsilon_1},-\frac{1}{\epsilon_1},\frac{\epsilon_2}{\epsilon_1})\, \hat Z_{pert}(\frac{t_\alpha}{\epsilon_2},\frac{m}{\epsilon_2},\frac{\epsilon_1}{\epsilon_2},-\frac{1}{\epsilon_2})\bigg) \\
&&\hspace{.1in} \sim {\hat Z_{pert}(t_\alpha,m,\epsilon_1,\epsilon_2) \hat Z_{pert}(\frac{t_\alpha}{\epsilon_1},\frac{m}{\epsilon_1},\frac{1}{\epsilon_1},\frac{\epsilon_2}{\epsilon_1})\, \hat Z_{pert}(\frac{t_\alpha}{\epsilon_2},\frac{m}{\epsilon_2},\frac{\epsilon_1}{\epsilon_2},\frac{1}{\epsilon_2})}.
\eea
\newline

\noindent Finally, $ Z_{inst} $ also consists of three factors:
\bea
Z_{inst} = Z_{inst}^{( a )}\, Z_{inst}^{( b )}\, Z_{inst}^{( c )},
\eea
where
\bea\label{eq:Zinstc}
Z_{inst}^{( c )} = \sum_{k=0}^\infty e^{2\pi i \tau k}\sum_{|\nu_1|+\dots+|\nu_N|=k} Z_{\vec\nu}^{( c )}(\tau,t_\alpha,m,\epsilon_1,\epsilon_2);
\eea
the two other factors $ Z_{inst}^{( a ),( b )} $ in the instanton piece are obtained by permuting $ (a,b,c) $, which corresponds respectively to taking $ (\epsilon_1,\epsilon_2,\tau,\dots)\to (1/\epsilon_1,\epsilon_2/\epsilon_1,\tau/\epsilon_1,\dots) $ and $ (\epsilon_1,\epsilon_2,\tau,\dots)\to (\epsilon_1/\epsilon_2,1/\epsilon_2,\tau/\epsilon_2) $.  Eq.\eqref{eq:Zinstc} involves a sum over collections of $ N $ Young diagrams $ \nu_1,\dots, \nu_N $ such that the total number of boxes is $ k $, and with a little effort one can rewrite the expression for $ Z_{\vec\nu}^{( c )} $ that appears in \cite{Kim:2012qf} as
\bea\nn
Z_{\vec\nu}^{( c )} &=& \prod_{\alpha,\beta=1}^N \prod_{(i,j)\in \nu_\alpha}\frac{(1+Q_m^{-1}Q_{\alpha \beta}\,q^{\nu_{\alpha,i}-j+1/2}t^{\nu^t_{\beta,j}-i+1/2})(1+Q_mQ_{\alpha \beta}\,q^{\nu_{\alpha,i}-j+1/2}t^{\nu^t_{\beta,j}-i+1/2})}{(1-Q_{\alpha \beta}\,q^{\nu_{\alpha,i}-j}t^{\nu^t_{\beta,j}-i+1})(1-Q_{\alpha \beta}\,q^{\nu_{\alpha,i}-j+1}t^{\nu^t_{\beta,j}-i})},\\
\eea
where $ Q_{\alpha \beta} = e^{2\pi i(t_\alpha-t_\beta)} $, provided that we shift $ \lambda_\alpha \to -i\lambda_\alpha $ in equation (2.48) of \cite{Kim:2012qf}. This can be rewritten as
\bea\nn
(-Q_m\sqrt{t/q})^{-Nk}\prod_{\alpha,\beta=1}^N \bigg[ &&\prod_{(i,j)\in \nu_\beta}\frac{1+Q_mQ_{\alpha \beta}\, q^{-(\nu_{\beta,i}-j+1/2)}t^{-(\nu^t_{\alpha,j}-i+1/2)}}{1-Q_{\alpha \beta}\,q^{-(\nu_{\beta,i}-j)}t^{-(\nu^t_{\alpha,j}-i+1)}}\\
&&\prod_{(i,j)\in\nu_\alpha}\frac{1+Q_mQ_{\alpha \beta}\,q^{\nu_{\alpha,i}-j+1/2}t^{\nu^t_{\beta,j}-i+1/2}}{1-Q_{\alpha \beta}\,q^{\nu_{\alpha,i}-j+1}t^{\nu^t_{\beta,j}-i}}\bigg].
\eea
It is easy to see that Eq.\eqref{eq:Zinstc} is identical to Eq.(\ref{eq:ZUNN}), provided that in Eq.(\ref{eq:ZUNN}) we take $y = Q_m\sqrt{t/q}$ and $\widetilde Q = Q_\tau y^{-N}$ and use invariance of the Nekrasov partition function under $ (\epsilon_1,\epsilon_2)\to(\epsilon_2,\epsilon_1) $. The two equations are in complete agreement once we shift $ \epsilon_1 \to \epsilon_1 + 1 $ in the Nekrasov partition function. Thus we find that $ \hat Z_{pert} $ and $ Z_{inst}^{( c )} $ precisely combine into $ Z_{top}(\tau,t_\alpha,m,\epsilon_1+1,\epsilon_2) $, and indeed the partition function we compute agrees with the one in \cite{Kim:2012qf}, up to an overall factor of $ 1/N!$ which is simply explained by the choice of region of integration.

\section{Directions for Future Research}
In this paper we have shown how to compute the supersymmetric partition function for M-strings on $T^2$.  We have found
a number of interesting structures and insights about the nature of M-strings.  In particular we have seen the similarities
and the differences between  the theory seen by $N$ copies of M-strings as compared to the N-fold symmetric product of $\mathbb{R}^4$.  We have computed
the partition function of domain walls induced by M5 branes which separates a number of M2 branes.  We have also
seen how M-strings can form new bound states when they wind around a circle.

There are a number of remaining issues that need to be better understood:  The theory of M-strings should enjoy a $(4,4)$
supersymmetry.  However we have not been able to find a $(4,4)$ supersymmetric theory which for generic choices
of twistings along cycles of $T^2$ leads to the elliptic genus of M-strings.  Instead we have seen that there is a $(4,0)$
theory which can accomplish this task. On the other hand, we have shown that, as expected, on a codimension one subspace
of twistings along cycles of $T^2$ the anticipated $(4,4)$ quiver theory does lead to the correct answer
for the elliptic genus.

The $(4,0)$ theory that we found was obtained by a duality involving the sigma model of instanton moduli spaces of $U(1)^{N-1}$ quiver $A_{N-1}$ gauge theories in 6 dimensions.  It would be interesting to see if the same duality works for 6 dimensional D and E $(2,0)$ theories as well, leading
to a D and E quiver theory in six dimensions whose instanton moduli space leads to a sigma model description of the $(4,0)$ theory of these strings.  This should be computable using \cite{Katz:1997eq} or instanton calculus \cite{Nekrasov:2012xe}.

The M2 branes wrapped on $T^2$ lead to a 1-dimensional theory.  We have also studied
the partition function of M-strings from this viewpoint.  In this context, the domain walls become
operators acting on the Hilbert space, where we view the one dimension as time.  In this way, the partition
function of M-strings gets translated to a computation in a quantum mechanical theory, where the Hilbert space
is identified with Young diagrams and the Hamiltonian is the number of boxes.  It would be interesting
to better understand these domain wall theories \cite{Berman:2009xd}.  In particular it would be useful
to further develop what these
domain wall theories are and how they couple to the ABJM theory \cite{Aharony:2008ug}.

The viewpoint of domain walls may have other applications.  In particular exceptional strings (E-strings), which can be viewed   \cite{Seiberg:1996vs,Ganor:1996mu}
 as M2 branes stretched between M5 branes
and M9 branes (the boundary of the space) should admit a similar decomposition, where we use two types
of domain wall operators:  one induced by M5 branes which we have studied here, combined with another one
induced from M9 branes (see in particular \cite{Sakai:2012zq}).  It would be interesting to develop this picture.

We have seen how the elliptic genus of M-strings can be used to compute the partition function of M5 branes
on $S^1\times S^5$ and $T^2\times S^4$.  It would be interesting to see if this can be extended to a stringy
definition of the M5 brane theory.  Can one compute arbitrary M5 brane amplitudes using M-strings?  We leave
this question to future research.

\section*{Acknowledgements}
We would like to thank G. Bonelli, A. Gadde, S. Gukov, S. Hohenegger, K. Hori, D. Jafferis, S. Katz, A. Klemm, M. Rocek, A. Tanzini, S. Vandoren and F. Yagi for valuable discussions.  C.K. would also like to thank the Harvard University Theoretical High Energy Physics/String Theory group for hospitality.

The work of B.H. is supported by DFG fellowship HA6096/1-1. A.I. was supported in part by a grant from the Higher Education Commission.  C.K. was partly supported by the INFN Research Project TV12. The work of G.L. is supported in part by the Department of Energy Office of Science Graduate Fellowship Program (DOE SCGF), made possible in part by the American Recovery and Reinvestment Act of 2009, administered by ORISE-ORAU under contract no. DE-AC05-06OR23100. The work of C.V. is supported in part by NSF grant PHY-0244821.

\appendix

\section{Useful identities}

\par{The refined topological vertex can be written in different bases of the ring of symmetric functions. In this paper, we use the representation of \cite{Iqbal:2007ii} which is based on the combinatorial interpretation of the vertex. Like the usual topological vertex, the refined one is labeled by three Young diagrams and can be written in terms of skew Schur functions $s_{\lambda/\eta}(\mathbf{x})$ and Macdonald polynomial $P_{\nu}(t^{-\rho};q,t)$ as}
\begin{align}\nn
C_{\lambda\,\mu\,\nu}(t,q)&=\Big(\frac{q}{t}\Big)^{\frac{\Arrowvert\mu\Arrowvert^2}{2}}\,t^{\frac{\kappa(\mu)}{2}-\frac{\Arrowvert\nu^{t}\Arrowvert^{2}}{2}}\,q^{\frac{\Arrowvert\nu\Arrowvert^2}{2}}\,P_{\nu}(t^{-\rho};q,t)\\
&\times\sum_{\eta}\Big(\frac{q}{t}\Big)^{\frac{|\eta|+|\lambda|-|\mu|}{2}}\,s_{\lambda^{t}/\eta}(t^{-\rho}\,q^{-\nu})\,s_{\mu/\eta}(t^{-\nu^t}\,q^{-\rho})\, ,
\end{align}
where $\rho$ is used for $\rho=\{-\frac{1}{2},-\frac{3}{2},-\frac{5}{2},\cdots\}$. The Macdonald polynomial with the special arguments appearing in the refined vertex can be expressed in terms of Young diagrams
\bea
P_{\nu}(t^{-\rho};q,t)=t^{\frac{\Arrowvert\nu^t\Arrowvert^2}{2}}\,\widetilde{Z}_{\nu}(t,q),
\eea
where we have defined the following function
\bea
\widetilde{Z}_{\nu}(t,q)&=&\prod_{(i,j)\in \nu}\Big(1-q^{\nu_{i}-j}\,t^{\nu^{t}_{j}-i+1}\Big)^{-1}.
\eea
In our computations, we have explicitly used the function $\widetilde{Z}_{\nu}(t,q)$. The refined vertex has the following form in terms of it:

\begin{align}
C_{\lambda\,\mu\,\nu}(t,q)&=&\Big(\frac{q}{t}\Big)^{\frac{\Arrowvert\mu\Arrowvert^2}{2}}\,t^{\frac{\kappa(\mu)}{2}}\,q^{\frac{\Arrowvert\nu\Arrowvert^2}{2}}\,\widetilde{Z}_{\nu}(t,q)
\sum_{\eta}\Big(\frac{q}{t}\Big)^{\frac{|\eta|+|\lambda|-|\mu|}{2}}\,s_{\lambda^{t}/\eta}(t^{-\rho}\,q^{-\nu})\,s_{\mu/\eta}(t^{-\nu^t}\,q^{-\rho})\, .
\end{align}

We have made use of the following identities in our computations of the topological string partition functions
\begin{align}\label{n1}
&n(\lambda)\equiv\sum_{i=1}^{\ell(\lambda)}(i-1)\lambda_{i}=\frac{1}{2}\sum_{i=1}^{\ell(\lambda)}\lambda^{t}_{i}(\lambda^{t}_{i}-1)=\sum_{(i,j)\in\lambda}(\lambda_{j}^{t}-i)=\frac{\Arrowvert\lambda^{t}\Arrowvert^{2}}{2}-\frac{|\lambda|}{2}\,,\\\label{n2}
&n(\lambda^{t})\equiv\sum_{i=1}^{\ell(\lambda^{t})}(i-1)\lambda^{t}_{i}=\frac{1}{2}\sum_{i=1}^{\ell(\lambda^{t})}\lambda_{i}(\lambda_{i}-1)=\sum_{(i,j)\in\lambda}(\lambda_{i}-j)=\frac{\Arrowvert\lambda\Arrowvert^{2}}{2}-\frac{|\lambda|}{2}\,,
\end{align}
with $\ell(\lambda)$ being the number of non-zero $\lambda_{i}$'s. We have also used $\Arrowvert\lambda\Arrowvert^{2}=\sum_{i=1}^{\ell(\lambda)}\lambda_{i}^{2}$. The hook length $h(i,j)$ and the content $c(i,j)$ are defined as
\bea
h(i,j)=\nu_{i}-j+\nu_{j}^{t}-i+1,\qquad c(i,j)=j-i\,,
\eea
which satisfy
\begin{align}
&\sum_{(i,j)\in\lambda}h(i,j)=n(\lambda^{t})+n(\lambda)+|\lambda|,\\
&\sum_{(i,j)\in\lambda}c(i,j)=n(\lambda^{t})-n(\lambda)=\frac{1}{2}\Arrowvert\lambda\Arrowvert^{2}-\frac{1}{2}\Arrowvert\lambda^{t}\Arrowvert^{2}\equiv \frac{1}{2}\kappa(\lambda).
\end{align}
We also have made use of the following identities
\begin{align}\label{taki}
&\sum_{(i,j)\in\nu}\mu_{j}^{t}=\sum_{(i,j)\in\mu}\nu_{j}^{t}\qquad\mbox{\cite{Taki:2007dh}}\\
&\sum_{(i,j)\in\nu}\nu_{j}^{t}=\Arrowvert\nu^{t}\Arrowvert^{2}
\end{align}
The following sum rules are essential for vertex computations \cite{macdonald}
\begin{align}
&\sum_{\eta}s_{\eta/\lambda}(\mathbf{x})s_{\eta/\mu}(\mathbf{y})=\prod_{i,j=1}^{\infty}(1-x_{i}y_{j})^{-1}\sum_{\tau}s_{\mu/\tau}(\mathbf{x})s_{\lambda/\tau}(\mathbf{y})\,.\\
&\sum_{\eta}s_{\eta^{t}/\lambda}(\mathbf{x})s_{\eta/\mu}(\mathbf{y})=\prod_{i,j=1}^{\infty}(1+x_{i}y_{j})\sum_{\tau}s_{\mu^{t}/\tau}(\mathbf{x})s_{\lambda^{t}/\tau^{t}}(\mathbf{y})\,.
\end{align}
We have normalised the topological string amplitudes. Both the open and closed amplitudes are infinite series in the K\"{a}hler parameters (\textit{e.g} $W_{\nu_{m}^{t}\nu_{m+1}}(Q_{\tau},Q_{m},t,q)$), however, their ratio is finite (\textit{e.g.} $D_{\nu_{m}^{t}\nu_{m+1}}(Q_{\tau},Q_{m},t,q)$) as a result of the following identity
\begin{align}
\prod_{i,j=1}^{\infty}\frac{1-Q\,q^{\nu_{i}-j}t^{\mu_{j}^{t}-i+1}}{1-Q\,q^{-j}t^{-i+1}}=\prod_{(i,j)\in\nu}\Big(1-Q\,q^{\nu_{i}-j}t^{\mu_{j}^{t}-i+1}\Big)\prod_{(i,j)\in\mu}\Big(1-Q\,q^{-\mu_{i}+j-1}t^{-\nu_{j}^{t}+i}\Big),
\end{align}
and its specializations
\begin{align}
&\prod_{i,j=1}^{\infty}\frac{1-Q\,t^{\nu^{t}_{j}-i+\frac{1}{2}}\,q^{-j+\frac{1}{2}}}{1-Q\,t^{-i+\frac{1}{2}}\,q^{-j+\frac{1}{2}}}=\prod_{(i,j)\in \nu}\Big(1-Q\,q^{-j+\frac{1}{2}}\,t^{i-\frac{1}{2}}\Big),\\
&\prod_{i,j=1}^{\infty}\frac{1-Q\,q^{\nu_{i}-j+\frac{1}{2}}\,t^{-i+\frac{1}{2}}}{1-Q\,q^{-j+\frac{1}{2}}\,t^{-i+\frac{1}{2}}}=\prod_{(i,j)\in \nu}\Big(1-Q\,q^{j-\frac{1}{2}}\,t^{-i+\frac{1}{2}}\Big),\\
&\prod_{i,j=1}^{\infty}\frac{1-Q\,t^{\nu^{t}_{j}-i}\,q^{\nu_{i}-j+1}}{1-Q\,t^{-i}\,q^{-j+1}}=\prod_{(i,j)\in \nu}(1-Q\,t^{\nu^{t}_{j}-i}\,q^{\nu_{i}-j+1})(1-Q\,t^{-\nu^{t}_{j}+i-1}\,q^{-\nu_{i}+j}).
\end{align}
We have also written the partition functions in terms of the $\theta$-functions. We follow the following definitions in our computations. The (first) $\theta$-function and the Dedekind $\eta$-function are defined by
\bea\nn
\theta_{1}(\tau;z)&=&-i\,e^{\frac{i\pi\,\tau}{4}}\,e^{i\pi z}\prod_{k=1}^{\infty}\Big[(1-e^{2\pi\,i\,k\tau})(1-e^{2\pi\,i\,k\tau}\,e^{2\pi i\,z})(1-e^{2\pi\,i\,(k-1)\tau}\,e^{-2\pi i\,z})\Big]\\\nn
&=&-i\,e^{\frac{i\pi\,\tau}{4}}\,(e^{i\pi z}-e^{-i\pi z})\prod_{k=1}^{\infty}\Big[(1-e^{2\pi\,i\,k\tau})(1-e^{2\pi\,i\,k\tau}\,e^{2\pi i\,z})(1-e^{2\pi\,i\,k\tau}\,e^{-2\pi i\,z})\Big]\\\nn
\eta(\tau)&=&e^{\frac{i\pi \tau}{12}}\,\prod_{k=1}\Big(1-e^{2\pi i\,k\tau}\Big)
\eea
They satisfy the following modular transformations
\bea
\theta_{1}(\tau+1;z)&=&\theta_{1}(\tau;z)\,,\,\,\,\,\,\,
\theta_{1}\left(-\frac{1}{\tau};\frac{z}{\tau}\right)=-i(-i\tau)^{\frac{1}{2}}\,\mbox{exp}\left(\frac{i\pi z^2}{\tau}\right)\theta_{1}(\tau; z),\\\nn
\eta(\tau+1)&=&e^{\frac{i\,\pi}{12}}\eta(\tau)\,,\,\,\,\eta\left(-\frac{1}{\tau}\right)=\sqrt{-i\tau}\,\eta(\tau).
\eea


\section{Derivation of the building block $W_{\nu_{m}^{t}\nu_{m+1}}$}
\par{In this appendix, we will represent the derivation of the topological string partition function of the building blocks $W_{\nu_{m}^{t}\nu_{m+1}}(Q_{\tau},Q_{m},t,q)$ that we used to compute the partition function of the geometry engineering the ${\cal N}=2^{*}$ $SU(N)$ theory. The corresponding toric diagram is shown in \figref{sunbl}}.
\begin{figure}[h]
  \centering
  \includegraphics[width=1in]{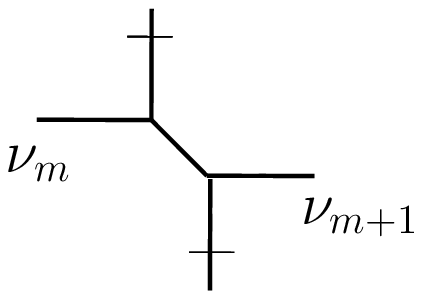}\\
  \caption{The building block of the $SU(N)$ geometry. The dashes represent the partial compactification of the resolved conifold.}\label{sunbl}
\end{figure}
The partition function is in the following generic form when the refined topological vertex is used to compute it

\bea
G(x,y,w,z)\equiv\sum_{\lambda,\mu,\eta_{1},\eta_{2}} Q^{|\lambda|} \rho^{|\mu|}s_{\lambda/\eta_{1}}(x)s_{\lambda^{t}/\eta_{2}}(y)s_{\mu^{t}/\eta_{1}}(w)s_{\mu/\eta_{2}}(z).
\eea
We can perform the sums twice and show that $G(x,y,w,z)$ satisfies the following recursion relationship,
\bea
G(x,y,w,z)=\prod_{i,j}\frac{(1+Q\,x_{i}y_{j})(1+\rho\,w_{i}z_{j})}{(1-Q\rho\,x_{i}w_{j})(1-Q\rho\,y_{i}z_{j})}G(Qx,\rho y,\rho w,Qz).
\eea
Note that the argument in each four factors scale with $Q\rho$,
\bea
x_{i}y_{j}&\mapsto& (Q\rho)\, x_{i}y_{j}\\\nn
w_{i}z_{j}&\mapsto& (Q\rho)\, w_{i}z_{j}\\\nn
x_{i}w_{j}&\mapsto& (Q\rho)\, x_{i}w_{j}\\\nn
y_{i}z_{j}&\mapsto& (Q\rho)\, y_{i}z_{j}\\\nn
\eea
Following the conventions in the rest of the paper, we call $Q_{\tau}=Q\rho$ and write down the $n^{th}$ iterative step,
\begin{align}
G(x,y,w,z)=\prod_{k=1}^{n}\prod_{i,j}\frac{(1+Q_{\tau}^{k}\rho^{-1}\,x_{i}y_{j})(1+Q_{\tau}^{k-1}\rho\,w_{i}z_{j})}{(1-Q_{\tau}^{k}\,x_{i}w_{j})(1-Q_{\tau}^{k}\,y_{i}z_{j})}G(Q^{n}x,\rho^{n} y,\rho^{n} w,Q^{n}z).
\end{align}
Under the assumption that $Q^{n},\rho^{n}\rightarrow 0$ as $n\rightarrow \infty$ (this assumption is the same as the one used by Macdonald \cite{macdonald} to prove the summation rules we are using), we need to take the following limit,
\bea
\lim_{n \to \infty}G(Q^{n}x,\rho^{n} y,\rho^{n} w,Q^{n}z).
\eea
It is easy to show that
\begin{align}
G(Q^{n}x,\rho^{n} y,\rho^{n} w,Q^{n}z)=\sum_{\lambda,\mu,\eta_{1},\eta_{2}} Q^{|\lambda|} \rho^{|\mu|}s_{\lambda/\eta_{1}}(Q_{\tau}^{n}x)s_{\lambda^{t}/\eta_{2}}(y)s_{\mu^{t}/\eta_{1}}(w)s_{\mu/\eta_{2}}(Q_{\tau}^{n}z).
\end{align}
The only surviving terms in the $n\rightarrow\infty$ limit are when $\lambda=\eta_{1}$ and $\mu=\eta_{2}$. Therefore, four sums reduces to two sums
\bea
\sum_{\eta_{1},\eta_{2}}Q^{|\eta_{1}|} \rho^{|\eta_{2}|}s_{\eta_{1}^{t}/\eta_{2}}(y)s_{\eta_{2}^{t}/\eta_{1}}(z).
\eea
The only non-zero terms in these resulting sums are when $\eta_{1}^{t}\succ\eta_{2}$, and simultaneously, when $\eta_{2}^{t}\succ\eta_{1}$. These last two conditions require $\eta_{1}^{t}=\eta_{2}$. A further reduction of the sums to a single one occurs and we end up with the following sum which is known how to perform to get an infinite product (notice it is proportional to the Dedekind $\eta$-function )
\bea
\sum_{\eta}Q_{\tau}^{|\eta|}=\prod_{k=1}^{\infty}(1-Q_{\tau}^{k})^{-1}.
\eea
All in all, $G(x,y,w,z)$ can be written as the following triple infinite product
\bea
G(x,y,w,z)=\prod_{k=1}^{\infty}(1-Q_{\tau}^{k})^{-1}\prod_{i,j}\frac{(1+Q_{\tau}^{k}\rho^{-1}\,x_{i}y_{j})(1+Q_{\tau}^{k-1}\rho\,w_{i}z_{j})}{(1-Q_{\tau}^{k}\,x_{i}w_{j})(1-Q_{\tau}^{k}\,y_{i}z_{j})}.
\eea
Having performed all the sums and found a product formula we can now replace $x,\,y,\,w$ and $z$ with what we have in the vertex computation
\begin{equation}
\begin{array}{ll}
x=t^{\rho-\frac{1}{2}}q^{\nu_{m+1}+\frac{1}{2}}\qquad& y=q^{\rho-\frac{1}{2}}t^{\nu_{m}^{t}+\frac{1}{2}} \\
w=t^{\nu_{m+1}^{t}}q^{\rho} & z=q^{\nu_{m}}t^{\rho}.
\end{array}
\end{equation}
The building block therefore takes the following form
\begin{align}\nn
&W_{\nu_{m}^{t}\nu_{m+1}}(Q_{\tau},Q,t,q)=t^{-\frac{\Arrowvert\nu^{t}_{m+1}\Arrowvert^2}{2}}\,q^{-\frac{\Arrowvert\nu_{m}\Arrowvert^2}{2}}\,\widetilde{Z}_{\nu_{m}^{t}}(q^{-1},t^{-1})\widetilde{Z}_{\nu_{m+1}}(t^{-1},q^{-1})\,Q_{m}^{-\frac{|\nu_m|+|\nu_{m+1}|}{2}}\\
&\times\prod_{k=1}^{\infty} \left(1-Q_{\tau}^k\right)^{-1} \prod_{i,j=1}^{\infty}\frac{\left(1-Q_{\tau}^{k}Q_{m}^{-1}\,q^{\nu_{m+1,i}-j+\frac{1}{2}}t^{\nu^{t}_{m,j}-i+\frac{1}{2}}\right)
\left(1-Q_{\tau}^{k-1}Q_{m}\,q^{\nu_{m,i}-j+\frac{1}{2}}t^{\nu^{t}_{m+1,j}-i+\frac{1}{2}}\right)}
{\left(1-Q_{\tau}^{k}\,q^{\nu_{m+1,i}-j+1}t^{\nu^{t}_{m+1,j}-i}\right)\left(1-Q_{\tau}^{k}\,q^{\nu_{m,i}-j}t^{\nu_{m,j}^{t}-i+1}\right)}.
\end{align}

\section{The $SU(N)$ partition function in terms of $\theta$-function}
\par{In this section, we want to collect few details of our computation how to express the $SU(N)$ partition function in terms of $\theta$-functions. In the previous section of the Appendix we demonstrated our derivation of the building blocks that we use to compute the topological string partition function for the geometries engineering $SU(N)$ theories. Although the individual building blocks are modular only in the unrefined case (after the non-holomorphic extension), the blocks are not modular in the refined case. However, factors appearing in the neighbouring building blocks combine in a very nice way into $\theta$-functions. Let us look at the gluing along the $m^{th}$ internal leg and collect all the infinite products including $\nu_{m}$:}

\begin{align}\nn
\prod_{k=1}^{\infty}\prod_{(i,j)\in\nu_{m}}&\frac{(1-Q_{\tau}^{k}Q_{m}^{-1}\,q^{\nu_{m,i}-j+\frac{1}{2}}t^{\nu_{m-1,j}^{t}-i+\frac{1}{2}})(1-Q_{\tau}^{k-1}Q_{m} \,q^{-\nu_{m,i}+j-\frac{1}{2}}t^{-\nu_{m-1,j}^{t}+i-\frac{1}{2}})}{\underbracket{(1-Q_{\tau}^{k}\,q^{\nu_{m,i}-j+1}t^{\nu_{m,j}^{t}-i})}_{I}\underbracket{(1-Q_{\tau}^{k-1}\,q^{-\nu_{m,i}+j}t^{-\nu_{m,j}^{t}+i-1})}_{II}}\\
\times&\frac{(1-Q_{\tau}^{k}Q_{m}^{-1}\,q^{-\nu_{m,i}+j-\frac{1}{2}}\,t^{-\nu_{m+1,j}^{t}+i-\frac{1}{2}})
(1-Q_{\tau}^{k-1}Q_{m}\,q^{\nu_{m,i}-j+\frac{1}{2}}\,t^{\nu_{m+1,j}^{t}-i+\frac{1}{2}})}{\underbracket{(1-Q_{\tau}^{k}\,q^{\nu_{m,i}-j}\, t^{\nu_{m,j}^{t}-i+1})}_{II}\underbracket{(1-Q_{\tau}^{k-1}\,q^{-\nu_{m,i}+j-1}\,t^{-\nu_{m,j}^{t}+i})}_{I}},
\end{align}
where we have written the factors including $\nu_{m}$ from $D_{\nu_{m-1}^{t}\nu_{m}}(\tau,m,\epsilon_{1},\epsilon_{2})$ in the first line and the ones from $D_{\nu_{m}^{t}\nu_{m+1}}(\tau,m,\epsilon_{1},\epsilon_{2})$ in the second line. It is clear from the above expression that the factors in the numerator originating from the same block can be combined into $\theta$-functions, however, only the underlined factors from the neighbouring blocks combine into $\theta$-functions.
The definition of the $\theta_{1}(\tau;z)$, in addition to the infinite products in the above expansion, includes the factor $-ie^{i\pi\,\tau/2}e^{i\pi\,z}(1-Q_{\tau}^{k})$. We can multiply the numerator and denominator by the $\tau$-dependent pieces without anything else needed, however, $e^{i\pi\,z}$ requires a little bit more attention. Let us separately treat the numerator and the denominator and start with the easier one, the denominator: we will have the following factors
\begin{align}
\prod_{(i,j)\in\nu_{m}}\frac{1}{(qt)^{-\frac{1}{2}}q^{-\nu_{m,i}+j} t^{-\nu_{m,j}^{t}+i}}=q^{\frac{\Arrowvert\nu_{m}\Arrowvert^{2}}{2}}t^{\frac{\Arrowvert\nu_{m}^{t}\Arrowvert^{2}}{2}}
\end{align}
where we have made use of Eq.(\ref{n1}) and Eq.(\ref{n2}). These factors will cancel against the factors appearing in the definition of $D_{\nu_{m-1}^{t}\nu_{m}}(\tau,m,\epsilon_{1},\epsilon_{2})$ and $D_{\nu_{m}^{t}\nu_{m+1}}(\tau,m,\epsilon_{1},\epsilon_{2})$. In the numerator we end up with
\begin{align}
\prod_{(i,j)\in\nu_{m}}Q_{m}\,t^{\nu_{m+1,j}^{t}-\nu_{m-1,j}^{t}},
\end{align}
where we take the product over the Young diagram $\nu_{m}$ of quantities which depend on the neighbouring Young diagrams $\nu_{m-1}$ and $\nu_{m+1}$. We do not know the closed form expressions for these products. However, we can use the Eq.(\ref{taki}) to show that these factors all disappear if we consider the $SU(N)$ as a whole.

After gluing all these blocks for the $SU(N)$ theory we end up with the following sum (we take $\nu_{0}=\nu_{N}=\emptyset$)

\begin{align}\nn
&\sum_{(i,j)\in\nu_{1}}(-\nu_{0,j}^{t}+\nu_{2,j}^{t})+\sum_{(i,j)\in\nu_{2}}(-\nu_{1,j}^{t}+\nu_{3,j}^{t})+\mathellipsis+\sum_{(i,j)\in\nu_{N-1}}(-\nu_{N-2,j}^{t}+\nu_{N,j}^{t})\\
=&\left(\sum_{(i,j)\in\nu_{1}}\nu_{2,j}^{t}-\sum_{(i,j)\in\nu_{2}}\nu_{1,j}^{t}\right)+\mathellipsis+\left(\sum_{(i,j)\in\nu_{N-2}}\nu_{N-1,j}^{t}-\sum_{(i,j)\in\nu_{N-1}}\nu_{N-2,j}^{t}\right)=0
\end{align}
The partition function of the $SU(N)$ theory can be those written only in terms of $\theta$-functions without any other factors.


\section{Spin content of $SU(2)$ theory}
In this section, we want to tabulate the spin content of the BPS states which we have obtained by isolating curves in the topological string free energy for the $SU(2)$ theory, in other words we have determined $N_{j_{L},j_{R}}(\Sigma)$ for some of the low degree curves. Higher degree curves can in principle be computed as well with the increasing need for computational power.
\begin{figure}[h]
\begin{center}
\includegraphics[scale=0.7]{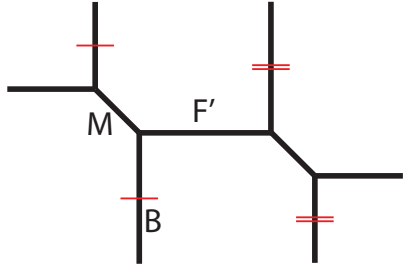}
\caption{The toric diagram for ${\cal N}=2^{*}$ SU(2) theory with the curve classes labeled by $M$, $B$ and $F'$, after mass, base and fiber.}\label{basis}
\end{center}
\end{figure}
The topological string free energy is the generating function for the BPS states and is a positive power expansion in the homology classes $M$, $B$ and $F'$. One interesting property of the $SU(2)$ partition function is invariance under the following transformation:}
\bea\label{t1}
(M,B,F')\mapsto (-M,B,F')\,.
\eea
This transformation is nothing but realisation of flop transition on the curve in the class $M$. Since the geometry is the same after flop transition (toric diagram is shown in \figref{td}) the partition function is expected to be invariant. It is easy to see from the Eq.(\ref{su2pf}) that the partition function is indeed invariant under transformation given in Eq.(\ref{t1}). Flop transition can also be carried out with respect to the curve with parameter $E-M$.

\begin{figure}[h]
  \centering
  \includegraphics[width=2in]{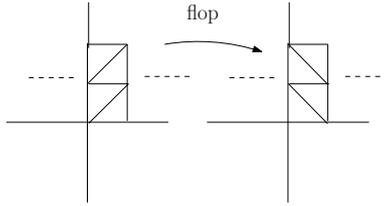}\\
  \caption{Toric diagram of $SU(2)$ geometry.}\label{td}
\end{figure}

The invariance under Eq.(\ref{t1}) implies that the spin content of the curve $k\,F'+n\,B+r\,M$ is the same as that of $k\,F'+n\,B+(2k+2n-r)\,M$. This also implies that curves for which $r>2k+2n$ are not holomorphic and do not contribute. Thus for a given $k$ and $n$ the curves which contribute are,
\bea
kF'+nB+rM\,,\,\,r=0,\mathellipsis, 2(k+n)\,.
\eea
If we change the basis of the second homology of our target space the symmetries and properties of the BPS content becomes more manifest. Therefore we will use the basis when we represent
\begin{align}
F&=F'+M\\
E&=B+M
\end{align}
referring to fiber and elliptic curves, respectively. In our computations we denoted the corresponding K\"{a}hler parameters for $F$ with $Q_{f}$, for $E$ with $Q_{\tau}$ and for $M$ with $Q_{m}$. In this new basis only the following curves will contribute,
\bea
kF+nE+rM\,,\,\,r=-(k+n),\mathellipsis, (k+n)\,.
\eea
As shown below, for $k=1,n=0$ we will only have $F-M$, $F$ and $F+M$, and for $k=n=1$ we will only have $F+E-2M,F+E-M,F+E,F+E+M,F+E+2M$. Another interesting observation based on the BPS content we computed is that the state corresponding to the curve $C$ are included in $C+nE$ for any positive $n$:
\begin{align}
{\cal H}_{C}\subset {\cal H}_{C+n E}
\end{align}
In the $ (F,E,M) $ basis the consequence of flop invariance is that the spin content is the same for all curves belonging to the same orbit of the group $ G = \mathbb{Z}_2\times\mathbb{Z}_2 $. The action of its generators on the collection of curves $ \{f F + e E + m M\} $ is given by
\bea\nn
r:&& (f,e,m)\to (f,e,-m) \qquad\qquad\qquad\quad\; \text{(flop transition on curve in class $M$)}\\\nn
s:&& (f, e,m)\to (f,f+e+m,-2f-m) \quad \text{(flop transition on curve in class $E - M$).}
\eea
It is straightforward to check that $ r^2 = s^2= id $ and that the stabilizer of a generic curve is trivial, the only exceptions being curves in class $ (f, e, 0) $ and $ (f, e,-f) $, which are respectively fixed points under the action of $ r $ and $ s $.

\newpage
\begin{tabular}{|l|p{11.5cm}|}
\hline
$F-M$& $(0,0)$ \\\hline
$F$& (0,1/2) \\\hline
$F+M$ &$(0,0)$ \\
\hline \hline
$F+E-2M$ &$(0,1/2)$ \\\hline
$F+E-M$&$(0,1)\oplus(1/2,1/2)\oplus(0,0)$\\\hline
$F+E$&$(1/2,1)\oplus(1/2,0)\oplus2(0,1/2)$\\\hline
$F+E+M$&$(0,1)\oplus(1/2,1/2)\oplus(0,0)$\\\hline
$F+E+2M$&$(0,1/2)$\\
\hline \hline
$F+2E-3M$&$(0,0)$\\\hline
$F+2E-2M$&$(1/2,1)\oplus(1/2,0)\oplus2(0,1/2)$\\\hline
$F+2E-M$&$(1/2,3/2)\oplus(1,1)\oplus2(0,1)\oplus3(1/2,1/2)\oplus4(0,0)$\\\hline
$F+2E$&$(1,3/2)\oplus(1,1/2)\oplus3(1/2,1)\oplus(0,3/2)\oplus3(1/2,0)\oplus5(0,1/2)$\\\hline
$F+2E+M$&$(1/2,3/2)\oplus(1,1)\oplus2(0,1)\oplus3(1/2,1/2)\oplus4(0,0)$\\\hline
$F+2E+2M$&$(1/2,1)\oplus(1/2,0)\oplus2(0,1/2)$\\\hline
$F+2E+3M$&$(0,0)$\\
\hline \hline
$F+3E-3M$&$(0,1)\oplus(1/2,1/2)\oplus(0,0)$\\\hline
$F+3E-2M$&$(1,3/2)\oplus(1,1/2)\oplus3(1/2,1)\oplus(0,3/2)\oplus3(1/2,0)\oplus5(0,1/2)$\\\hline
$F+3E-M$&$(1,2)\oplus(3/2,3/2)\oplus3(1/2,3/2)\oplus3(1,1)\oplus2(1,0)\oplus7(0,1)\oplus9(1/2,1/2)\oplus7(0,0)$\\\hline
$F+3E$&$(3/2,2)\oplus(3/2,1)\oplus3(1,3/2)\oplus(1/2,2)\oplus4(1,1/2)\oplus9(1/2,1)\oplus3(0,3/2)\oplus8(1/2,0)\oplus12(0,1/2)$\\\hline
$F+3E+M$&$(1,2)\oplus(3/2,3/2)\oplus3(1/2,3/2)\oplus3(1,1)\oplus2(1,0)\oplus7(0,1)\oplus9(1/2,1/2)\oplus7(0,0)$\\\hline
$F+3E+2M$&$(1,3/2)\oplus(1,1/2)\oplus3(1/2,1)\oplus(0,3/2)\oplus3(1/2,0)\oplus5(0,1/2)$\\\hline
$F+3E+3M$&$(0,1)\oplus(1/2,1/2)\oplus(0,0)$\\
\hline \hline
$2F+E-2M$&$(0,3/2)$\\\hline
$2F+E-M$&$(1/2,3/2)\oplus(0,2)\oplus(0,1)$\\\hline
$2F+E$&$(1/2,2)\oplus(1/2,1)\oplus2(0,3/2)$\\\hline
$2F+E+M$&$(1/2,3/2)\oplus(0,2)\oplus(0,1)$\\\hline
$2F+E+2M$&$(0,3/2)$\\
\hline\hline
$2F+2E-3M$&$(1/2,3/2)\oplus(0,2)\oplus(0,1)$\\\hline
$2F+2E-2M$&$(1,5/2)\oplus(1,3/2)\oplus3(1/2,2)\oplus(0,5/2)\oplus3(1/2,1)\oplus4(0,3/2)\oplus2(0,1/2)$\\\hline
$2F+2E-M$&$(3/2,5/2)\oplus(1,3)\oplus3(1,2)\oplus3(1/2,5/2)\oplus8(1/2,3/2)\oplus6(0,2)\oplus2(1,1)\oplus7(0,1)\oplus3(1/2,1/2)\oplus(0,0)$\\\hline
$2F+2E$&$(3/2,3)\oplus(3/2,2)\oplus3(1,5/2)\oplus(1/2,3)\oplus4(1,3/2)\oplus8(1/2,2)\oplus3(0,5/2)\oplus(1,1/2)\oplus8(1/2,1)\oplus10(0,3/2)\oplus(1/2)\oplus5(0,1/2)$\\\hline
$2F+2E+M$&$(3/2,5/2)\oplus(1,3)\oplus3(1,2)\oplus3(1/2,5/2)\oplus8(1/2,3/2)\oplus6(0,2)\oplus2(1,1)\oplus7(0,1)\oplus3(1/2,1/2)\oplus(0,0)$\\\hline
$2F+2E+2M$&$(1,5/2)\oplus(1,3/2)\oplus3(1/2,2)\oplus(0,5/2)\oplus3(1/2,1)\oplus4(0,3/2)\oplus2(0,1/2)$\\\hline
$2F+2E+3M$&$(1/2,3/2)\oplus(0,2)\oplus(0,1)$\\
\hline
\end{tabular}
\begin{tabular}{|l|p{11.5cm}|}
\hline
$2F+3E-4M$&$(1/2,2)\oplus(1/2,1)\oplus2(0,3/2)$\\\hline
$2F+3E-3M$&$(3/2,5/2)\oplus(1,3)\oplus3(1,2)\oplus3(1/2,5/2)\oplus8(1/2,3/2)\oplus6(0,2)\oplus2(1,1)\oplus7(0,1)\oplus3(1/2,1/2)\oplus(0,0)$\\\hline
$2F+3E-2M$&$(2,7/2)\oplus(2,5/2)\oplus3(3/2,3)\oplus(1,7/2)\oplus4(3/2,2)\oplus10(1,5/2)\oplus4(1/2,3)\oplus(3/2,1)\oplus12(1,3/2)\oplus20(1/2,2)\oplus7(0,5/2)\oplus5(1,1/2)\oplus20(1/2,1)\oplus24(0,3/2)\oplus4(1/2,0)\oplus11(0,1/2)$\\\hline
$2F+3E-M$&$(5/2,7/2)\oplus(2,4)\oplus3(2,3)\oplus3(3/2,7/2)\oplus11(3/2,5/2)\oplus11(1,3)\oplus2(1/2,7/2)\oplus2(2,2)\oplus26(1,2)\oplus23(1/2,5/2)\oplus4(0,3)\oplus7(3/2,3/2)\oplus2(3/2,1/2)\oplus46(1/2,3/2)\oplus31(0,2)\oplus19(1,1)\oplus4(1,0)\oplus36(0,1)\oplus23(1/2,1/2)\oplus9(0,0)$\\\hline
$2F+3E$&$(5/2,4)\oplus(5/2,3)\oplus3(2,7/2)\oplus(3/2,4)\oplus4(2,5/2)\oplus10(3/2,3)\oplus4(1,7/2)\oplus(2,3/2)\oplus14(3/2,2)\oplus26(1,5/2)\oplus12(1/2,3)\oplus(0,7/2)\oplus6(3/2,1)\oplus31(1,3/2)\oplus48(1/2,2)\oplus18(0,5/2)\oplus(3/2,0)\oplus14(1,1/2)\oplus48(1/2,1)\oplus50(0,3/2)\oplus12(1/2,0)\oplus26(0,1/2)$\\\hline
$2F+3E+M$&$(5/2,7/2)\oplus(2,4)\oplus3(2,3)\oplus3(3/2,7/2)\oplus11(3/2,5/2)\oplus11(1,3)\oplus2(1/2,7/2)\oplus2(2,2)\oplus26(1,2)\oplus23(1/2,5/2)\oplus4(0,3)\oplus7(3/2,3/2)\oplus2(3/2,1/2)\oplus46(1/2,3/2)\oplus31(0,2)\oplus19(1,1)\oplus4(1,0)\oplus36(0,1)\oplus23(1/2,1/2)\oplus9(0,0)$\\\hline
$2F+3E+2M$&$(2,7/2)\oplus(2,5/2)\oplus3(3/2,3)\oplus(1,7/2)\oplus4(3/2,2)\oplus10(1,5/2)\oplus4(1/2,3)\oplus(3/2,1)\oplus12(1,3/2)\oplus20(1/2,2)\oplus7(0,5/2)\oplus5(1,1/2)\oplus20(1/2,1)\oplus24(0,3/2)\oplus4(1/2,0)\oplus11(0,1/2)$\\\hline
$2F+3E+3M$&$(3/2,5/2)\oplus(1,3)\oplus3(1,2)\oplus3(1/2,5/2)\oplus8(1/2,3/2)\oplus6(0,2)\oplus2(1,1)\oplus7(0,1)\oplus3(1/2,1/2)\oplus(0,0)$\\\hline
$2F+3E+4M$&$(1/2,2)\oplus(1/2,1)\oplus2(0,3/2)$\\
\hline
\end{tabular}

\bibliography{references}

\end{document}